\documentclass[floatfix]{emulateapj}

\usepackage{amsmath}
\usepackage[colorlinks = true,linkcolor = blue, citecolor = blue]{hyperref}
\usepackage{natbib}
\usepackage{multirow}
\usepackage{enumitem}
\usepackage{placeins}

\newcommand{\Lsolar}{\mbox{$L_{\odot}\,$}}

\shorttitle{Chemically Distinct Nuclei in Arp\,220}

\begin{document}

\title{Chemically Distinct Nuclei and Outflowing Shocked Molecular Gas in Arp\,220}

\author{R. Tunnard\altaffilmark{1}, T.\,R. Greve\altaffilmark{1}, S. Garcia-Burillo\altaffilmark{2}, J. Graci\'{a} Carpio\altaffilmark{3}, J. Fischer\altaffilmark{4}, A. Fuente\altaffilmark{2}, E. Gonz\'{a}lez-Alfonso\altaffilmark{5}, S. Hailey-Dunsheath\altaffilmark{3}, R. Neri\altaffilmark{5}, E. Sturm\altaffilmark{3}, A. Usero\altaffilmark{2} and P. Planesas\altaffilmark{2}}
\altaffiltext{1}{Department of Physics and Astronomy, University College London, Gower Street, London WC1E 6BT, UK; \email{richard.tunnard.13@ucl.ac.uk}}
\altaffiltext{2}{Observatorio Astron\'{o}mico Nacional, Observatorio de Madrid, Alfonso XII, 3, 28014 Madrid, Spain}
\altaffiltext{3}{Max-Planck-Institute for Extraterrestrial Physics (MPE), Giessenbachstra\ss e 1, 85748 Garching, Germany}
\altaffiltext{4}{Naval Research Laboratory, Remote Sensing Division, 4555 Overlook Ave SW, Washington, DC 20375, USA}
\altaffiltext{5}{Universidad de Alcal\'{a} de Henares, Departamento de F\'{i}sica y Matem\'{a}ticas, Campus Universitario, 28871 Alcal\'{a} de Henares, Madrid, Spain}
\altaffiltext{6}{IRAM, 300 rue de la Piscine, Domaine Universitaire, 38406 Saint Martin d'H\`{e}res Cedex, France}


\begin{abstract}
We present the results of interferometric spectral line observations of Arp\,220 at 3.5\,mm and 1.2\,mm from the Plateau de Bure Interferometer (PdBI), imaging the two nuclear disks in H$^{13}$CN$(1 - 0)$ and $(3 - 2)$,  H$^{13}$CO$^+(1 - 0)$ and $(3 - 2)$, and HN$^{13}$C$(3 - 2)$ as well as SiO$(2 - 1)$ and $(6 - 5)$, HC$^{15}$N$(3 - 2)$, and SO$(6_6 - 5_5)$. The gas traced by SiO$(6 - 5)$ has a complex and extended kinematic signature including a prominent P Cygni profile, almost identical to previous observations of HCO$^+(3 - 2)$. Spatial offsets $0.1''$ north and south of the continuum centre in the emission and absorption of the SiO$(6 - 5)$ P Cygni profile in the western nucleus (WN) imply a bipolar outflow, delineating the northern and southern edges of its disk and suggesting a disk radius of $\sim40$\,pc, consistent with that found by ALMA observations of Arp\,220. We address the blending of SiO$(6 - 5)$ and H$^{13}$CO$^+(3 - 2)$ by considering two limiting cases with regards to the H$^{13}$CO$^+$ emission throughout our analysis. Large velocity gradient (LVG) modelling is used to constrain the physical conditions of the gas and to infer abundance ratios in the two nuclei. Our most conservative lower limit on the [H$^{13}$CN]/[H$^{13}$CO$^+$] abundance ratio is 11 in the WN, cf.\ 0.10 in the eastern nucleus (EN). Comparing these ratios to the literature we argue on chemical grounds for an energetically significant AGN in the WN driving either X-ray or shock chemistry, and a dominant starburst in the EN.
\end{abstract}
  
\FloatBarrier


\section{Introduction}\label{sec:intro}

Local ultra luminous infrared galaxies (ULIRGs, defined as $L_{\rm 8-1000\mu m} > 10^{12}\,\Lsolar$) as revealed by IRAS \citep[][see also \citealp{Sanders1996,lonsdale2006ultraluminous}]{Soifer1987} have been a topic of great interest since their discovery as they are the products of gas rich major mergers with extreme dust obscurations, star formation rates ($\sim100$\,M$_{\odot}$\,yr$^{-1}$) and molecular gas masses relative to other populations in the local Universe. 

Early suspicions that the majority of ULIRG luminosity $(30\% - 100\%)$ arises from extremely compact structures on the scale of 100 - 300\,pc were confirmed by interferometric CO observations and high resolution mid-IR imaging \citep[e.g.][]{Downes1998,Soifer2000}, as well as from the analysis of molecular absorption lines in the far-IR \citep[][and references therein, hereafter GA12]{Gonzalez-Alfonso2012}. In many cases this compact source is found to be a dust enshrouded active galactic nuclei (AGN) \citep[e.g.][]{Soifer1999,lonsdale2006ultraluminous}. Indeed, in Seyfert I type ULIRGs the AGN contribution to the bolometric luminosity can be as high as 75\% \citep{Veilleux2009}. AGN have been known for some time to play a pivotal role in galaxy evolution from the observed correlation between the mass of the central supermassive black hole and that of its host spheroid \citep{Magorrian1998,Ferrarese2000,Tremaine2002}. AGN feedback is thought to be responsible for the observed sharp cutoff at the upper end of the galactic mass function \citep{Silk1998}. While AGN provide an excess of energy for quenching star formation \citep[$\sim80\times$ the gravitational binding energy of their host galaxies:][]{Fabian2012}, it is uncertain how this energy is coupled to the ISM and surrounding dust, and in particular whether there is sufficient momentum coupling to evacuate the central regions of gas. Nevertheless, massive molecular outflows from AGN residing in local ULIRGs have recently been observed, \citep[][and references therein]{Cicone2014}, while ionised outflows have been found to be not only ubiquitous, but to exist with sufficient outflow rates to shape galaxy evolution \citep{Harrison2014}. These results have confirmed that AGN can drive outflows in excess of the host starburst, but that frequently most of the mass in these outflows is moving slower than escape velocity, perhaps leading to future bursts of star formation as the gas fountains back into the host. Feedback from star formation in the form of stellar winds and supernovae also has a significant role to play in galaxy evolution, although this is predominantly in low mass galaxies where it is responsible for the low-mass turnover in the the galactic mass function. Nevertheless, it may significantly contribute to feedback and outflows in massive galaxies by heating of the ISM \citep{Efstathiou2000}.

\begin{figure*}[!htb]
	\begin{center}
		\includegraphics[width=\textwidth]{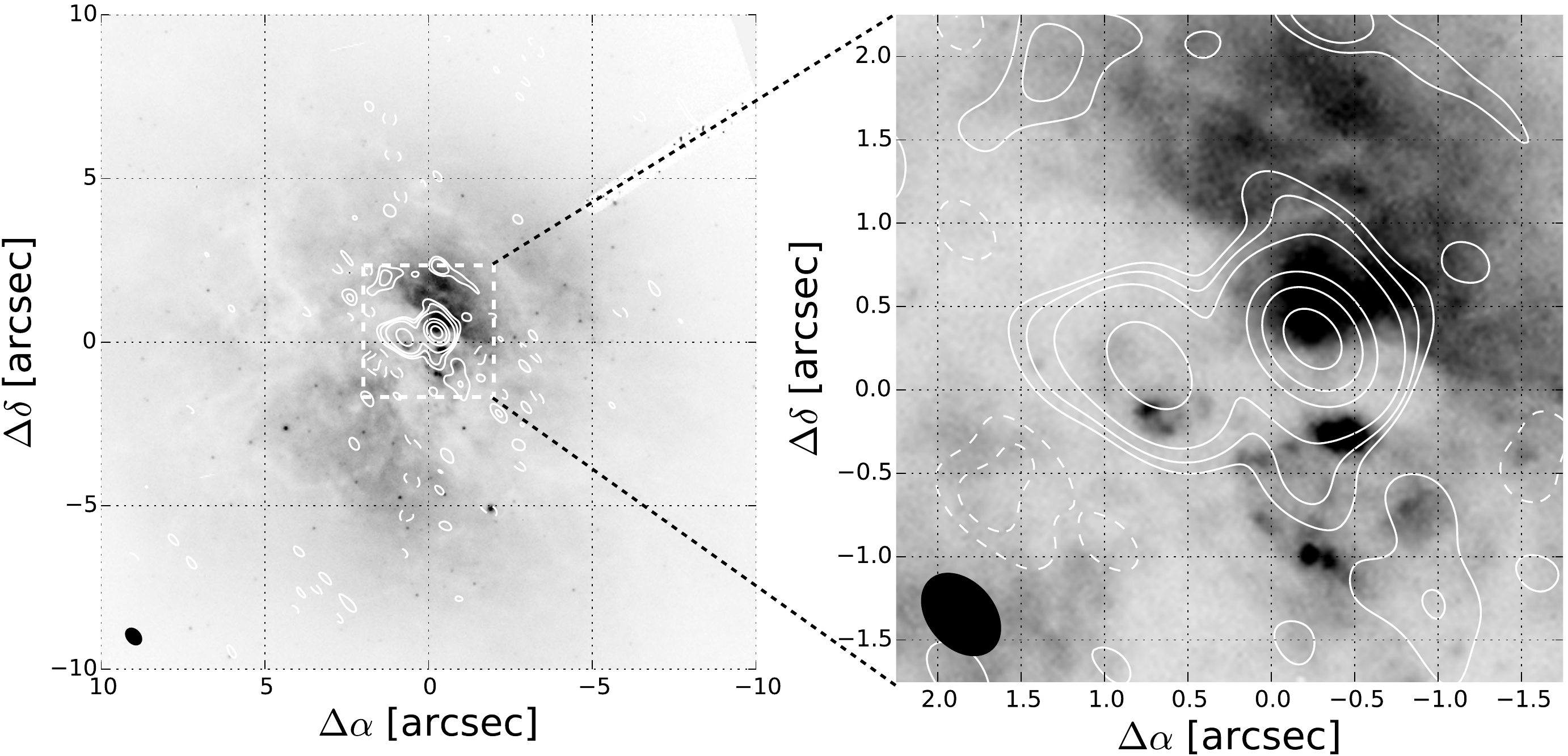}
		\caption{Hubble ACS/HRC F814W\footnote{F814W is a Hubble wideband filter centred on 814\,nm covering the $I$ band.} image with the 1.2\,mm continuum contours overlaid. The compact red emission demonstrates the highly obscured nature of the central regions of Arp\,220. The right hand image shows a $4''\times4''$ extract centring on the millimetre continua. {The 1.2\,mm continuum contours (in white) are overlaid at $-5$, $-3$, 3, 5, 10, 50, 100, 200 and 300$\sigma$ and the 1.2\,mm beam $(0.52''\times0.37'')$ is shown in the bottom left hand corners.} The HST image was created with the help of the ESA/ESO/NASA FITS Liberator, and has a nominal astrometric accuracy of $0.3''$.}\label{fig:master}
	\end{center}
\end{figure*}

The prototypical ULIRG Arp\,220 $(z = 0.01818)$ has been extensively studied across the entire EM-spectrum, revealing an extraordinarily dense and massive galactic merger \citep{Soifer1984,Emerson1984,Young1984,Scoville1991,Mcdowell2003}. It is a late-stage merger with two high column density ($N_{\text{H}_2}\gtrsim 1\times10^{25}$\,cm$^{-2}$), counter-rotating disks (separation $\simeq400$\,pc along the east-west axis) orbiting within a much larger gas disk (radius $\simeq0.25-1.5$\,kpc, average molecular gas density $\sim10^3$\,cm$^{-3}$), potentially with an extended galactic halo \citep{Sakamoto1999,Mundell2001,Sakamoto2008,Engel2011,Gonzalez-Alfonso2012}. Arp\,220 is extremely dusty with $A_\text{V}>1000$\,mag towards the central nuclear region \citep{Fischer1997}, and possesses at least three distinct stellar populations: a 10\,Myr old starburst, a $\gtrsim1$\,Gyr old population and a nascent, $\lesssim7$\,Myr old non-starburst component, all distributed across the kiloparsec scale merger\citep{Wilson2006,RodriguezZaurin2008,Engel2011}. 

There is mounting evidence for an AGN in the western nucleus (WN) of Arp\,220 \citep[and references therein]{Downes2007} as well as a nuclear starburst revealed with Very Long Baseline Interferometry (VLBI) \citep{RodriguezZaurin2008,Batejat2011}. In particular, the mass, luminosity and dust temperatures have previously been used in attempts to identify an AGN. \citet{Sakamoto2008} and \citet{Engel2011} found dynamical masses in the WN and EN from CO and stellar kinematics respectively. Within 100\,pc of the continuum centres of each nuclei they agree that the masses are on the order of $\sim1\times10^9\,M_{\odot}$, although \citet{Engel2011} found masses closer to $\sim5\times10^9\,M_{\odot}$. When \citet{Wilson2014} combined these results with their own high resolution ALMA observations of the nuclear continua, they obtained luminosity-mass ratios $540\,L_{\odot}\,M_{\odot}^{-1}$ and $30\,L_{\odot}\,M_{\odot}^{-1}$ for the WN and EN respectively: they conclude that the high WN ratio could be an AGN, or a hot starburst. Dust temperatures in the WN have been found from millimetre continuum measurements to be $\sim180-200$\,K, once accounting for optical depth effects \citep{Downes2007,Wilson2014}, while far-IR absorption lines suggest temperatures as high as 400\,K (GA12). Such temperatures are usually associated with dusty toroids around AGN \citep[e.g.][]{Weiss2007}. Attempts to identify an AGN from X-ray observations have also been unsuccessful, but due to the Compton thick H$_2$ column density $(\sim1\times10^{25}$\,cm$^{-2})$ towards the WN any X-ray emission would be almost completely attenuated \citep{Wilson2014}. Molecular outflows have been observed in OH$^+$, H$_2$O$^+$ and H$_2$O by \citet{Rangwala2011,Gonzalez-Alfonso2012} and in CO$(3 - 2)$, HCO$^+(4 - 3)$ and $(3 - 2)$  by \citet{Sakamoto2009} (hereafter Sa09), although these could be driven by either an AGN or a starburst. \citet{Mangum2013} confirmed a hot, dense gas phase through NH$_3$ thermometry, finding solutions including $T_{\text{k}}\sim300$\,K at $n_{\text{H}_2}\sim10^6$\,cm$^{-3}$.

Sa09 observed the HCO$^+(3 - 2)$, $(4 - 3)$ and CO$(3 - 2)$ lines at high resolution ($\sim0.3''$) with the {Sub-Millimetre Array} (SMA), identifying prominent P Cygni profiles in both nuclei in the HCO$^+$ lines as well as absorption about the CO$(3 - 2)$ line centre. All of their WN lines displayed redshifted emission out to  $+500$\,km\,s$^{-1}$, while the HCO$^+$ lines showed maximum absorption around $-100$\,km\,s$^{-1}$, but with some signs of blueshifted absorption out to $-500$\,km\,s$^{-1}$, mixed with some blueshifted emission. The HCO$^+(4 - 3)$ lines in both nuclei presented much cleaner P Cygni profiles than the $(3 - 2)$ lines, although they still showed extended redwards emission. 

The warm, dense dust in the WN (Sa09 found $T_{\text{d}} \simeq 130$\,K, \citealp{Matsushita2009} suggested $T_{\text{d}}$ could be as high as 310\,K and optically thick at 435\,$\mu$m while GA12 found $T_{\text{d}} \simeq 90 - 120$\,K and that the dust could be optically thick {out} to 2.4\,mm) provides a powerful radiation source against which absorption in molecular outflows can be clearly seen. In particular Sa09 and \citet{Papadopoulos2010} noted that the unusually high dust temperature favours the observation of higher frequency rotational transitions in absorption, and in the case of CO$(6 - 5)$ \citet{Papadopoulos2010} found that the optical thickness of the dust in Arp\,220 suppresses emission in higher transitions. 

In this work, we present a P Cygni profile in SiO($6 - 5$) towards the western nucleus (WN) of Arp\,220. We also report PdBI observations of SiO$(2 - 1)$, H$^{13}$CN and H$^{13}$CO$^+$ $(3 - 2)$ and $(1 - 0)$; HN$^{13}$C and HC$^{15}$N $(3 - 2)$ and SO$(6_6 - 5_5)$. The lower optical depths of the $^{13}$C isotopologues provide a better measure of abundances than the $^{12}$C isotopologues, while the millimetre observations allow us to see through much of the dust enshrouding the nuclear disks of Arp\,220. This combination of effects allows us to probe deeper into the driving regions of Arp\,220 than has previously been possible. The observations are outlined in section \ref{sec:obs} and the continuum and line identification are discussed in section \ref{sec:lines}. We outline the details of the lines in each nuclei in sections \ref{sec:wn} and \ref{sec:en}, before entering into a more complete analysis in section \ref{sec:analysis}. Finally we discuss the significance of our results in section \ref{sec:discuss}.

We assume a flat universe with the 2013 Planck cosmological parameters \citep{Planck2013}: $h$ = 0.673, $\Omega_m$ = 0.315 and $\Omega_{\Lambda}$ = 0.685. For Arp\,220 at $z_{\text{co}}=0.01818$ ($V_{\text{co}}(\text{WN}) = 5355\pm15$\,km\,s$^{-1}$, $V_{\text{co}}(\text{EN}) = 5415\pm15$\,km\,s$^{-1}$, radio, LSR, Sa09) this gives a scale of 384\,pc/$''$ and a luminosity distance $D_{\text{L}}=82.1$\,Mpc \citep{Wright2006}. Our longest baseline for the 1.2\,mm data was 382.5 kilowavelengths {and 218.1 kilowavelengths at 3.5\,mm}, giving a maximum resolution $\simeq0.37''$, or $\sim$140\,pc at 1.2\,mm and $\simeq0.70''$, or $\sim$270\,pc at 3.5\,mm.

\section{Observations and Data Reduction}\label{sec:obs}

We obtained two datasets from the IRAM Plateau de Bure Interferometer (PdBI) at 3.5\,mm (2007 February 1 and 2007 March 14 in the A configuration (beam $1.44''\times0.70''$, $PA=22^{\circ}$, maximum angular scale $\sim5.7''$)) and one at 1.2\,mm (2011 January 11 in the B configuration (beam $0.52''\times0.37''$, $PA=44^{\circ}$, maximum angular scale $\sim9.1''$)). All six antennas were available for all observations. Total on source time for 3.5\,mm was 17550\,s and 21690\,s for 1.2\,mm. 

For both 2007 observations the bandpass was classed as excellent, with excellent and good amplitude calibrations in the February and March observations respectively. The correlator was centred on 85.0474\,GHz sky frequency, corresponding to 85.2783\,GHz in the $z_{\text{co}}=0.01818$ rest frame, and had a bandwidth of 965\,MHz. Bandpass calibrators were 3C273 and 3C454.3, amplitude and phase calibration used 1611+343 and radio seeing was 0.30$''$. The data were reduced in CLIC, and the UV tables produced were analysed and mapped using the MAPPING program in the IRAM proprietary software GILDAS\footnote{\url{http://www.iram.fr/IRAMFR/GILDAS}} \citep{Guilloteau2000}. {These observations targeted SiO$(2 - 1)$, H$^{13}$CN$(1 - 0)$ and H$^{13}$CO$^+(1 - 0)$ in both nuclei.}

The 1.2\,mm data used the full 3.6\,GHz bandwidth of the WideX correlator, centred on 255.127\,GHz sky frequency, corresponding to 259.765\,GHz in the $z_{\text{co}}=0.01818$ rest frame, with spectral channels of width 3.9\,MHz ($\sim$4.59\,km\,s$^{-1}$), with the aim of mapping H$^{13}$CN$(3 - 2)$, H$^{13}$CO$^+(3 - 2)$ and SiO$(6 - 5)$. The observations were phase centred at RA: $15^{\text{h}}34^{\text{m}}57^{\text{s}}24$, Dec: $23^{\circ}30'11''2$. The data were imaged from UV tables using natural weighting. 

For both frequencies we estimate a 10\% uncertainty in the absolute flux calibration. The Hogbom CLEAN algorithm \citep{Hogbom1974} was used to produce data cubes, cleaning down to a flux level of $\sim$0.5\,$\sigma$. We obtain rms.\ noise levels of 3.3\,mJy\,beam$^{-1}$ at 1.2\,mm with channels of 4.59\,km\,s$^{-1}$ and 1.60\,mJy\,beam$^{-1}$ at 3.5\,mm with channels of 8.81\,km\,s$^{-1}$. 

\section{Results}\label{sec:lines}
\subsection{Continuum}

Interpretation of the line profiles and implied kinematics requires knowledge of the dust properties of the source, and in particular to what extent dust is obscuring our observations. The continuum emission is dominated by dust, so provides a means of studying the dust properties in Arp\,220. Furthermore, we must confirm that our detected continuum is consistent with previous observations to ensure that the sub-continuum absorption is real. The continuum analysis is summarised below, and additional details are included in {appendix \ref{sec:cont_details}.}

The 1.2\,mm continuum was obtained from the central 443\,km\,s$^{-1}$ wide line free region from $259.388 - 259.765$\,GHz, WN rest frame, and the 3.5\,mm continuum was found from the central 511\,km\,s$^{-1}$ wide line free region from $85.925 - 86.623$\,GHz, WN rest frame. These regions are line free in both nuclei. Continuum contours for the 1.2\,mm nuclei are shown in Fig.\ \ref{fig:master}. {The two nuclei are at best marginally resolved, so we worked in the UV plane to constrain the continuum parameters of the two nuclei. Fitting elliptical gaussians with the GILDAS task UV\_FIT} we found 1.2\,mm continuum centres and sizes, and fluxes (RA=$15^{\text{h}}34^{\text{m}}57^{\text{s}}220$, Dec=$23^{\circ}30'11''50$, FWHM=$0.19''\times0.15''$, $S_{\rm 1.2\,mm}=200\pm20$\,mJy and $S_{\rm 3.5\,mm}=17\pm2$\,mJy for the WN and RA=$15^{\text{h}}34^{\text{m}}57^{\text{s}}291$, Dec=$23^{\circ}30'11''34$, FWHM=$0.26''\times0.18''$, $S_{\rm 1.2\,mm}=71\pm7$\,mJy and $S_{\rm 3.5\,mm}=11\pm1$\,mJy for the EN). {These results and comparable measurements from the literature are included in appendix \ref{sec:cont_details} (Table \ref{tab:continuum}).} Our fitted continuum centres at 1.2\,mm are precisely in agreement with those that \citet{Wilson2014} found at much higher resolution $(0.363''\times0.199")$ with ALMA. Our 1.2\,mm flux densities are consistent with the 1.1\,mm flux densities of Sa09 who found 200\,mJy and 70\,mJy with 15\% uncertainties. Subtracting synchrotron emission we find flux densities $196\pm20$\,mJy and $69\pm7$\,mJy for the WN and EN at 1.2\,mm and $8\pm2$\,mJy and $6\pm1$\,mJy for the WN and EN at 3.5\,mm.

The {global (i.e., WN and EN, combined)} dust optical depth and spectral index found from SED fitting of our continuum measurements and data from the literature were $\tau_{258{\rm GHz}}= 0.06$ and $\beta = 1.5$, {with a turnover frequency $\sim1600$\,GHz. This is similar to the global fit of \citet{Rangwala2011}, who found a turnover frequency of 1277\,GHz and  $\beta = 1.8$. However, compared with the four component SED fits of GA12, who for} $\beta_{\text{WN}}=1 - 2$ predict $\tau_{258{\rm GHz}}= 2 - 0.4$ and $\tau_{86{\rm GHz}}= 0.7 - 0.04$ in the WN, our value is low, most likely due to fitting only two components (GA12 used four). Due to the more sophisticated modelling of GA12 we use their dust optical depths.

\begin{turnpage}
	\begin{figure*}
		\centering
		\includegraphics[width=1.33\textwidth,height=0.7\textheight]{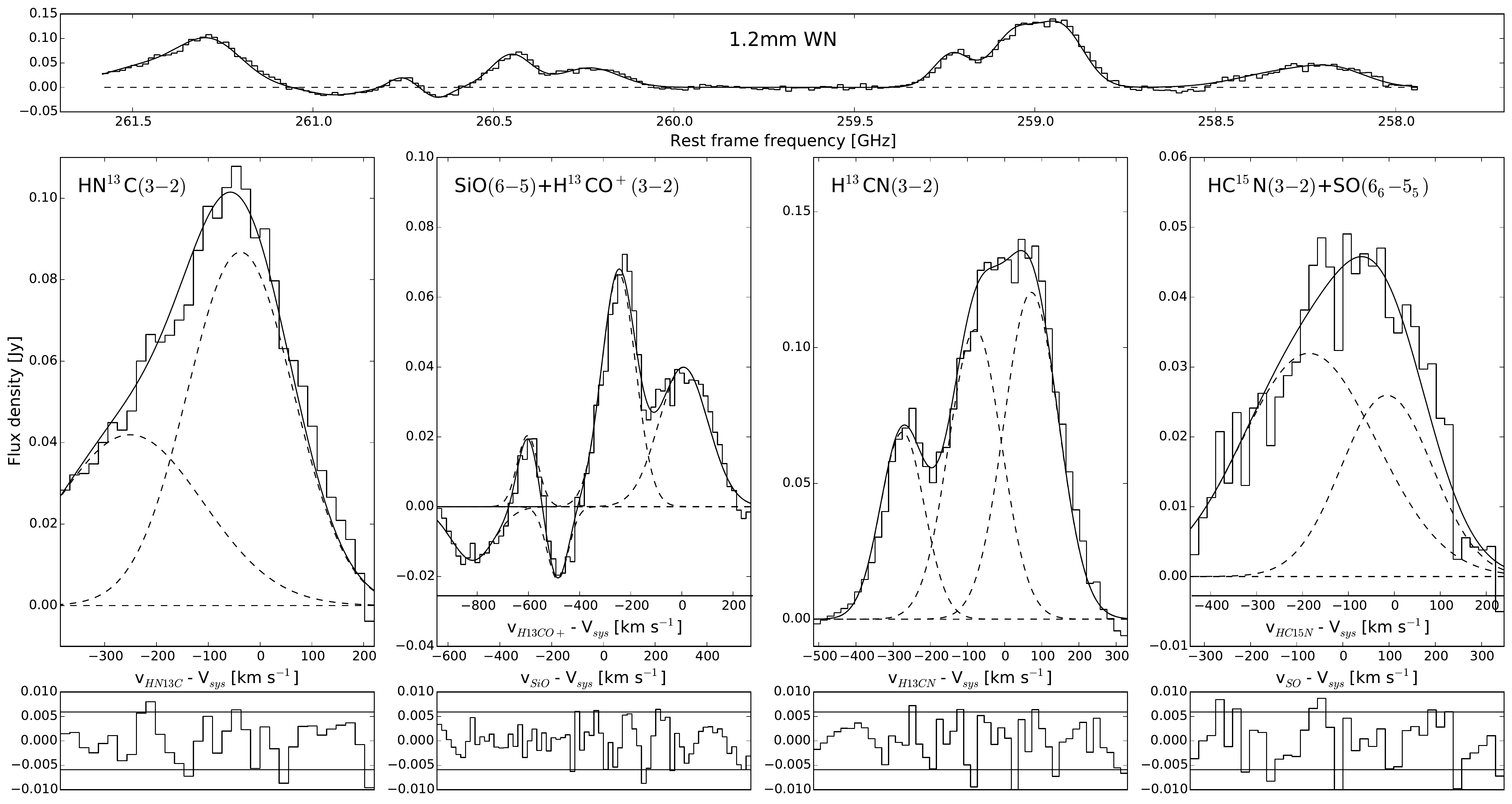}
		\caption{Continuum subtracted spectrum of Arp\,220 WN extracted using a $1''$ diameter circular apertures centred on the continuum centre. Top: the whole spectrum with the fitted lines overlaid. Frequencies are rest frame with $z=0.01818$. Middle: multiple component Gaussian fits to the observed lines, with velocities shown relative to the line centres. Bottom: residuals from the fitting. The horizontal lines indicate the $\pm1\sigma$ rms.\ noise level (5.9\,mJy). Reduced $\chi^2$ values range from 0.25 to 0.7. Full details of fitting components are included in the appendix.}\label{fig:spec1}
	\end{figure*}
\end{turnpage}

\begin{turnpage}
	\begin{figure*}
		\centering
		\includegraphics[width=1.33\textwidth,height=0.7\textheight]{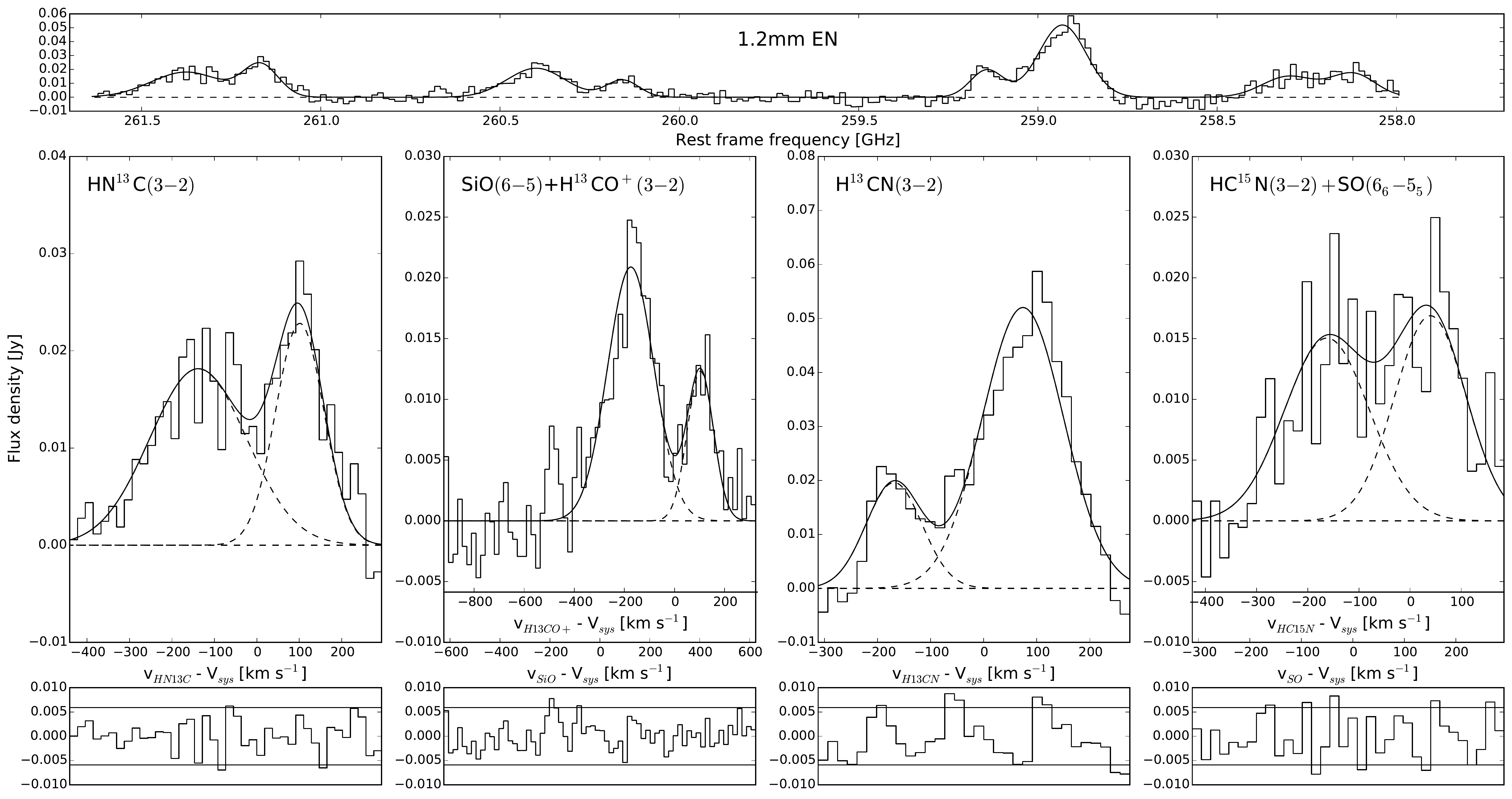}
		\caption{Continuum subtracted spectrum of Arp\,220 EN extracted using a $1''$ diameter circular apertures centred on the continuum centre. Top: the whole spectrum with the fitted lines overlaid. Frequencies are rest frame with $z=0.01839$. Middle: multiple component Gaussian fits to the observed lines, with velocities shown relative to the line centres. Bottom: residuals from the fitting. The horizontal lines indicate the $\pm1\sigma$ rms.\ noise level (5.9\,mJy). Reduced $\chi^2$ values range from 0.25 to 0.7. Full details of fitting components are included in the appendix.}\label{fig:spec2}
	\end{figure*}
\end{turnpage}

\begin{figure*}
	\begin{center}
		\includegraphics[width=\textwidth]{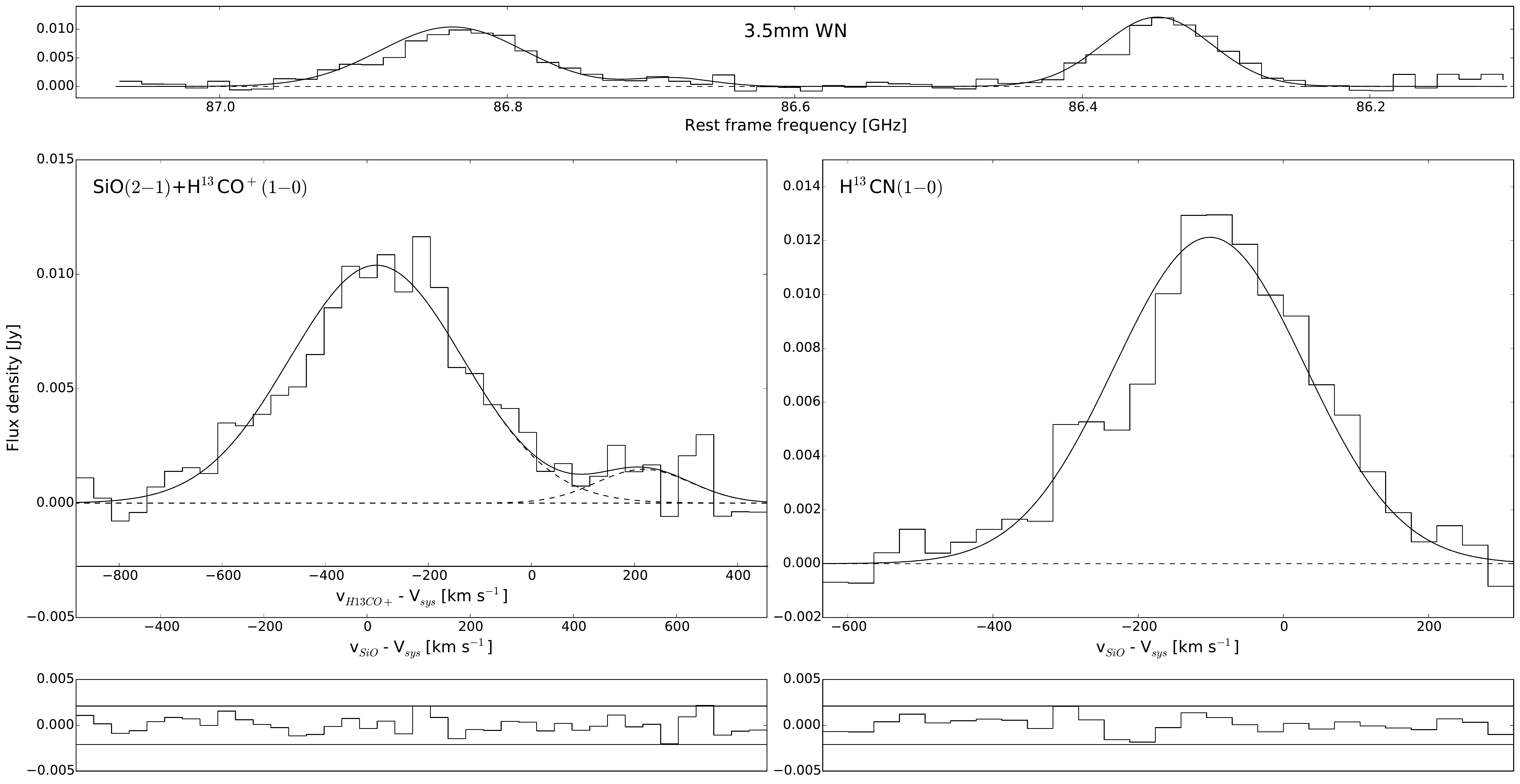}
		\caption{Continuum subtracted spectrum of Arp\,220 WN extracted using elliptical apertures with major and minor axes equal to scaled FWHMs of our 3.5\,mm beam, centred so as to minimise inter-nuclear contamination (RA: $15^{\text{h}}34^{\text{m}}57^{\text{s}}226$, Dec: $23^{\circ}30'11''613$, scaling = 1.8). Top: the whole spectrum with the fitted lines overlaid. Frequencies are rest frame with $z=0.01818$. Middle: multiple component Gaussian fits to the observed lines, with velocities shown relative to the line centres. Bottom: residuals from the fitting. The horizontal lines indicate the $\pm1\sigma$ rms.\ noise level (2.1\,mJy). Reduced $\chi^2$ values range from 0.4 to 0.5. Full details of fitting components are included in the appendix.}\label{fig:spec3}
		\includegraphics[width=\textwidth]{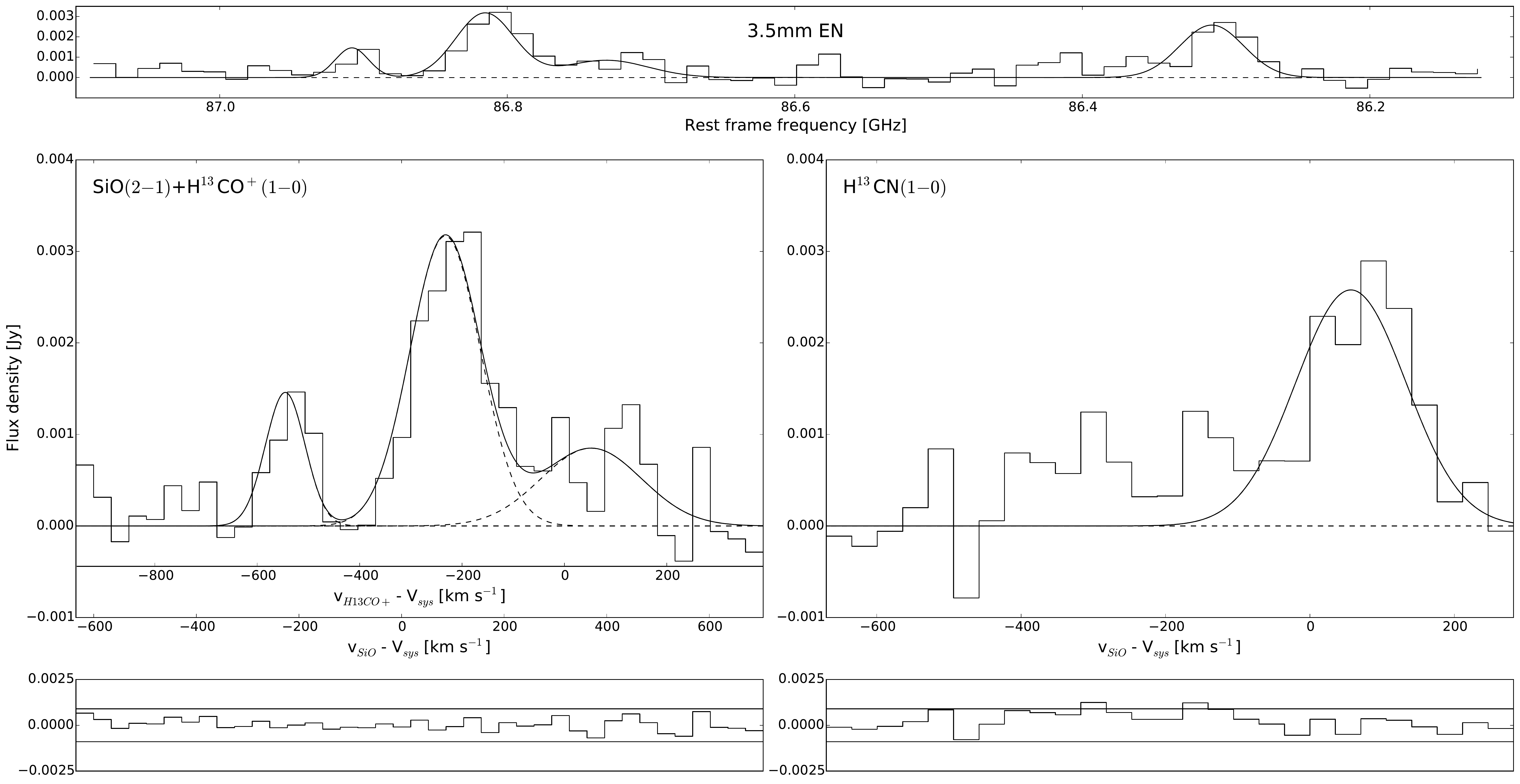}
		\caption{Continuum subtracted spectrum of Arp\,220 EN extracted using elliptical apertures with major and minor axes equal to scaled FWHMs of our 3.5\,mm beam, centred so as to minimise inter-nuclear contamination (RA: $15^{\text{h}}34^{\text{m}}57^{\text{s}}302$, Dec: $23^{\circ}30'11''415$, scaling = 1.2). Top: the whole spectrum with the fitted lines overlaid. Frequencies are rest frame with $z=0.01839$. Middle: multiple component Gaussian fits to the observed lines, with velocities shown relative to the line centres. Bottom: residuals from the fitting. The horizontal lines indicate the $\pm1\sigma$ rms.\ noise level (0.9\,mJy). Reduced $\chi^2$ values range from 0.2 to 0.5. Full details of fitting components are included in the appendix.}\label{fig:spec4}
	\end{center}
\end{figure*}

\subsection{Line Identification}\label{subsec:blending}

Within the bandwidth of the WideX correlator we detected a plethora of wide, high S/N lines in emission towards both nuclei of Arp\,220, {the frequencies and upper level energies of which are given in Table \ref{tab:line-info}. The continuum subtracted spectra, including the multiple component Gaussian line fits}, are shown in Figs.\ \ref{fig:spec1} to \ref{fig:spec4}. The {fitting derived} line parameters are {given} in Table \ref{tab:line-data}, and an {outline of the fitting procedure, including a full list of the fitted components, is included in appendix \ref{app:B}}. 

Lines were identified using the Cologne Database for Molecular Spectroscopy (CDMS) \citep{CDMS} and the Spectral Line Atlas of Interstellar Molecules (SLAIM) \citep{SLAIM}, as well as the SMA line survey of \citet{Martin2011}. Our final line list consists of H$^{13}$CN$(1 - 0)$ and $(3 - 2)$, HN$^{13}$C$(3 - 2)$, SiO$(2 - 1)$ and $(6 - 5)$, H$^{13}$CO$^+(1 - 0)$ and $(3 - 2)$, SO$(6_6 - 5_5)$ and HC$^{15}$N$(3 - 2)$. The WN HN$^{13}$C$(3 - 2)$ line extends beyond the end of our spectrum. We see the blue edge of what is likely the SO$(2_2 - 1_1)$ in the 3.5\,mm spectra, but almost the entire line is beyond the red edge of the spectrum. The relative contributions of SO$(6_6 - 5_5)$ and HC$^{15}$N$(3 - 2)$ in the 1.2\,mm spectrum are highly uncertain. 

\begin{table}
\centering
\caption{Observed Lines}\label{tab:line-info}
\begin{tabular}{l c c}\hline\hline
\multirow{2}{*}{Line} & $\nu$\footnote{Rest frame frequency.} & E$_\text{u}/$k \\
& [GHz] & [K] \\ \hline\hline
H$^{13}$CN$(1 - 0)$ & 86.340 & 4.1\\
H$^{13}$CO$^+(1 - 0)$ & 86.754 & 4.2\\
SiO$(2 - 1)$ & 86.847 & 6.3\\
HC$^{15}$N$(3 - 2)$ & 258.157 & 24.8 \\
SO$(6_6 - 5_5)$ & 258.256 & 56.5\\
H$^{13}$CN$(3 - 2)$ & 259.012 & 24.9\\
H$^{13}$CO$^+(3 - 2)$ & 260.255 & 25.0\\
SiO$(6 - 5)$ & 260.518 & 43.8\\
HN$^{13}$C$(3 - 2)$ & 261.263 & 25.1 \\\hline
\end{tabular}
\end{table}

\bigskip

Vibrationally excited molecular absorption and emission (including HCN) was detected in Arp\,220 by \citet{Salter2008} and \citet{Martin2011} respectively, suggesting that we may have some contribution from the vibrationally excited H$^{13}$CN$(v_2=1, J=3 - 2, l = 1e \text{ \& } 1f)$ lines at 258.936\,GHz and 260.225\,GHz respectively.  However, the ratios of the predicted line intensities for the $v_2=0$ to $v_2=1$ transitions at 75\,K and 300\,K from the CDMS database \citep{CDMS} are $1:10^{-6}$ and $1:35$ respectively. This suggests that even with considerable mid-IR pumping \citep{Aalto2007,Sakamoto2010,Rangwala2011} the vibrational lines should be below our noise level. Whether the effect of mid-IR pumping on the $v_{1,2}=0$ rotational transitions is significant is  still unclear \citep{Rangwala2011, Gonzalez-Alfonso2012}.

\subsubsection{SiO and H$^{13}$CO$^+$}\label{subsubsec:sio-h13co}

There is significant uncertainty regarding the identification of SiO and H$^{13}$CO$^+$ in our spectra. At {both observing frequencies and in both nuclei} there is the possibility that all of the line components shown in the SiO$+$H$^{13}$CO$^+$ boxes in Figs.\ \ref{fig:spec1}$-$\ref{fig:spec4} are due to SiO, with no significant H$^{13}$CO$^+$ detection. {Therefore, we examine the kinematic, spatial and line brightness evidence in an attempt to constrain the contribution of H$^{13}$CO$^+$ at 3.5\,mm and at 1.2\,mm. While Sa09 provide a comprehensive benchmark with their HCO$^+(3 - 2)$ observations there are no published observations of HCO$^+(1 - 0)$ in Arp\,220, so that the following analysis of H$^{13}$CO$^+$ contribution focusses on the $(3 - 2)$ transition in the 1.2\,mm data.}

\smallskip

\begin{figure}[h!]
\begin{center}
\includegraphics[width=0.5\textwidth]{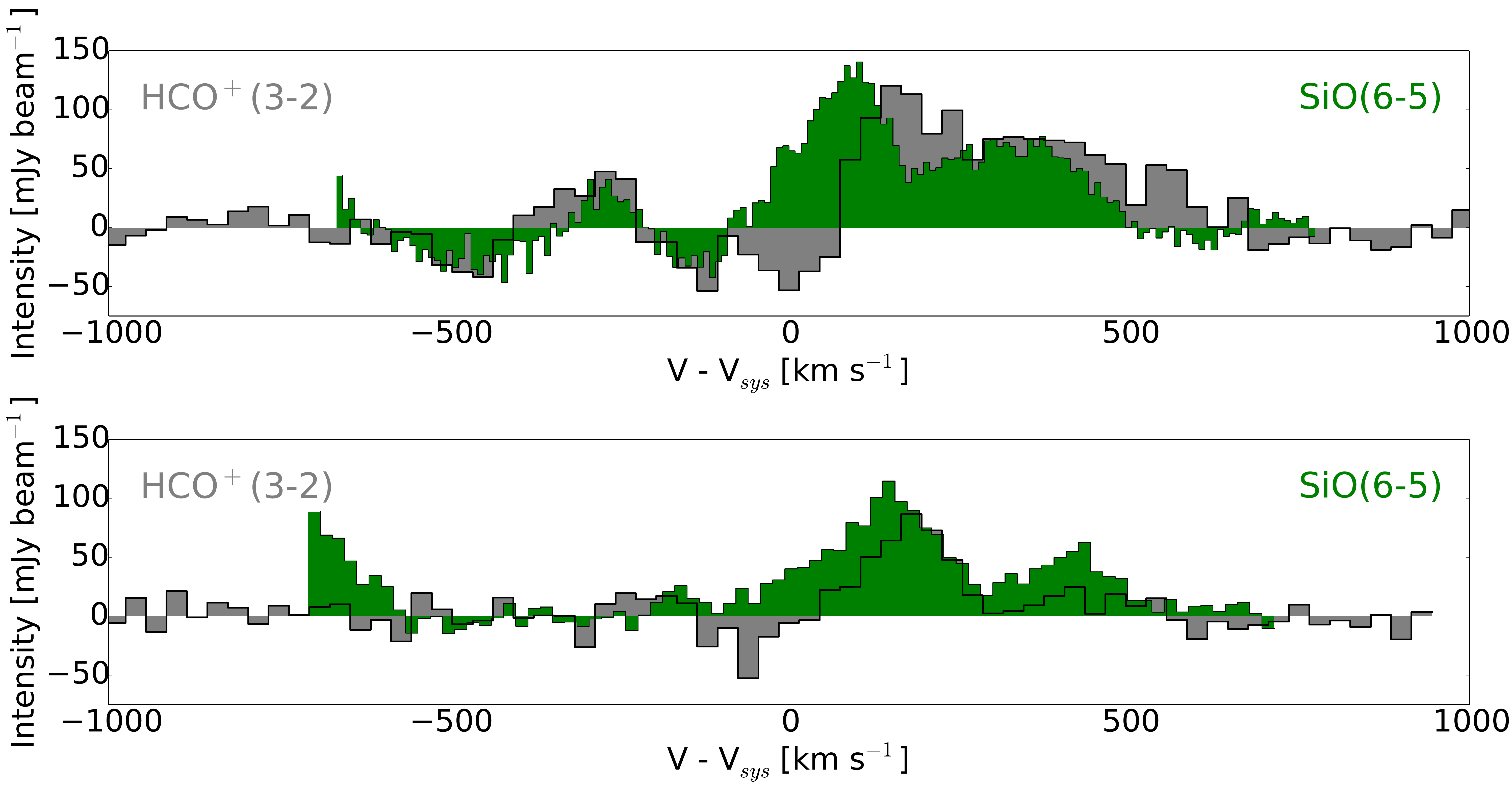}
\caption{The HCO$^+(3 - 2)$ spectra from Sa09, with our SiO$(6 - 5)$-H$^{13}$CO$^+(3 - 2)$ spectra scaled by a factor of 8 and overlaid. The WN lines (top) are centred on $V_{\text{sys}}(\text{WN})=5355$\,km\,s$^{-1}$, the EN (bottom) on $V_{\text{sys}}(\text{EN})=5415$\,km\,s$^{-1}$. The similarity between the profiles, in particular in the WN, suggests that our spectrum may be predominantly SiO$(6 - 5)$, with no more than a minor contribution from the blended H$^{13}$CO$^+(3 - 2)$ component at $+317$\,km\,s$^{-1}$. The H$^{13}$CO$^+(3 - 2)$ line centre is $+$317\,km\,s$^{-1}$ redwards of SiO$(6 - 5)$, so cannot be the dominant contributor given the HCO$^+(3 - 2)$ profile.}\label{fig:sakacomp}
\end{center}
\end{figure}

{Kinematically, a comparison with the HCO$^+(3 - 2)$ spectra of Sa09, shown in Fig.\ \ref{fig:sakacomp}, reveals startlingly similarities between the line profiles of their HCO$^+(3 - 2)$ and those of our SiO$(6 - 5) - $H$^{13}$CO$^+(3 - 2)$ line complex, in both nuclei. Particularly in the WN, the line profiles are extremely similar, with all five of our fitted Gaussian line components in Fig.\ \ref{fig:spec1} having an equivalent in the HCO$^+(3 - 2)$ spectrum.} We rule out that the lines are entirely H$^{13}$CO$^+(3 - 2)$ based on the significant ($\sim320$\,km\,s$^{-1}$) velocity offset from the observed systemic velocity that this would require{, but the kinematics cannot exclude the possibility that the lines are entirely SiO$(6 - 5)$}. 

{In our 3.5\,mm spectra there are $\sim1\sigma$ fitted components, currently attributed to H$^{13}$CO$^+(1 - 0)$ (line centre $+250$\,km\,s$^{-1}$ in the WN and line centre $+50$\,km\,s$^{-1}$ in the EN, see Figs.\ \ref{fig:spec3} and \ref{fig:spec4}). Given the low significance of these components, as well as their considerable offset from their line centres, it is highly uncertain whether they are (a) real detections and (b) not an extended red component of the the SiO$(2 - 1)$ lines.}

\smallskip

\begin{figure*}
\begin{center}
\leavevmode
\includegraphics[width=\textwidth]{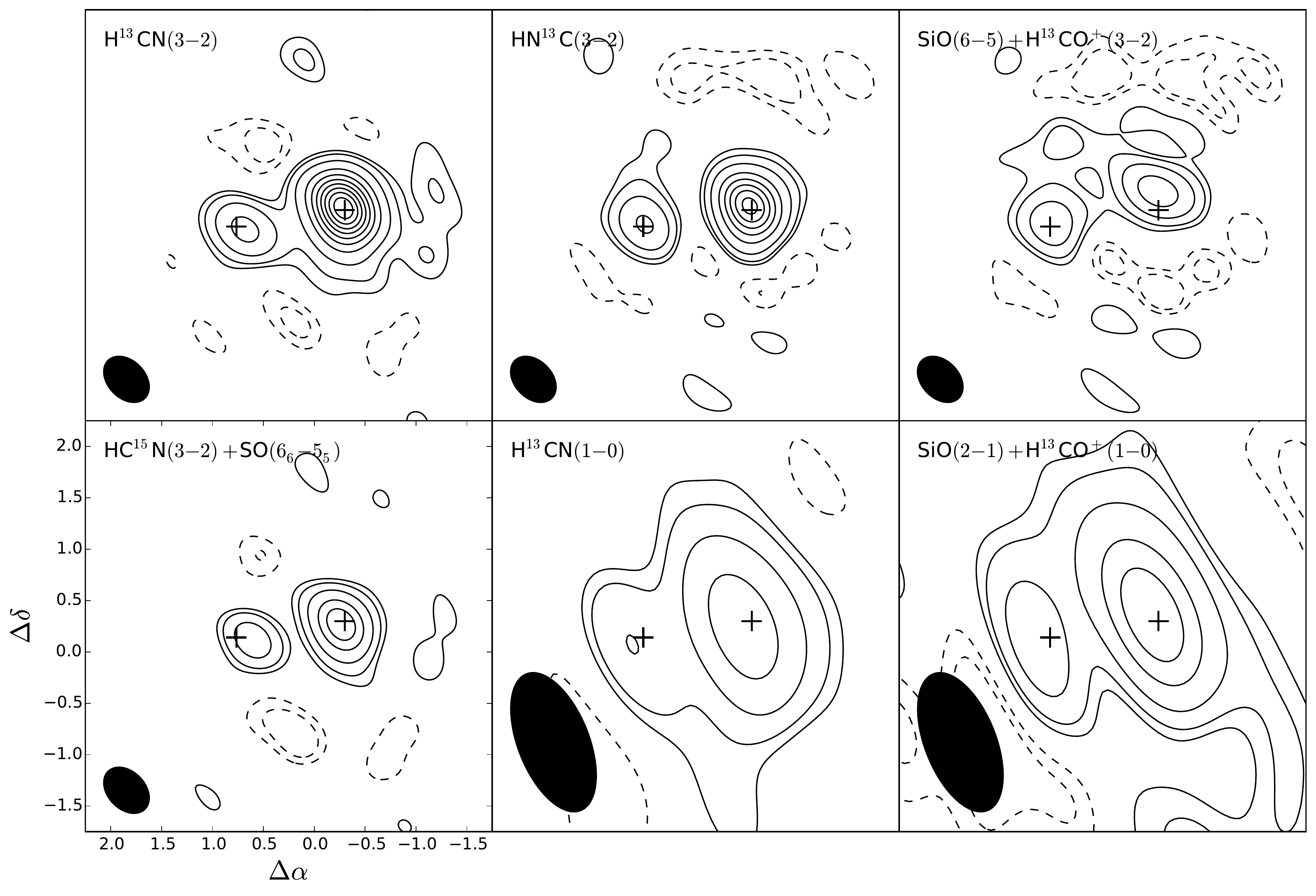}
\caption{Line integrated contour plots of the observed species. Contours are at $-5$, $-3$, 3, 5 and $10n\sigma$, for $n=1,2...10$. The synthesised beams are $0.52''\times0.37''$ and $1.44''\times0.70''$ and are shown in the bottom left corners of the plots. Crosses mark the continuum centres. {The $5\sigma$ sidelobe contours suggest that our observations are dynamic range limited.} }
\label{fig:Overlays}
\end{center}
\end{figure*}

{To search for spatial clues we produced line integrated velocity maps for all of our lines. Shown in Fig.\ \ref{fig:Overlays}, we see that all of the 1.2\,mm lines, except for H$^{13}$CO$^+(3 - 2)$ and SiO$(6 - 5)$ are tightly bound around the nuclear continua, while the SiO$(6 - 5) - $H$^{13}$CO$^+(3 - 2)$ line complex is strongly asymmetric about the WN continuum centre, highlighting it as a line of interest. The 3.5\,mm lines are observed at too low a spatial resolution for us to derive any solid conclusions on the spatial distribution of the emission. Spatially, Sa09 found the line integrated HCO$^+(3 - 2)$ emission to be tightly collocated with the continuum emission, just as we see for our H$^{13}$CN$(3 - 2)$ and HN$^{13}$C$(3 - 2)$ lines, but contrary to what we see for the SiO$(6 - 5)- $H$^{13}$CO$^+(3 - 2)$ line complex, which in the WN is offset north of the continuum centre (Fig.\ \ref{fig:Overlays}). While it is not impossible that H$^{13}$CO$^+(3 - 2)$ is tracing very different regions from HCO$^+(3 - 2)$, possibly due to optical depth effects, it is far more likely, especially in light of the kinematic evidence presented above, that the explanation for this unique spatial arrangement is that the line integrated emission is dominated by SiO$(6 - 5)$.}

\smallskip

As well as the kinematic {and spatial} comparisons discussed above, we also consider the brightness temperature ratios implied with detection and non-detection of H$^{13}$CO$^+$, {using simple LTE estimations and HCO$^+(3 - 2)$ data from Sa09 to identify whether non-detections of H$^{13}$CO$^+(3 - 2)$ are physically viable. Details of the calculations are included in appendix \ref{app:h13co}.} Sa09 report peak HCO$^+(3 - 2)$ $T_\text{B}\sim20$\,K, with nuclear averaged values closer to $13.5$\,K for the WN and $10$\,K for the EN. Assuming that the component centred on the H$^{13}$CO$^+(3 - 2)$ line centre in Fig.\ \ref{fig:spec1} is entirely H$^{13}$CO$^+(3 - 2)$, we have corresponding temperatures of 2.5\,K and 1\,K {in the WN and EN respectively,} giving $T_\text{B}$ ratios {in the WN} between 8 and 13.5. Assuming a $[^{12}$C]/[$^{13}$C$]$ abundance ratio of 60 \citep[e.g.\ ][]{Langer1993}, we find a minimum optical depth $\tau_{12}\geq4.5$ (with $\tau_{13}<<1$), {where $\tau_{12}$ and $\tau_{13}$ denote the optical depths of the $^{12}$C and $^{13}$C HCO$^+(3 - 2)$ isotopologue lines respectively. }For a $T_\text{B}$ ratio of 8 we find $\tau_{12}\geq7$. This is almost 20 times higher than the optical depth measured in absorption {in the WN} by Sa09. If we instead use the optical depth {for the WN} from Sa09 (0.36), we find a predicted H$^{13}$CO$^+(3 - 2)$ $T_\text{B}\sim0.27$\,K, approximately at the $1\sigma$ noise level. {In this case, the H$^{13}$CO$^+(3 - 2)$ WN line would be completely blended with SiO$(6 - 5)$, its contribution wholly undetectable, and the SiO$(6 - 5)$ would extend redwards to $+500$\,km\,s$^{-1}$, almost far as HCO$^+(3 - 2)$. We interpret this as weak evidence for an insignificant H$^{13}$CO$^+(3 - 2)$ contribution in the WN.} 

{In the EN, Sa09 observed a much higher optical depth in absorption $\tau_{12}=3.7$. This predicts an H$^{13}$CO$^+(3 - 2)$ $T_\text{B}\sim1.0$\,K, exactly as is observed in our data. }

{While these estimates are extremely uncertain, including multiple simplifying assumptions, they do offer the tantalising possibility that while the H$^{13}$CO$^+(3 - 2)$ line in the EN is real, and accurately determined, the line that is suspected to be H$^{13}$CO$^+(3 - 2)$ in the WN is almost entirely SiO$(6 - 5)$, with H$^{13}$CO$^+(3 - 2)$ contributing at the same level as random noise.}

\bigskip

{The evidence from the above kinematic, spatial and line brightness analyses casts doubt on the identification of H$^{13}$CO$^+(3 - 2)$ in the WN, but we do not believe that it is sufficient to rule it out entirely. It seems much more likely that H$^{13}$CO$^+(3 - 2)$ has been observed in the EN, but this conclusion rests upon numerous assumptions. As we cannot rule clearly for one side or another, in either nucleus, we therefore proceed working with two limiting cases which can apply to each nucleus independently:}

\begin{itemize}
\item \textbf{Case 1}: we have no detection of H$^{13}$CO$^+$ in {either the $(1 - 0)$ or $(3 - 2)$ transition}. 
\item \textbf{Case 2}: the fitted components{ near the H$^{13}$CO$^+(1 - 0)$ and $(3 - 2)$ line centres} are purely H$^{13}$CO$^+$, with no contamination (other than that from the other fitted components). 
\end{itemize}

	\begin{table*}
		\begin{center}
		\leavevmode
		\caption{Line parameters in the two nuclei of Arp\,220.}
	\label{tab:line-data}
	\begin{tabular}{ l c c c r@{$\pm$}l r@{$\pm$}l}
		\multicolumn{8}{c}{WN: RA 15:34:57.220, Dec 23:30:11.50} \\ \hline \hline
		Line & Flux\footnote{Fluxes are estimated from the Gaussian fitting components in Figs.\ \ref{fig:spec1} to \ref{fig:spec4}{, which are listed in appendix \ref{app:B} in Table \ref{tab:line_fits}}. Uncertainties in the fluxes and $L'$ are estimated from the fitting covariance matrices and the absolute flux calibration (10\%). While the total flux in the HC$^{15}$N$(3 - 2)+$SO$(6_6 - 5_5)$ line blend is relatively well constrained, the contribution of each species is extremely uncertain, indicated by their $>200\%$ uncertainties.} & $L^{\prime}$\footnote{For consistency with HCN single dish data used in the analysis we include source integrated $L'$ values:
	\begin{equation}
	L' = 3.25\times10^7\left(\frac{\nu_{\text{obs}}}{\text{GHz}}\right)^{-2}\left(\frac{D_L}{\text{Mpc}}\right)^2(1+z)^{-3}\left(\frac{\int_{\Delta v}S_v\text{d}v}{\text{Jy\,km\,s}^{-1}}\right),
	\end{equation}
	\citep{Solomon1992}} & $T_{\text{B, peak}}\footnote{Peak of the line profile extracted from single spaxels towards the continuum centres. Relative errors on 1.2\,mm data are $\pm0.21$\,K, on the 3.5\,mm data are $\pm0.11$\,K, but in almost all cases these are insignificant compared to the absolute 10\% calibration uncertainty.}$ &\multicolumn{2}{c}{FWZI} & \multicolumn{2}{c}{FWHM\footnote{Where multiple Gaussian components are required to fit the line, we use a non-parametric FWHM derived from the intensity weighted mean width inspired by \citet{Bothwell2013}, defined in the discreet form as:
\begin{equation}
\text{FWHM}_{\text{iwm}} = 2\sqrt{2\ln2}\left(\frac{\sum\nolimits_i\left(v-\bar{v}\right)^2|I_i|\Delta v}{\sum\nolimits_i|I_i|\Delta v}\right)^{0.5}.
\end{equation}
Note that this method is very sensitive to both the location of the line centre (and hence to the assumed redshift) and to noise at the ends of the fainter lines: this method becomes increasingly susceptible to noise at low S/N.}}\\
		& [Jy\,km\,s$^{-1}$] & [$10^8$\,K\,km\,s$^{-1}$\,pc$^2$] & [K] & \multicolumn{2}{c}{[km\,s$^{-1}$]} &\multicolumn{2}{c}{[km\,s$^{-1}$]} \\ \hline \hline
		H$^{13}$CN$(3 - 2)$\dotfill & $52\pm5$  & $1.66\pm0.24$ & $11.2\pm1.1$ &775 & 25 &  464 & 5 \\[0.1cm]
		H$^{13}$CN$(1 - 0)$\dotfill & $4.0\pm0.4$ & $1.15\pm0.12$ &$1.94\pm0.22$  &800 & 50 & 285 & 27 \\[0.1cm]
		HC$^{15}$N$(3 - 2)$\dotfill & $6\pm14$ & $0.2\pm0.5$ & $1.86\pm0.28$ & \multicolumn{2}{c}{$--$} & 220 & 40 \\[0.1cm]
		SO$(6_6 - 5_5)$\dotfill & $12\pm16$ & $0.4\pm0.6$ & $2.8\pm0.4$ &\multicolumn{2}{c}{$--$} & 320 & 270  \\[0.1cm]
		SiO$(6 - 5)$\footnote{Case 1: no H$^{13}$CO$^+$.}\dotfill & $17.1\pm2.5$ & $0.54\pm0.08$ & $5.75\pm0.61$ & 1200 & 100 &  730 & 30\\[0.1cm]
		SiO$(6 - 5)$\footnote{Case 2: with H$^{13}$CO$^+$.}\dotfill & $7.6\pm2.0$ & $0.24\pm0.06$ & $5.75\pm0.61$ & 950 & 50 & \multicolumn{2}{c}{$--$}\\[0.1cm]
		SiO$(2 - 1)^\text{e}$\dotfill & $4.8\pm0.5$ & $1.38\pm0.14$ & $1.62\pm0.20$ & 803 & 22 & 440 & 50\\[0.1cm]
		SiO$(2 - 1)^\text{f}$\dotfill & $4.5\pm0.5$ & $1.28\pm0.13$ & $1.62\pm0.20$ & \multicolumn{2}{c}{$--$}& 370 & 40\\[0.1cm]
		H$^{13}$CO$^+(3 - 2)^\text{f}$\dotfill & $9.5\pm1.1$&$0.30\pm0.03$&$2.77\pm0.29$&500&50& 205 & 21 \\[0.1cm]
		H$^{13}$CO$^+(1 - 0)^\text{f}$\dotfill & $0.35\pm0.08$ & $0.100\pm0.022$ &$0.22\pm0.11$ & 350& 50 &200&200\\[0.1cm]
		HN$^{13}$C$(3 - 2)$\dotfill & $37\pm4$ &$1.16\pm0.13$& $7.72\pm0.80$ &\multicolumn{2}{c}{$--$}& 350 & 100\\ \hline \\[0.1cm] 

		\multicolumn{8}{c}{EN: RA 15:34:57.291, Dec 23:30:11.34} \\ \hline \hline
		Line & Flux$^{\text{a}}$ & ${L^{\prime}}^{\text{b}}$  & ${T_{\text{B, peak}}}^{\text{c}}$ &\multicolumn{2}{c}{FWZI} & \multicolumn{2}{c}{FWHM$^\text{d}$}\\
		& [Jy\,km\,s$^{-1}$] & [$10^8$\,K\,km\,s$^{-1}$\,pc$^2$] & [K] & \multicolumn{2}{c}{[km\,s$^{-1}$]} &\multicolumn{2}{c}{[km\,s$^{-1}$]} \\ \hline \hline
		H$^{13}$CN$(3 - 2)$\dotfill & $12.7\pm1.3$ & $0.41\pm0.04$ &$4.91\pm0.53$&550 & 50 & 523 & 10\\[0.1cm]
		H$^{13}$CN$(1 - 0)$\dotfill & $0.50\pm0.05$ & $0.144\pm0.015$ &$0.57\pm0.12$ &450 & 100  & 170 & 40\\[0.1cm]
		HC$^{15}$N$(3 - 2)$\dotfill & $3.1\pm1.2$ & $0.10\pm0.04$ & $1.50\pm0.26$ &\multicolumn{2}{c}{$--$} & 160 & 60 \\[0.1cm]
		SO$(6_6 - 5_5)$\dotfill & $3.2\pm1.1$ & $0.10\pm0.04$ &$1.10\pm0.24$& \multicolumn{2}{c}{$--$}& 186& 26\\[0.1cm]
		SiO$(6 - 5)^{\text{e}}$\dotfill & $6.5\pm0.7$ & $0.206\pm0.029$ &$2.48\pm0.33$& 740 & 80  & 410 & 20\\[0.1cm]
		SiO$(6 - 5)^{\text{f}}$\dotfill & $4.9\pm0.5$ & $0.155\pm0.017$ &$2.48\pm0.33$& 600 & 50 & 200 & 30\\[0.1cm]
		SiO$(2 - 1)^\text{e}$\dotfill & $0.92\pm0.14$ & $0.26\pm0.04$ & $0.72\pm0.13$ & 800 & 100 & 539 & 11\\[0.1cm]
		SiO$(2 - 1)^\text{f}$\dotfill & $0.70\pm0.17$ & $0.20\pm0.06$ & $0.72\pm0.13$ & 500 & 50 & \multicolumn{2}{c}{$--$}\\[0.1cm]
		H$^{13}$CO$^+(3 - 2)^\text{f}$\dotfill & $1.63\pm0.25$ & $0.052\pm0.008$ & $1.15\pm0.24$ &300 & 50 & 110 & 40\\[0.1cm]
		H$^{13}$CO$^+(1 - 0)^\text{f}$\dotfill & $0.22\pm0.06$ & $0.062\pm0.018$ & $0.16\pm0.11$ & 450 & 50 & 220 & 190\\[0.1cm]
		HN$^{13}$C$(3 - 2)$\dotfill & $8.4\pm1.0$ &$0.26\pm0.03$ & $2.22\pm0.31$&700 & 50 & 408 & 10\\ \hline 		
	
	\end{tabular}
	\end{center}
	\end{table*}


	\section{Western Nucleus}\label{sec:wn}
			
	The WN line profiles all share similar features, although there is a significant variation between the 3.5\,mm and 1.2\,mm spectra. The 1.2\,mm lines show a distinct asymmetry, peaking just redwards of the line centre but with the majority of the velocity integrated emission bluewards of the line. Of particular interest are H$^{13}$CN and SiO, which are not only the brightest lines, but those with the most complex line profiles.

\begin{figure*}
\begin{center}
\includegraphics[width=0.9\textwidth]{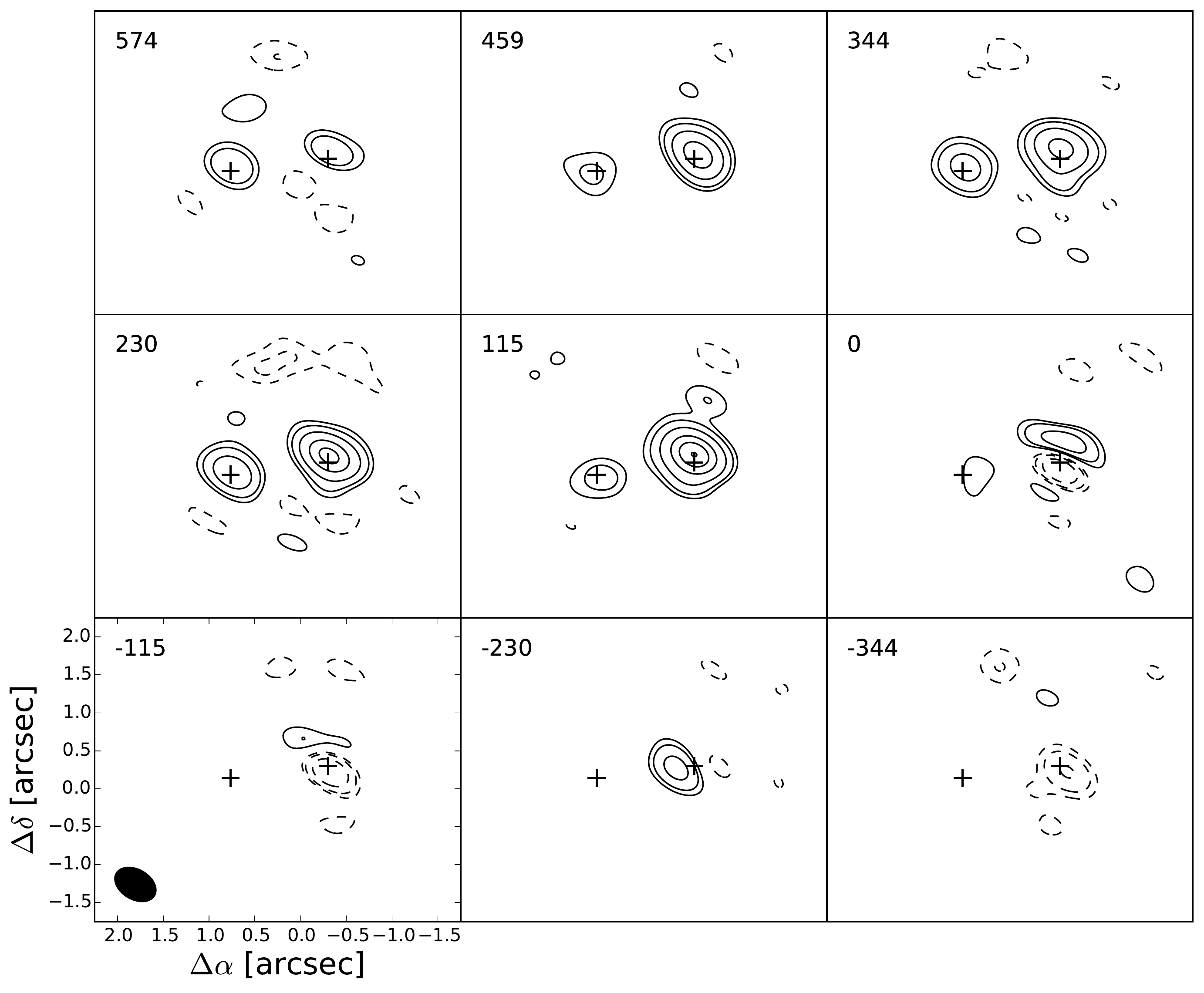}
\caption{Channel maps for the SiO$(6 - 5)$ line complex, including the subsidiary emission (at $-230$\,km\,s$^{-1}$) and absorption (at $-344$\,km\,s$^{-1}$). {Channels have been averaged to 115\,km\,s$^{-1}$.} {Contours are at $-10$, $-5$, $-3$, 3, 5, and $10n\sigma$ intervals $(1\sigma =0.63$\,mJy\,beam$^{-1})$}. The mean velocity of each channel relative to the line centre is given in the top left corners of the plots. Note the eastwards offset of the emission at $-230$\,km\,s$^{-1}$ and the colocation of the sub-continuum absorption red and bluewards of this subsidiary peak $(-115$\,km\,s$^{-1}$ and $-344$\,km\,s$^{-1})$. The $0.52''\times0.37''$ synthesised beam is shown in the bottom left panel. The crosses mark the continuum centres.}\label{fig:siomaps}
\end{center}
\end{figure*}

	The SiO$(6 - 5)$ line is especially complex, both spectrally (requiring 5 Gaussian components including the possible H$^{13}$CO$^+(3 - 2)$ component), and spatially, with multiple components in the WN. {The intriguing spatial distribution seen in Fig.\ \ref{fig:Overlays} is explored in Fig.\ \ref{fig:siomaps} where we present the SiO$(6 - 5)$ channel maps. The overall peak emission and absorption are $\sim0.1''$ north and south of the continuum centre and in the $+115$\,km\,s$^{-1}$ and $-115$\,km\,s$^{-1}$ channels respectively, while the {full} SiO$(6 - 5)$ data cube has $1\sigma$ continuum absorption extending as far north as the peak of the SiO$(6 - 5)$ emission. The peak emission and absorption in the $0$\,km\,s$^{-1}$ channel map only has an even greater separation, almost $0.4''$, clearly indicating that the offset is real, and not due to radio seeing effects.} The north-western offset of the SiO$(6 - 5)$ emission peak from the continuum centre was also seen in CO$(2 - 1)$ observations by \citet{Downes2007}. Fractional absorption against the continuum implies a lower limit for the optical depth, $\tau_{\text{SiO}(6 - 5)}\geq$ 0.25 towards the absorption peak. 
	
	\smallskip
	
\begin{figure*}
	\begin{center}
	\includegraphics[width=\textwidth]{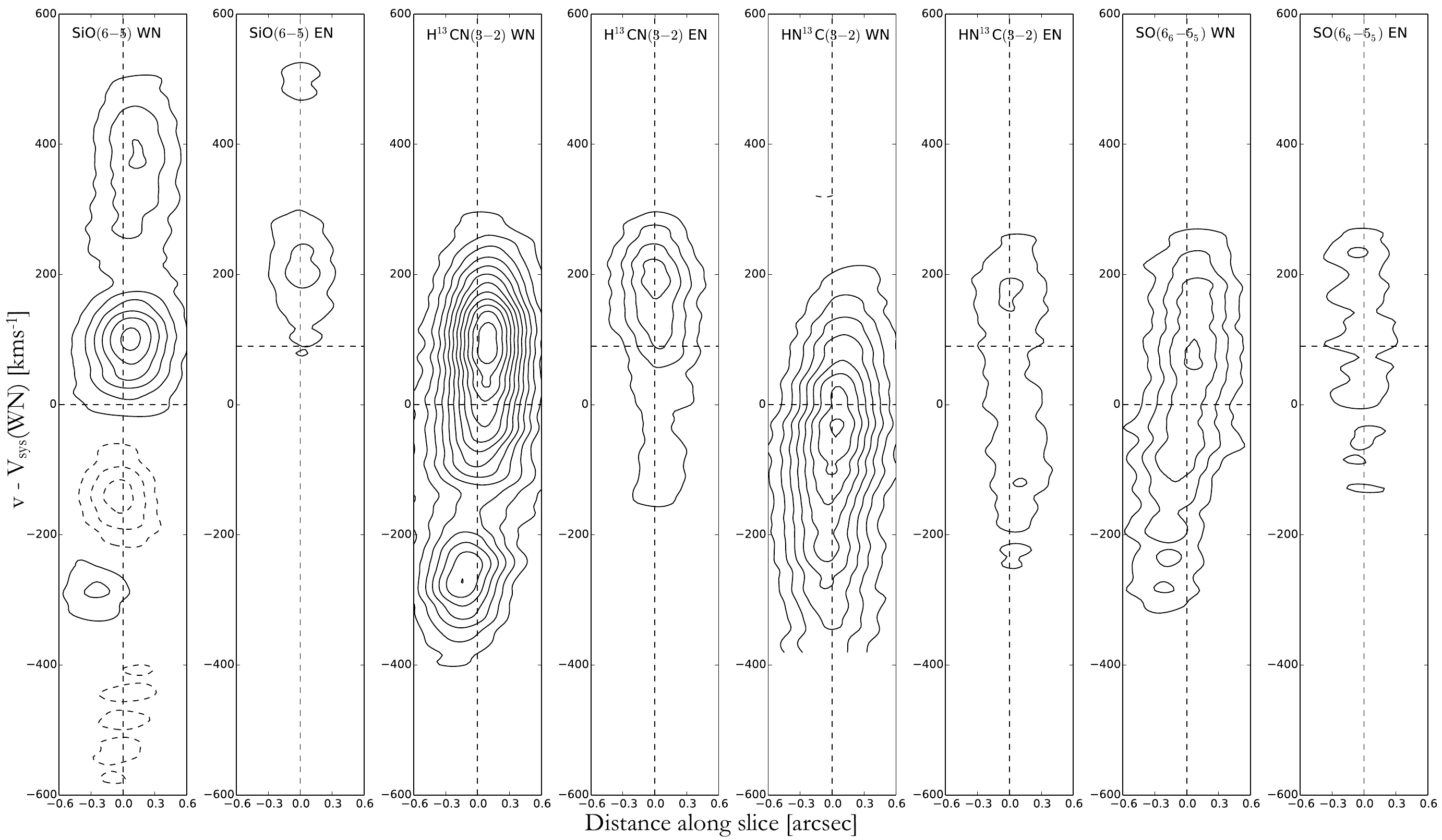}
	\caption{Velocity slices across the two nuclei of Arp\,220. Slices are centred on the systemic velocity (radio, LSR) of the WN, 5355\,km\,s$^{-1}$, to highlight the offsets between the two nuclei. Slices are 1.2$''$ long, centred on the continuum centres and taken at 270$^{\circ}$ across the WN (east to west) and 225$^{\circ}$ across the EN (south-east to north-west) to be consistent with Sa09. The horizontal axis corresponds to the angular distance along the slice. Contours are at $\pm3n\sigma$. The H$^{13}$CN and SO WN slices suggest disk like rotation. The SiO WN slice could be indicative of rotation, an outflow or a combination of the two. Its prominent P Cygni profile is clearly seen, but the emission peaks at $\sim100$\,km\,s$^{-1}$ and $\sim -300$\,km\,s$^{-1}$ and absorption trough at $\sim -150$\,km\,s$^{-1}$ are almost precisely coincident with the two peaks and the minimum in the H$^{13}$CN WN slice, suggesting they are tracing, at least in part, the same kinematics.}
	\label{fig:velocity_slices}
	\end{center}
\end{figure*}

		{In Fig.\ \ref{fig:velocity_slices} we show position-velocity slices across the two nuclei. Slices are 1.2$''$ long, centred on the continuum centres and taken at 270$^{\circ}$ across the WN (east to west) and 225$^{\circ}$ across the EN (south-east to north-west) to be consistent with Sa09. The velocity slices have been smoothed with a Gaussian kernel for clarity. While the EN signal is too weak to identify any clear trends, the WN H$^{13}$CN$(3 - 2)$ and SO$(6_6 - 5_5)$ lines both display evidence of disk-like rotation. The SiO$(6 - 5)$ line shows signs of a strong, kinematically extended outflow with the P Cygni like absorption at negative velocities mixed with an emission signature at $\sim-300$\,km\,s$^{-1}$, coincident with the peak blueshifted emission in the H$^{13}$CN$(3 - 2)$ line. There is a clear systemic velocity offset between the two nuclei, as reported in the literature, potentially as large as 90\,km\,s$^{-1}$ based on our data, but the asymmetry and width of the lines makes reliably identifying the precise value of the offset highly uncertain. We therefore adopt the 60\,km\,s$^{-1}$ CO derived offset of Sa09, which is reliably determined from narrow, on line centre, absorption in the CO lines.}
		
		{Fig.\ \ref{fig:velocity_slices} reveals the bluewards absorption extending all the way out to $-600$\,km\,s$^{-1}$, with no equivalent redshifted emission, even in the case 1 limit. However, if we temporarily exclude the blueshifted emission peak at $-300$\,km\,s$^{-1}$ the emission and absorption appear symmetric about a point $\sim-50$\,km\,s$^{-1}$. This profile is suggestive of a symmetric outflow, traced by SiO, merged with SiO emission from the nuclear disk itself.} 
		
		\smallskip
	
	In case 2 the reddest peak in the {SiO$(6 - 5)$} line profile is attributed entirely to H$^{13}$CO$^+(3 - 2)$. This case seems unlikely as the emission at these velocities is concentrated north of the continuum (Fig.\ \ref{fig:siomaps}), {whereas} the HCO$^+(3 - 2)$ emission of Sa09 is continuum centred and spatially very similar to our H$^{13}$CN$(3 - 2)$ line (Fig.\ \ref{fig:Overlays}). It is clear from Fig.\ \ref{fig:velocity_slices} that case 1 implies that SiO reaches higher velocities than H$^{13}$CN, but it is still not as extended in velocity as HCO$^+$ (Fig.\ \ref{fig:sakacomp}), which extends almost 200\,km\,s$^{-1}$ further redwards {in emission}.
	
\smallskip

{It is worth noting that while SiO$(6 - 5)$ shows significant absorption in the WN, SiO$(2 - 1)$ has no sign of continuum absorption.} As was noted in Sa09, the high brightness temperature of the dust in the WN {implies that} higher frequency transitions are more likely to be seen in absorption against the dust {continuum}, and indeed in the far-IR {almost all molecular} lines are {observed} in absorption \citep{Gonzalez-Alfonso2004,Gonzalez-Alfonso2012}, {so the lack of absorption in the SiO$(2 - 1)$ line could simply be due to the drop in dust SED between 1.2\,mm and 3.5\,mm}. Sa09 further noted that the Einstein coefficients for photo-absorption could be having an effect. Assuming detailed balance:
\begin{equation}
\text{B}_{lu} = \frac{g_u}{g_l}\text{A}_{ul}\left(\frac{2\text{h}\nu^3}{c^2}\right)^{-1},
\end{equation}
which for the rotational transitions of linear diatomic molecules may be written as:
\begin{equation}
\text{B}_{J, J+1} = \frac{2J+3}{2J+1}\text{A}_{J+1, J}\left(\frac{2\text{h}\nu^3}{c^2}\right)^{-1},
\end{equation}
gives for SiO B$_{56}$/B$_{12}\simeq 1$, with similar results in other lines (since A$_{ul}\propto \nu^3$): this effect is most significant when comparing similar frequency transitions of different molecules. This suggests that the absence of absorption in our SiO$(2 - 1)$ lines is more likely a dust brightness effect. It does not appear to be a beam size issue: the absorption is still clearly visible when the SiO$(6 - 5)$ cube is smoothed to the 3.5\,mm beam size. 
{Finally, we consider the possibility that there is an angularly extended ``screen'' of blueshifted emission in front of the WN which is being spatially filtered in the 1.2\,mm observations.} The minimum baselines were 22.6\,k$\lambda$ and 36.4\,k$\lambda$ at 1.2\,mm and 3.5\,mm respectively, so nominally we resolve out emission on angular scales greater than $9.1''$ and $5.7''$.  Therefore the baselines are not likely to be responsible for the absence of absorption: the 3.5\,mm observations will resolve out emission before the 1.2\,mm, whereas the line profiles could only be explained if the 1.2\,mm observations were resolving out large-scale, blueshifted emission.
	
\smallskip

	The H$^{13}$CN$(3 - 2)$ line is the brightest line observed at 1.2\,mm. We would expect this line to trace the same high density star forming environment traced by HCN$(1 - 0)$, $(3 - 2)$ and $(4 - 3)$ as observed by \citet{Greve2009} and we find the emission strongly correlated with the continuum emission (Fig.\ \ref{fig:Overlays}). The H$^{13}$CN slice (Fig.\ \ref{fig:velocity_slices}) suggests predominantly disk like rotation, although our spatial resolution is insufficient for successful kinematic modelling. However, comparison with the SiO WN slice with emission peaks at $\sim100$\,km\,s$^{-1}$ and $\sim -300$\,km\,s$^{-1}$ and the absorption trough at $\sim -150$\,km\,s$^{-1}$ shows that they are almost precisely coincident with the two peaks and the minimum in the H$^{13}$CN WN slice suggesting they are tracing, at least in part, the same gas. However, if they are tracing the same gas, it is curious that they have such different levels of absorption. {The simplest explanation is that while H$^{13}$CN$(3 - 2)$ is tracing both the dense, star forming nuclear disk and an outflow, presenting as a P Cygni profile embedded within a Gaussian line, while SiO$(6 - 5)$ traces the outflow with only a small component in the star forming nuclear disk.}
	
{We also examine the possibility that H$^{13}$CN$(3 - 2)$ and SiO$(6 - 5)$ are tracing similar regions, but that the different line profiles are driven by the different properties of the transitions.} The H$^{13}$CN $J=3$ and SiO $J=6$ levels have energies of 24.9\,K and 43.8\,K respectively \citep{SLAIM,CDMS}. Coupled with the fact that higher $J$ transitions are harder to excite, as they need the intermediate $J$ levels to be excited as well, the P Cygni profile in H$^{13}$CN$(3 - 2)$ could be {less prominent} because the H$^{13}$CN $J=3$ level is easier to excite, so is seen more in emission, while SiO$(6 - 5)$ is seen in absorption. However, the HCO$^+$ $J=3$ level has an energy of 25.7\,K \citep{SLAIM}, only 3\% higher than that of H$^{13}$CN $J=3$, and is seen with strong absorption (Sa09), suggesting that the energies alone cannot fully explain the differences in the line profiles. Furthermore, it does not appear to be a critical density effect, since the critical densities for H$^{13}$CN$(3 - 2)$, SiO$(6 - 5)$ and HCO$^+(3 - 2)$ are $2.9\times10^6$\,cm$^{-3}$, $2.0\times10^6$\,cm$^{-3}$ and $7.8\times10^5$\,cm$^{-3}$ respectively. It would appear therefore to be either an optical depth effect, or that the majority of the H$^{13}$CN$(3 - 2)$ emission is tracing the nuclear disks, with a small component in the outflow traced by SiO$(6 - 5)$. This would imply however that there should be the same $\pm40$\,pc emission and absorption offset seen in SiO$(6 - 5)$ in the H$^{13}$CN$(3 - 2)$ line, which would shift the integrated contour peak north (as can be seen for SiO$(6 - 5)$ in Fig.\ \ref{fig:Overlays}); this is not seen, but could simply be a sign that H$^{13}$CN in the outflow is too faint to significantly affect the integrated plots. 

\smallskip

	HN$^{13}$C$(3 - 2)$ emission extends beyond the end of our spectrum, making it impossible to reliably compare it with the H$^{13}$CN$(3 - 2)$ emission. The absorption present in H$^{13}$CN$(3 - 2)$ appears to be absent in HN$^{13}$C$(3 - 2)$. This could be due to errors at the edges of the bandpass, or it could be the same effect as has been seen in single dish spectra, where HCN$(3 - 2)$ shows absorption but HNC$(3 - 2)$ does not, possibly due to mid-IR pumping or maser emission in HNC \citep{Aalto2009}. Alternatively, it could be a sign of HCN enhancement (over HNC) in the shocks in the outflow \citep[e.g.][]{Mitchell1984,Bachiller1997,Martin2014}. This would be consistent with the case above, where H$^{13}$CN is {over abundant} in an outflow from the centre of the nucleus as well as {being prevalent throughout} the nuclear disk. Both isomers trace the circumnuclear disk (CND), but only H$^{13}$CN would be detectable in the outflow, imprinting a P Cygni profile on to the H$^{13}$CN line profile but not the HN$^{13}$C.

\smallskip

	Like SiO, SO in the Milky Way is a shock tracer \citep{Chernin1994}, ubiquitous in star forming regions. It is unclear whether these remain reliable shock tracers extragalactically, where excitation and abundance effects could be more significant. Given the uncertainty in the fitting for SO$(6_6 - 5_5)$ and HC$^{15}$N$(3 - 2)$ we cannot reliably separate the two lines and they are not focussed on in this work. Similarly, we see a trace of the SO$(2_2 - 1_1)$ line at the red end of the 3.5\,mm WN spectrum, but this is severely affected by edge effects in the bandpass and is not included in the analysis.

\section{Eastern Nucleus}\label{sec:en}

While none of the lines in the EN are spatially resolved, SiO$(2 - 1)$, SiO$(6 - 5)$, HN$^{13}$C$(3 - 2)$ and H$^{13}$CN$(3 - 2)$ all show doubly peaked emission with no significant sign of sub-continuum absorption (while in case 1 the SiO$(2 - 1)$ and $(6 - 5)$ spectra both show {possible} signs of the same absorption-emission-absorption structure seen in the WN, this is not spatially correlated across the channels and is below $1\sigma$). 

The SiO line profiles in the EN are very similar: the $(6 - 5)$ line has  major and minor peaks at $V_{\text{sys}} +140$\,km\,s$^{-1}$ and $V_{\text{sys}} - 175$\,km\,s$^{-1}$ respectively, cf.\ $V_{\text{sys}} +150$\,km\,s$^{-1}$ and $V_{\text{sys}} - 200$\,km\,s$^{-1}$ in the $(2 - 1)$ line. We do not see the prominent P Cygni profiles observed in the HCO$^+(3 - 2)$ and $(4 - 3)$ lines in the EN (Sa09), although the 3.5\,mm spectrum in particular does allow for some degree of bluewards absorption around $-100$\,km\,s$^{-1}$. Only the absolute scaling of the line profile changes across the EN. This may explain the lack of a clear P Cygni profile in the EN, since the resolution of Sa09 was almost double ours, we may be smearing out any sub-continuum absorption in the EN. If the $\sim1\sigma$ signs of continuum absorption in the EN are real, then this implies $\tau_{\text{SiO}(6 - 5)} \simeq 0.03\pm0.01$: $1/10$ the WN value.

\section{Analysis}\label{sec:analysis}

\begin{table*}
	\begin{center}
		\caption{Line brightness temperature ratios for the two nuclei.}
		\begin{tabular}{l c c c} \hline

		\multicolumn{4}{c}{$L'$ ratios}\\ \hline
		\multicolumn{2}{l}{Line Ratio} & \multicolumn{2}{c}{Ratio} \\ \hline \hline
		\multicolumn{2}{l}{$L'(\text{HCN}(3 - 2))/L'(\text{HCN}(1 - 0))$} & \multicolumn{2}{c}{$0.79\pm0.12$}\\[0.1cm]
		\multicolumn{2}{l}{$L'(\text{H}^{13}\text{CN}(3 - 2))/L'(\text{H}^{13}\text{CN}(1 - 0))$} & \multicolumn{2}{c}{$1.60\pm0.24$}\\[0.1cm]
		\multicolumn{2}{l}{$L'(\text{HCN}(3 - 2))/L'(\text{H}^{13}\text{CN}(3 - 2))$} & \multicolumn{2}{c}{$5.0\pm0.8$}\\[0.1cm]
		\multicolumn{2}{l}{$L'(\text{HCN}(1 - 0))/L'(\text{H}^{13}\text{CN}(1 - 0))$} & \multicolumn{2}{c}{$10.1\pm1.3$}\\ \hline \\[0.1cm]
		
		\multicolumn{4}{c}{$T_{\text{B}}{\rm d}v$ line integrated ratios\footnote{Based on the fitted Gaussian components, using only the emission components. The H$^{13}$CN$(3 - 2)$/$(1 - 0)$ and SiO$(6 - 5)$/$(2 - 1)$ ratios are quoted here without corrections for beam dilution.}}\\ \hline
		\multicolumn{2}{l}{Line Ratio} & WN & EN \\ \hline\hline
		\multicolumn{2}{l}{$\text{H}^{13}\text{CN}(3 - 2)/\text{H}^{13}\text{CN}(1 - 0)$} & $6.7\pm0.9$ & $5.8\pm0.9$ \\[0.1cm]
		\multicolumn{2}{l}{$\text{H}^{13}\text{CN}(3 - 2)/\text{HN}^{13}\text{C}(3 - 2)$} & $1.41\pm0.07$ & $1.51\pm0.10$ \\[0.1cm]
		\multicolumn{4}{c}{Case 1\footnote{Upper limits on H$^{13}$CO$^+$ contribution estimated $1\sigma$ Gaussians with FWHM of 200\,km\,s$^{-1}$.}: no H$^{13}$CO$^+$}\\
		\multicolumn{2}{l}{$\text{H}^{13}\text{CN}(3 - 2)/\text{SiO}(6 - 5)$} & $2.32\pm0.12$ & $2.0\pm0.3$\\[0.1cm]
		\multicolumn{2}{l}{$\text{H}^{13}\text{CN}(1 - 0)/\text{SiO}(2 - 1)$} & $0.826\pm0.016$ & $0.54\pm0.07$\\[0.1cm]
		\multicolumn{2}{l}{$\text{H}^{13}\text{CN}(3 - 2)/\text{H}^{13}\text{CO}^+(3 - 2)$} & $>41$ & $>10$\\[0.1cm]
		\multicolumn{2}{l}{$\text{H}^{13}\text{CN}(1 - 0)/\text{H}^{13}\text{CO}^+(1 - 0)$} & $>9$ & $>3$\\[0.1cm]
		\multicolumn{2}{l}{$\text{SiO}(6 - 5)/\text{SiO}(2 - 1)$} & $2.4\pm0.3$ & $1.61\pm0.22$\\[0.1cm]
		\multicolumn{4}{c}{Case 2: H$^{13}$CO$^+$}\\
		\multicolumn{2}{l}{$\text{H}^{13}\text{CN}(3 - 2)/\text{SiO}(6 - 5)$} & $4.0\pm0.5$ & $2.60\pm0.11$\\[0.1cm]
		\multicolumn{2}{l}{$\text{H}^{13}\text{CN}(1 - 0)/\text{SiO}(2 - 1)$} & $0.890\pm0.027$ & $0.71\pm0.16$\\[0.1cm]
		\multicolumn{2}{l}{$\text{H}^{13}\text{CN}(3 - 2)/\text{H}^{13}\text{CO}^+(3 - 2)$} & $4.78\pm0.23$ & $7.8\pm0.9$\\[0.1cm]
		\multicolumn{2}{l}{$\text{H}^{13}\text{CN}(1 - 0)/\text{H}^{13}\text{CO}^+(1 - 0)$} & $11.5\pm2.3$ & $2.3\pm0.6$\\[0.1cm]
		\multicolumn{2}{l}{$\text{SiO}(6 - 5)/\text{SiO}(2 - 1)$} & $1.47\pm0.23$ & $1.6\pm0.4$\\ \hline \\[0.1cm]

		\multicolumn{4}{c}{$T_{\text{B,peak}}$ ratios\footnote{Peak of the line profile extracted from single spaniels towards the continuum centres in the unsmoothed maps: these ratios do not account for beam dilution in ratios of 1.2\,mm and 3.5\,mm lines.}}\\ \hline
		\multicolumn{2}{l}{Line Ratio} & WN & EN \\ \hline\hline
		\multicolumn{2}{l}{$\text{H}^{13}\text{CN}(3 - 2)/\text{H}^{13}\text{CN}(1 - 0)$} &$5.8\pm0.9$ &$8.6\pm2.1$ \\[0.1cm]
		\multicolumn{2}{l}{$\text{SiO}(6 - 5)/\text{SiO}(2 - 1)$} & $3.4\pm0.6$ & $2.8\pm0.7$ \\[0.1cm]
		\multicolumn{2}{l}{$\text{H}^{13}\text{CN}(3 - 2)/\text{H}^{13}\text{CO}^+(3 - 2)$} &$4.1\pm0.3$ &$1.72\pm0.13$ \\[0.1cm]
		\multicolumn{2}{l}{$\text{H}^{13}\text{CN}(1 - 0)/\text{H}^{13}\text{CO}^+(1 - 0)$} & $9\pm5$&$5\pm3$ \\[0.1cm]
		\multicolumn{2}{l}{$\text{H}^{13}\text{CN}(3 - 2)/\text{SiO}(6 - 5)$} &$1.95\pm0.08$ & $1.98\pm0.03$\\[0.1cm]
		\multicolumn{2}{l}{$\text{H}^{13}\text{CN}(1 - 0)/\text{SiO}(2 - 1)$} & $1.19\pm0.11$ & $0.79\pm0.05$ \\[0.1cm]
		\multicolumn{2}{l}{$\text{HN}^{13}\text{C}(3 - 2)/\text{H}^{13}\text{CN}(3 - 2)$} & $0.689\pm0.023$ & $0.453\pm0.010$ \\ \hline
		
		\end{tabular}\label{tab:LRs}
	\end{center}
\end{table*}

	\subsection{LVG modelling}\label{subsec:LVG}
	The LVG code RADEX\footnote{A non-LTE code for radiative transfer, available here: \url{http://home.strw.leidenuniv.nl/~moldata/radex.html}.} was used with line ratios, derived from our earlier analysis and presented in Table \ref{tab:LRs}, to provide estimates of the molecular hydrogen density ($n_{\text{H}_2}$), molecular column density over velocity width ($N_{\text{x}}/\Delta v$) and kinetic temperature ($T_{\text{k}}$) for our observed species. The column densities provide abundance ratios [H$^{13}$CN]/[SiO] and [H$^{13}$CN]/[H$^{13}$CO$^+$]. These simple models do not account for the potentially significant mid-IR pumping in Arp\,220. Since our beam is averaging over $\sim200\text{\,pc}\times140$\,pc, we are certainly including a wide range of regions including giant molecular clouds (GMCs), hot cores, shocked regions, photon dominated regions (PDRs) and perhaps an XDR if there is an AGN at the centre of the WN, so the model parameters are at best only representative of the conditions of the dominant emission regions.
	
	\smallskip
	
	We first used HCN and H$^{13}$CN data from \citet{Greve2009} (which includes data from \citealp{Solomon1992} and \citealp{Krips2008}) and from our work to find the best fitting parameters for Arp\,220 as a whole, then used these to model abundance ratios for SiO and H$^{13}$CO$^+$ in the two nuclei independently.
	
	\subsubsection{HCN}
	 We recalculated the $L'$ values of HCN $3 - 2$ and $1 - 0$ as weighted averages of the data reported in \citet{Greve2009} using their quoted $S_{\nu}\text{d}v$ values and our adopted cosmological parameters, finding $L'_{\text{HCN}(3 - 2)} = 9.5\pm1.1\times10^8$\,K\,km\,s$^{-1}$\,pc$^2$ and $L'_{\text{HCN}(1 - 0)} = 12.1\pm1.2\times10^8$\,K\,km\,s$^{-1}$\,pc$^2$.
	
\begin{figure*}
	\begin{center}
	\includegraphics[width=0.49\textwidth]{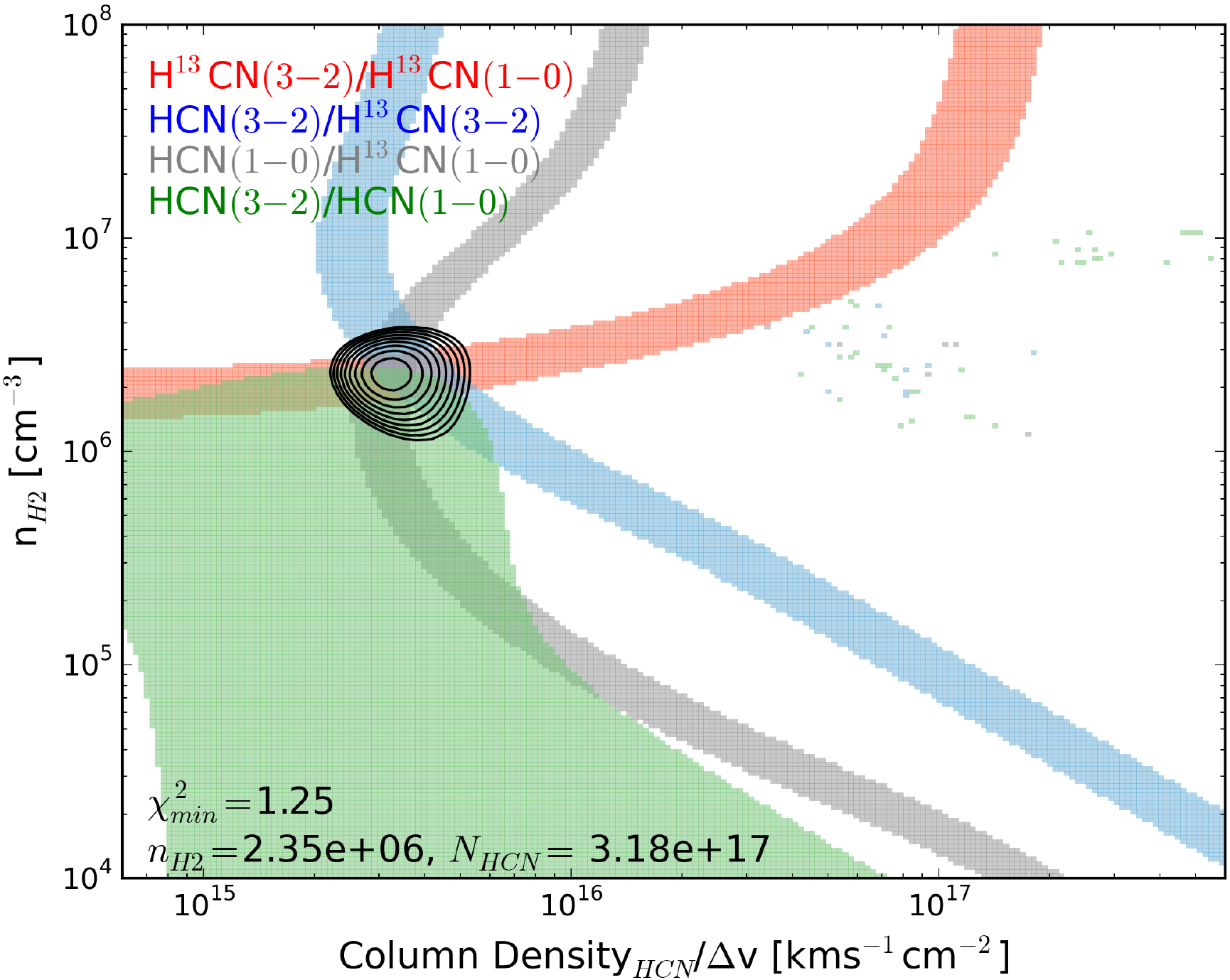}
	\hfill
	\includegraphics[width=0.49\textwidth]{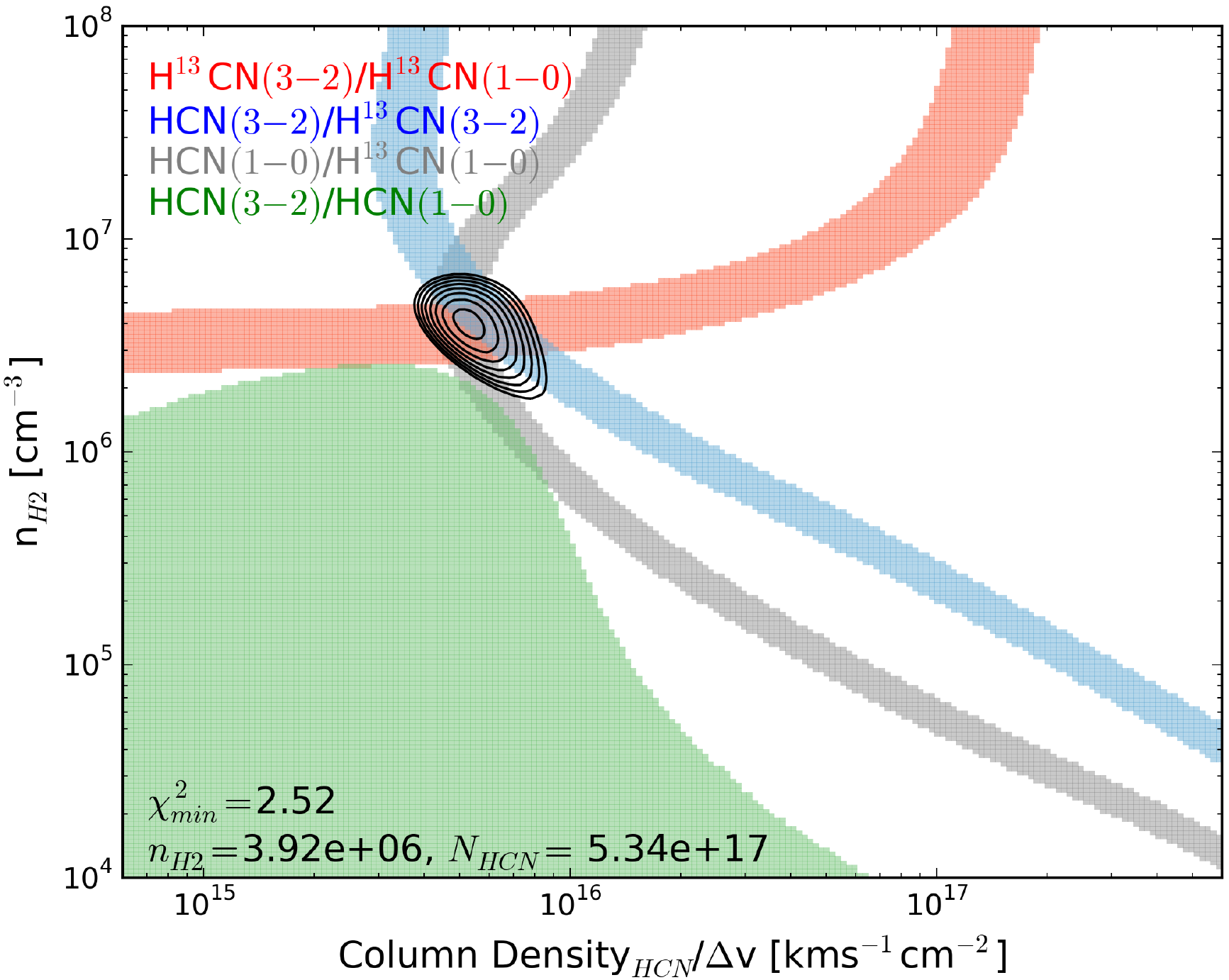}\\
	\includegraphics[width=0.49\textwidth]{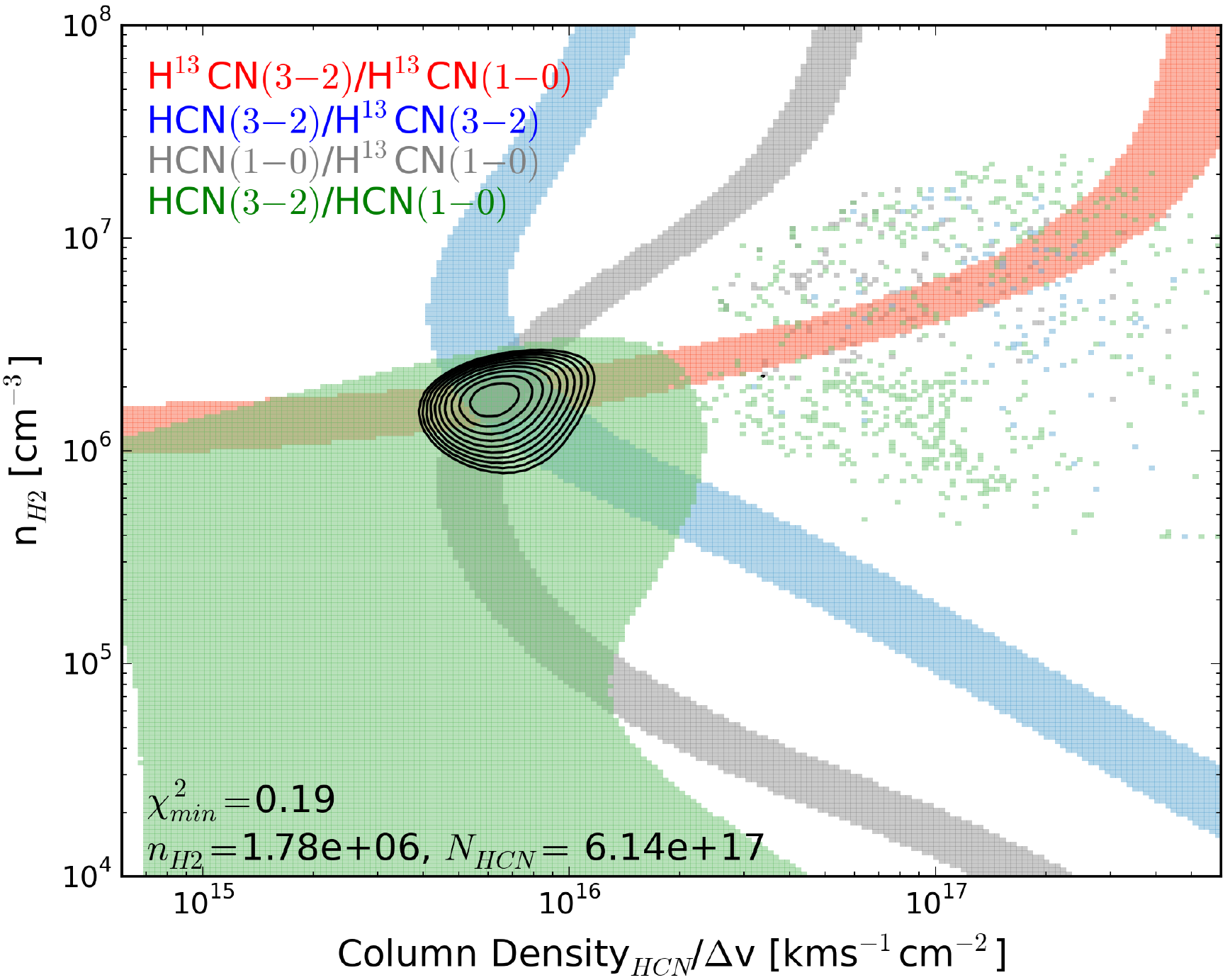}
	\hfill
	\includegraphics[width=0.49\textwidth]{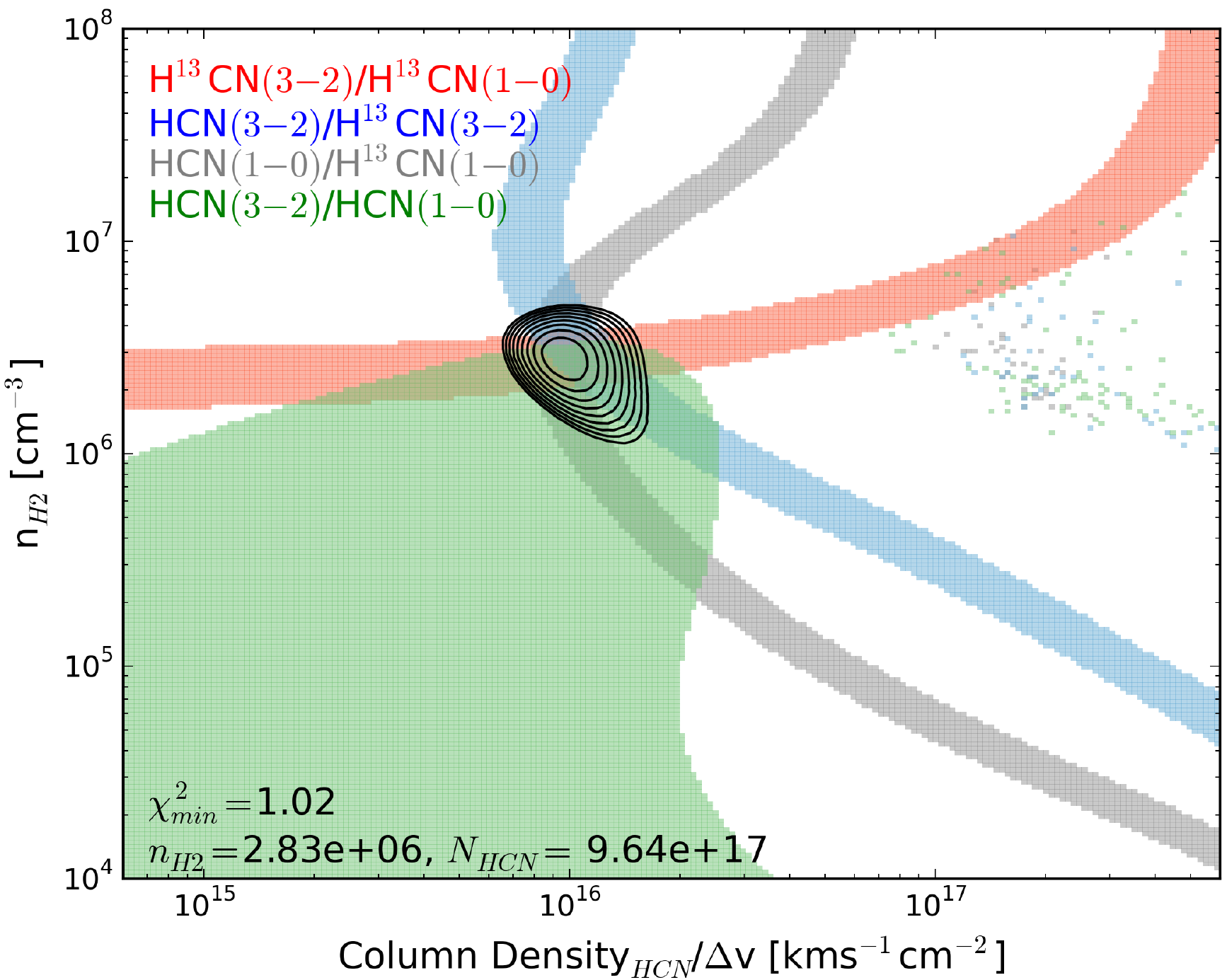}
	\caption{The best fitting parameters for the HCN and H$^{13}$CN ratios. Top: $T_\text{k}=150$\,K, bottom $T_\text{k}=300$\,K, left $T_\text{bg}=45$\,K, right $T_\text{bg}=110$\,K. The $1\sigma$ regions of individual ratios are highlighted, and the $\chi^2$ contours from 1 to 10 in steps of 1 are shown. The ``dotting'' in the upper right corner is due to localised non-convergences in the LVG code, and does not affect our results at these temperatures. These results imply $T_\text{k}\geq150$\,K and prefer $T_\text{bg}=45$\,K for all $T_\text{k}$, and as such we identify the $T_\text{k}=150$\,K, $T_\text{bg}=45$\,K as the best fit $(\chi^2_\text{min}=1.25$. The best fit parameters are found from weighted averages of four best fits shown above, and we find $n_{\text{H}_2} = 2.1\pm0.6\times10^6$\,cm$^{-3}$ and  $N_{\text{HCN}}/\Delta v = 6.2\pm1.6\times10^{15}$\,cm$^{-2}$\,km$^{-1}$\,s.}
	\label{fig:HCN_lvg_results}
	\end{center}
	\end{figure*}

	The background temperature ($T_{\text{bg}}$) was set to 45\,K and 110\,K (\citealp{Mangum2013} and GA12 respectively), and we modelled the kinetic gas temperature $(T_{\text{k}})$ as 70, 100, 150, 300, 350 and 450\,K. The molecular hydrogen density ($n_{\text{H}_2}$) was varied from $10^4 - 10^8$\,cm$^{-3}$ in $\log_{10}$ steps of 0.02, and $N_{\text{HCN}}$ was varied from $6\times10^{14} - 6\times10^{17}$\,cm$^{-2}$ in $\log_{10}$ steps of 0.015. We assume throughout that $\Delta v =  100$\,km\,s$^{-1}$. We further assumed [HCN]/[H$^{13}$CN] = 60 \citep[e.g.\ ][]{Langer1993}, and found $\chi^2$ values for the results grids {which are shown in Fig.\ \ref{fig:HCN_lvg_results}. }
		 
	 For both $T_{\text{bg}}$ values, the models exclude low kinetic temperatures ($T_{\text{k}} \lesssim 100$\,K), which can fit the H$^{13}$CN$(3 - 2)/(1 - 0)$ ratios alone but not the very optically thick HCN. For any given $T_{\text{k}}$ the models prefer the lower $T_{\text{bg}}=45$\,K. We find reduced $\chi^2 = 1.25$ at $T_{\text{k}}=150$\,K and $T_{\text{bg}}=45$\,K and this continues to decrease as $T_{\text{k}}$ increases towards 500\,K. Above this temperature we have no HCN transition data and our RADEX solutions begin to diverge rapidly. We adopt $T_{\text{k}}=150$\,K and $T_{\text{bg}}=45$\,K as our best fit solution, being wary of over-fitting, but note that the higher $T_{\text{k}}$ is consistent with the NH$_3$ thermometry of \citet{Mangum2013}. The green band {in Fig.\ \ref{fig:HCN_lvg_results}}, showing the parameters fitting the observed ratio to within $1\sigma$ for the $L'(\text{HCN}(3 - 2))/L'(\text{HCN}(1 - 0))$ line ratio, is much wider than for the other lines. This ratio of two optically thick lines is only weakly dependent upon $n_{\text{H}_2}$ and $N_{\text{HCN}}/\Delta v$ (provided $N_{\text{HCN}}/\Delta v$ is high enough for the line to be optically thick), and varies slowly across the explored parameter space. The four solutions shown in Fig.\ \ref{fig:HCN_lvg_results} are our four best fits below 350\,K. Given the lack of a clearly superior fit, we find the best fit $n_{\text{H}_2}$ and $N_{\text{HCN}}/\Delta v$ from $\chi^2_\nu$ weighted averages. We find $n_{\text{H}_2} = 2.1\pm0.6\times10^6$\,cm$^{-3}$ and $N_{\text{HCN}}/\Delta v = 6.2\pm1.6\times10^{15}$\,cm$^{-2}$\,km$^{-1}$\,s. 
	 
	 \smallskip
	 
	 Since the $\chi^2$ values for the HCN$ - $H$^{13}$CN LVG models continue to decrease at higher temperatures, we also consider the case that the HCN and H$^{13}$CN emission is dominated by shocks. We repeated the following H$^{13}$CN, H$^{13}$CO$^+$ and SiO analysis with $T_{\text{k}} =450$\,K and $T_{\text{bg}} =110$\,K, and find that while the column densities found are a dex higher for all species, the abundance {ratios} are only slightly changed and are consistent within uncertainties. The abundance ratios presented below are robustly determined, even if the kinetic temperature and background temperatures are not.

	 \smallskip
	 	
	Our best fit $T_{\text{k}}$ and $n_{\text{H}_2}$ are consistent with the maximum likelihood parameters of \citet{Rangwala2011}, which are based on both low J HCN emission and high J absorption seen with Herschel, while our $N_{\text{HCN}}$ is two orders of magnitude higher. They found $T_{\text{k}} = 320$\,K, $n_{\text{H}_2} = 2\times10^6$\,cm$^{-3}$ and $N_{\text{HCN}} = 2\times10^{15}$\,cm$^{-2}$. GA12 found, for $T_{\text{bg}}=110$\,K, $T_{\text{k}} \sim 150$\,K, $n_{\text{H}_2} = 4-8\times10^5$\,cm$^{-3}$ and $N_{\text{HCN}} = 6-4\times10^{17}$\,cm$^{-2}$. While our analysis prefers $T_{\text{bg}}=45$\,K, for the $T_{\text{bg}}$ and $T_{\text{k}}$ of GA12 we find $n_{\text{H}_2} = 3.6\pm1.5\times10^6$\,cm$^{-3}$ and $N_{\text{HCN}} = 6\pm3\times10^{17}$\,cm$^{-2}$. Our best fitting model with $T_{\text{bg}}=45$\,K and $T_{\text{k}} = 150$\,K gives $n_{\text{H}_2} = 2.1\pm0.6\times10^6$\,cm$^{-3}$ and $N_{\text{HCN}}/\Delta v = 6.2\pm1.6\times10^{15}$\,cm$^{-2}$\,km$^{-1}$\,s, consistent with GA12. There is some tension with earlier millimetre studies of HCN in Arp\,220, which found best fitting parameters of $45\text{\,K}<T_{\text{k}}<120$\,K and $n_{\text{H}_2}$ = $0.3\times10^6$\,cm$^{-3}$ \citep{Greve2009}. We note that while \citet{Rangwala2011,Greve2009} and this work all use simple LVG codes, GA12 used a more complex, multiphase model including dust attenuation. All of the above used HCN data averaged over the two nuclei, and only GA12 includes mid-IR pumping. Their results suggested that mid-IR pumping is not having a significant effect on the vibrational ground state, high$-J$ HCN population, but the effect on the low$-J$ transitions may be more significant.

\begin{figure*}
	\begin{center}
	\includegraphics[width=0.49\textwidth]{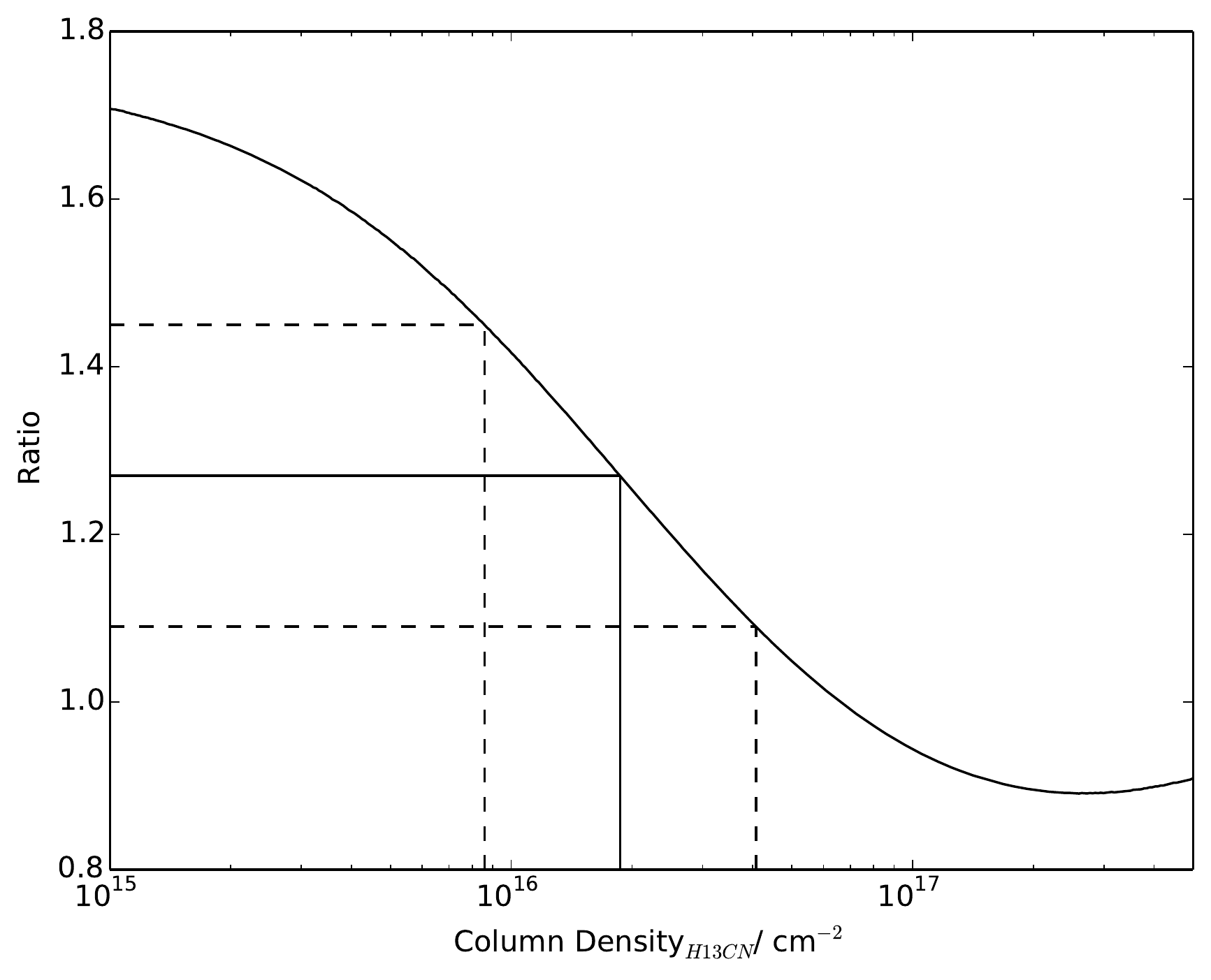}
	\hfill
	\includegraphics[width=0.49\textwidth]{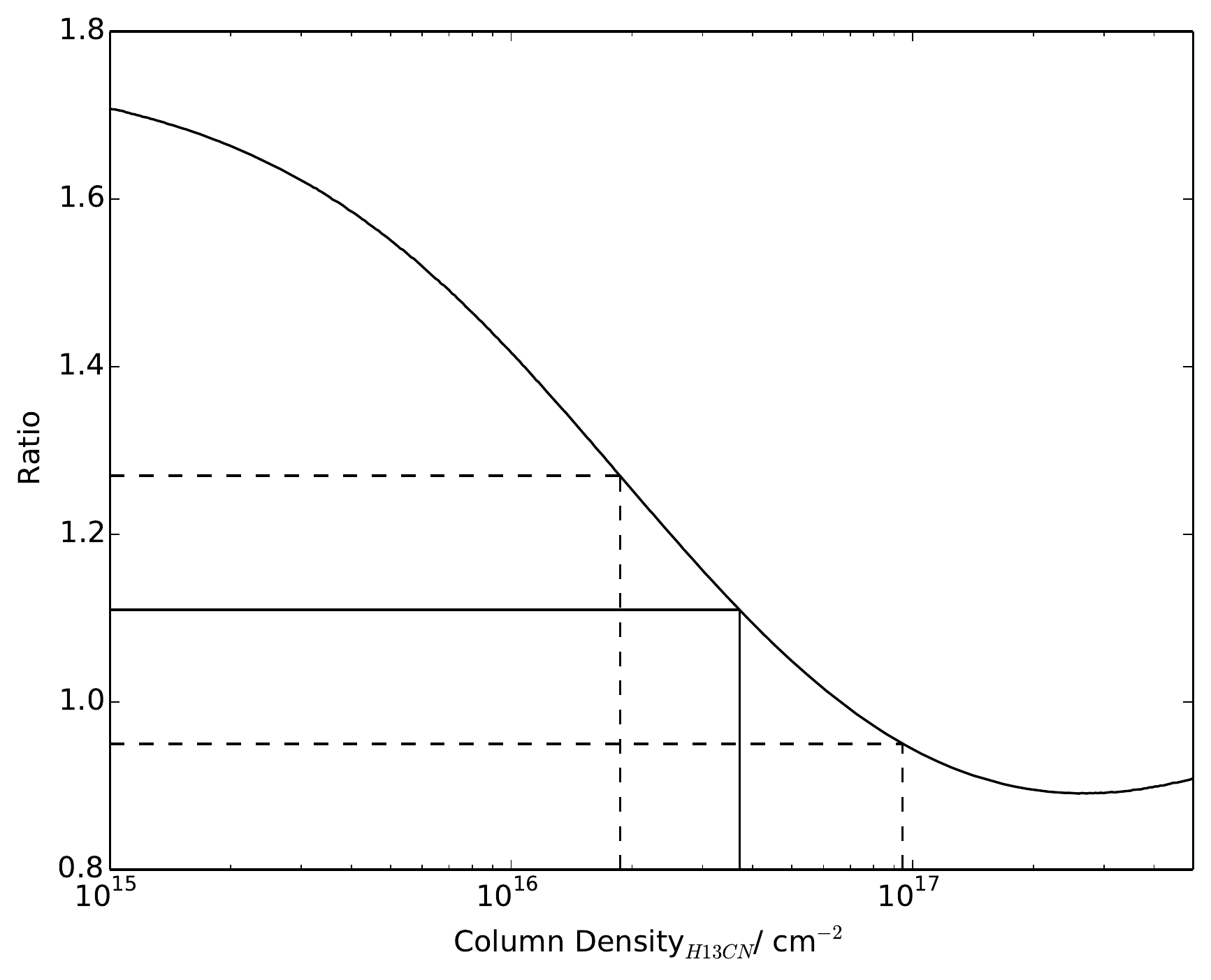}
	\caption{The H$^{13}$CN column densities found in the WN and EN (left and right respectively). The dashed lines indicate the error bounds. We find column densities $N_{\text{H}^{13}\text{CN}}/\Delta v =1.9^{+2.2}_{-1.0} \times10^{14}$\,cm$^{-2}$\,km$^{-1}$\,s and $N_{\text{H}^{13}\text{CN}}/\Delta v = 3.7^{+5.7}_{-1.8}\times10^{14}$\,cm$^{-2}$\,km$^{-1}$\,s in the WN and EN respectively.}\label{fig:h13cnColumn}
	\end{center}
	\end{figure*}

	Having found the ``average'' parameters for the HCN and H$^{13}$CN in Arp\,220, we use the high resolution of our observations to test for chemical differences between the nuclei. To this end, we ran RADEX again, fixing all parameters to the best fit HCN solution except for the H$^{13}$CN column density, and fitted this to the beam dilution corrected $T_\text{B}{\rm d}v$ ratios for H$^{13}$CN $(3 - 2)/(1 - 0)$, $1.27\pm0.18$ and $1.11\pm0.16$ in the WN and EN respectively. The results are shown in Fig.\ \ref{fig:h13cnColumn}. We find $N_{\text{H}^{13}\text{CN}}/\Delta v = 1.9^{+2.2}_{-1.0} \times10^{14}$\,cm$^{-2}$\,km$^{-1}$\,s in the WN and $N_{\text{H}^{13}\text{CN}}/\Delta v = 3.7^{+5.7}_{-1.8}\times10^{14}$\,cm$^{-2}$\,km$^{-1}$\,s in the EN. We note that these solutions are optically thick, with optical depths from 1 to 6. 
	
	\subsubsection{H$^{13}$CO$^+$}
	For H$^{13}$CO$^+$ we analyse the two nuclei separately, making use of the $T_{\text{B}}{\rm d}v$ ratios of H$^{13}$CN$(3 - 2)$/H$^{13}$CO$^+(3 - 2)$ and H$^{13}$CN$(1 - 0)$/H$^{13}$CO$^+(1 - 0)$, which do not require corrections for beam dilution (see Table \ref{tab:LRs}). We use a simple two phase model, where the H$^{13}$CN column and $n_{\text{H}_2}$ are fixed to the values found above for the WN and EN (with $\Delta v=100$\,km\,$^{-1}$). In case 1 we fix $n_{\text{H}_2}$ for H$^{13}$CO$^+$, as with only upper limits we cannot simultaneously constrain the hydrogen density and column density, while in case 2 $n_{\text{H}_2}$ and $N_{\text{H}^{13}\text{CO}^+}/\Delta v$ for H$^{13}$CO$^+$ are allowed to vary over the full range described previously for the HCN/H$^{13}$CN modelling. In both cases $T_{\text{k}}$ and $T_{\text{bg}}$ are fixed to the best fit values of 150\,K and 45\,K respectively from the previous section. We perform the analysis for both case 1 and case 2.
	We ran grids using the upper, middle and lower values for the H$^{13}$CN column density found above, and used the maximum and minimum values of the resultant H$^{13}$CO$^+$ column density to estimate uncertainties.

\begin{figure*}
	\begin{center}
	\includegraphics[width=0.49\textwidth]{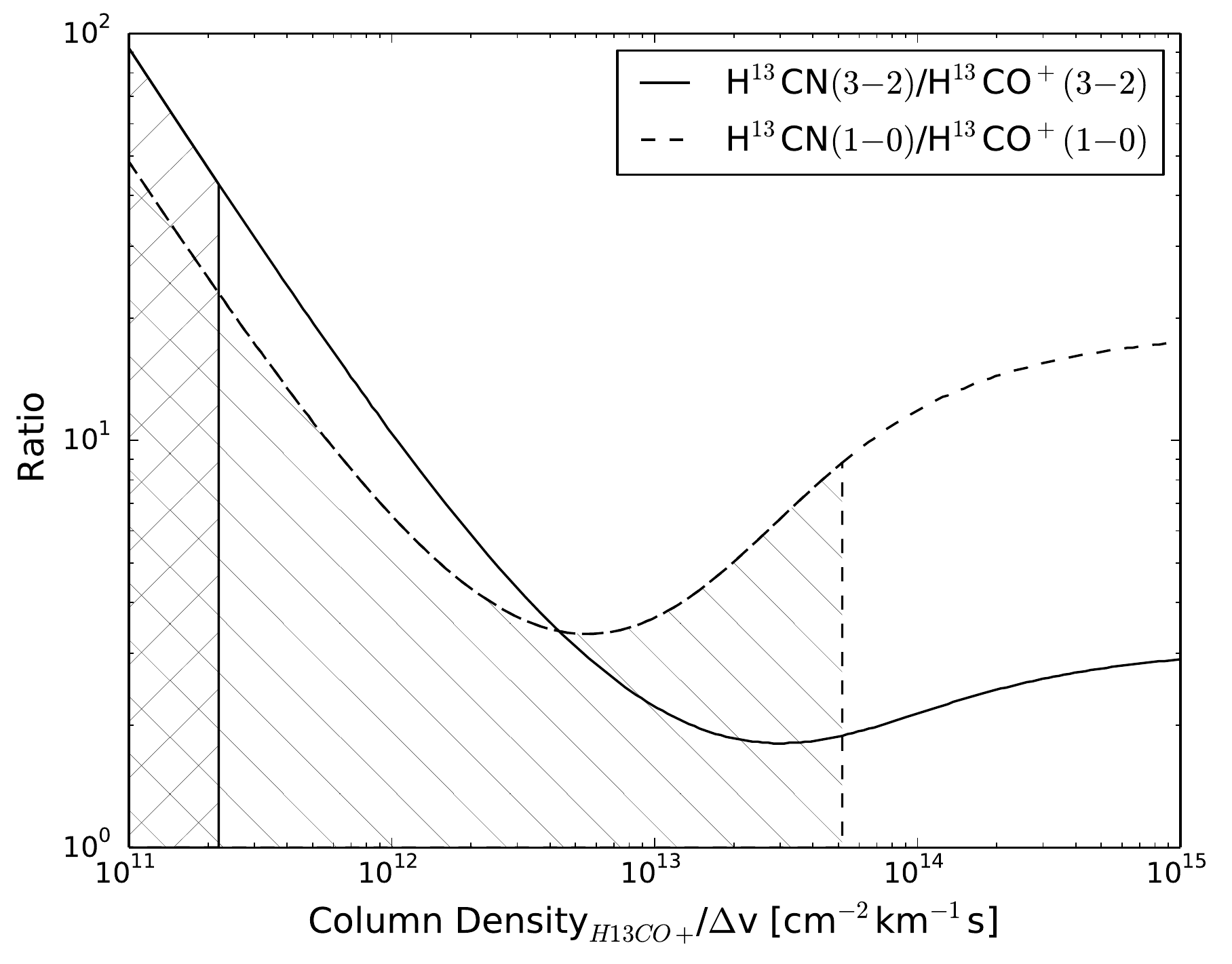}	
	\hfill
	\includegraphics[width=0.49\textwidth]{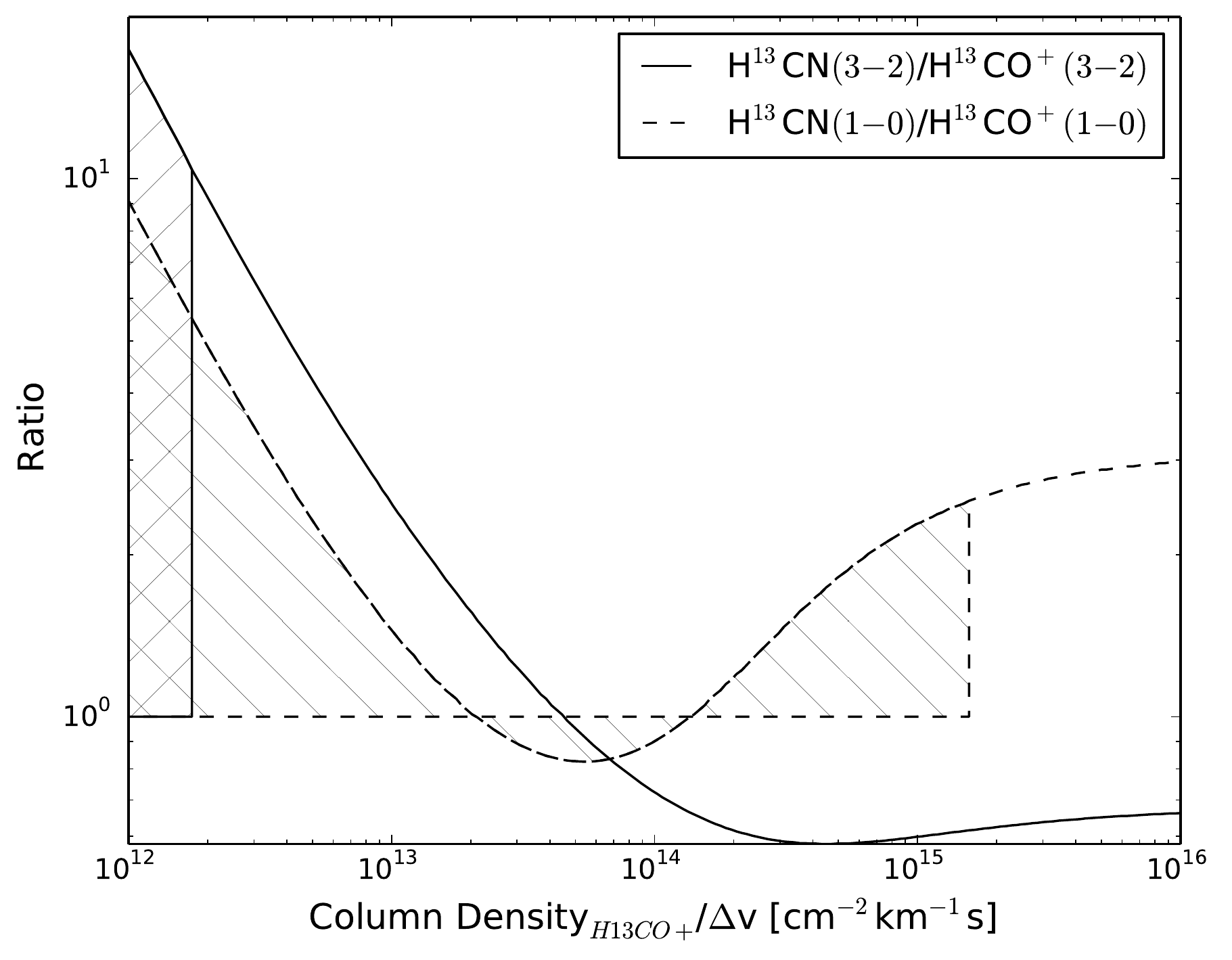}
	\caption{The regions of H$^{13}$CO$^+$ column density permitted by the lower limit ratios of case 1 in the WN (left) and EN (right). The hashed regions indicate the permitted columns for each ratio assuming $n_{\text{H}_2}=2.1\times10^6$\,cm$^{-3}$. Higher column densities are permitted for lower $n_{\text{H}_2}$ values. We find upper limits on the H$^{13}$CO$^+$ column density of $2\times10^{11}$\,cm$^{-2}$\,km$^{-1}$\,s  and $2\times10^{12}$\,cm$^{-2}$\,km$^{-1}$\,s in the WN and EN respectively.}\label{fig:h13co_lims}
	\end{center}
	\end{figure*}

	{Case 1:} The column densities permitted by the lower limit ratios are shown in Fig.\ \ref{fig:h13co_lims}. We find upper limits on the H$^{13}$CO$^+$ column density of $2\times10^{11}$\,cm$^{-2}$\,km$^{-1}$\,s  and $2\times10^{12}$\,cm$^{-2}$\,km$^{-1}$\,s in the WN and EN respectively. These correspond to lower limits on the abundance ratio [H$^{13}$CN]/[H$^{13}$CO$^+$] of 1000 and 200 in the WN and EN respectively. These extremely high limits are dramatically decreased by reducing $n_{\text{H}_2}$ in the H$^{13}$CO$^+$ phase, and at $n_{\text{H}_2}=5\times10^4$\,cm$^{-3}$ the limits are 44 and 60.
	
\begin{figure*}
	\begin{center}
	\includegraphics[width=0.49\textwidth]{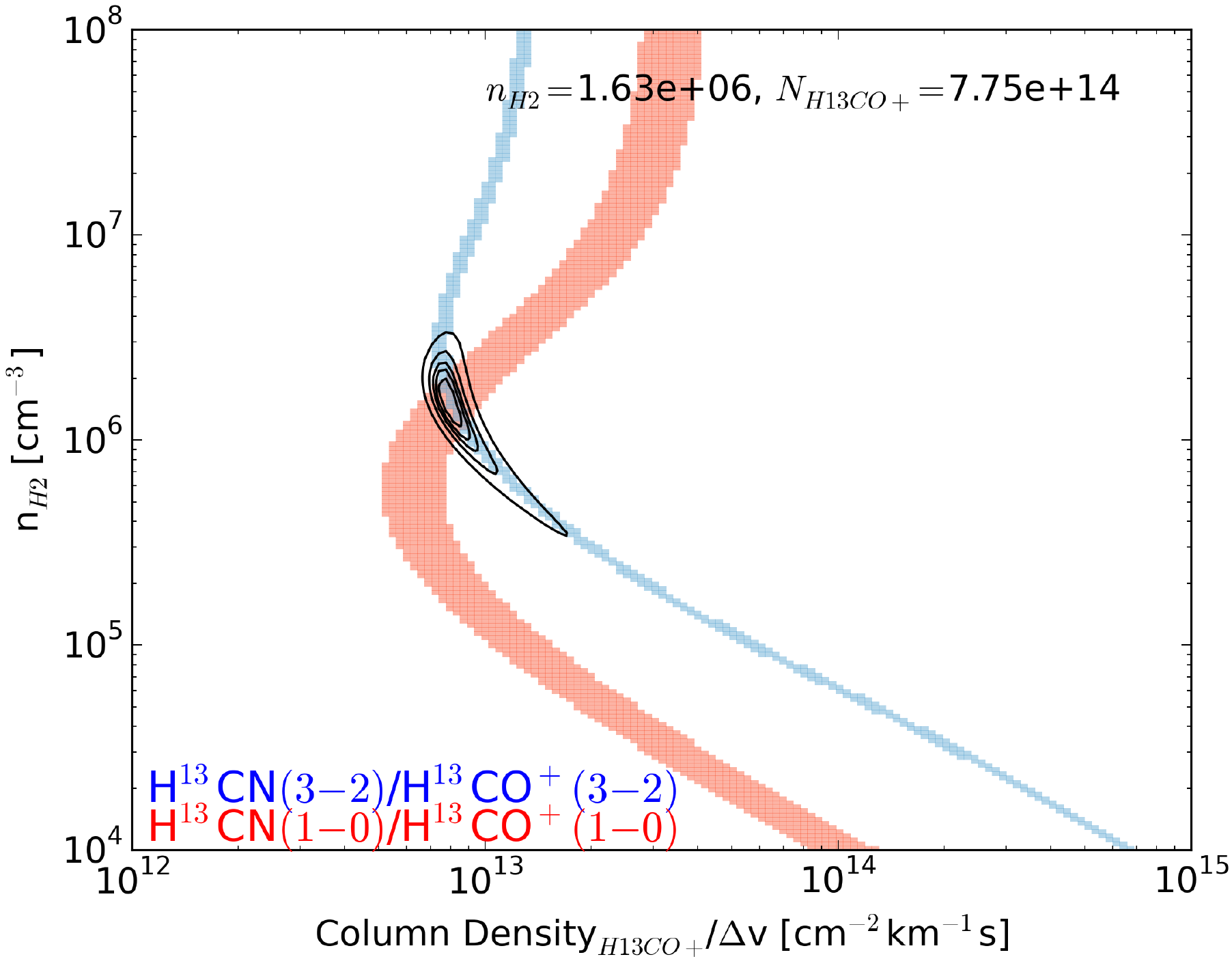}	
	\hfill
	\includegraphics[width=0.49\textwidth]{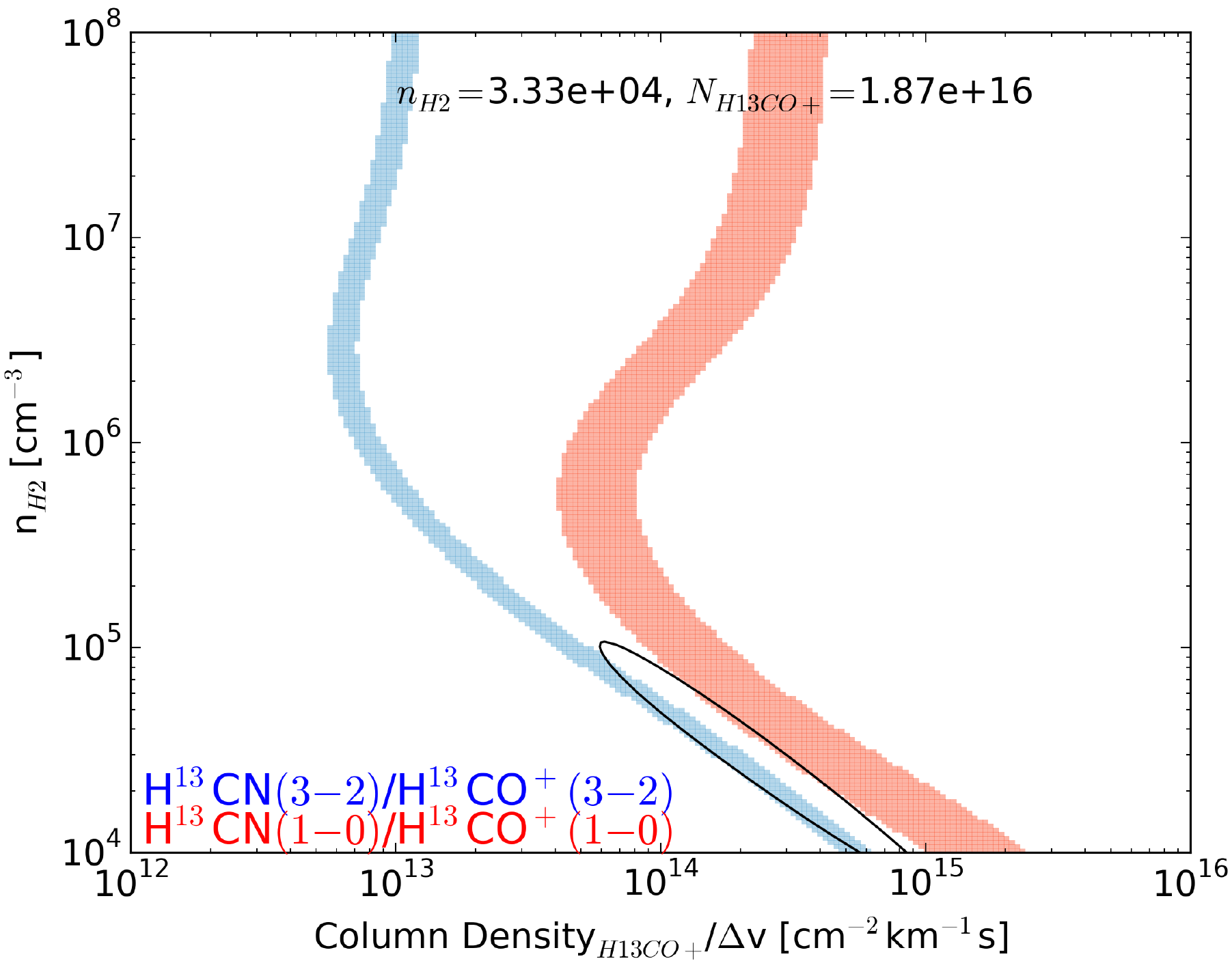}
	\caption{The error regions for the modelled H$^{13}$CO$^+$ line ratios in the WN (left) and EN (right). $\chi^2$ contours are overlaid at 1, 2, 3 ,5 and 10. The minimum $\chi^2$ locations are indicated in the top right corners of the plots. Best fits give an abundance ratio [H$^{13}$CN]/[H$^{13}$CO$^+$] of $25^{+16}_{-4}$ in the WN, and $2.0^{+4.2}_{-1.5}$ in the EN.}\label{fig:h13co_lvg}
	\end{center}
	\end{figure*}

	{Case 2:} We found best fit solutions in the two nuclei $N_{\text{H}^{13}\text{CO}^+}/\Delta v = 8^{+7}_{-4}\times10^{12}$cm$^{-2}$\,km$^{-1}$\,s in the WN and $N_{\text{H}^{13}\text{CO}^+}/\Delta v = 1.9^{+6.6}_{-1.3}\times10^{14}$cm$^{-2}$\,km$^{-1}$\,s in the EN. The grids for the best fit H$^{13}$CN column densities are shown in Fig.\ \ref{fig:h13co_lvg}. The uncertainties are {predominantly} driven by the uncertainty in the H$^{13}$CN column density, so that the abundance ratio [H$^{13}$CN]/[H$^{13}$CO$^+$] is relatively well determined as $25^{+16}_{-4}$ in the WN, and $2.0^{+4.2}_{-1.5}$ in the EN.
	
	\subsubsection{SiO}

\begin{figure*}
	\begin{center}
	\includegraphics[width=0.49\textwidth]{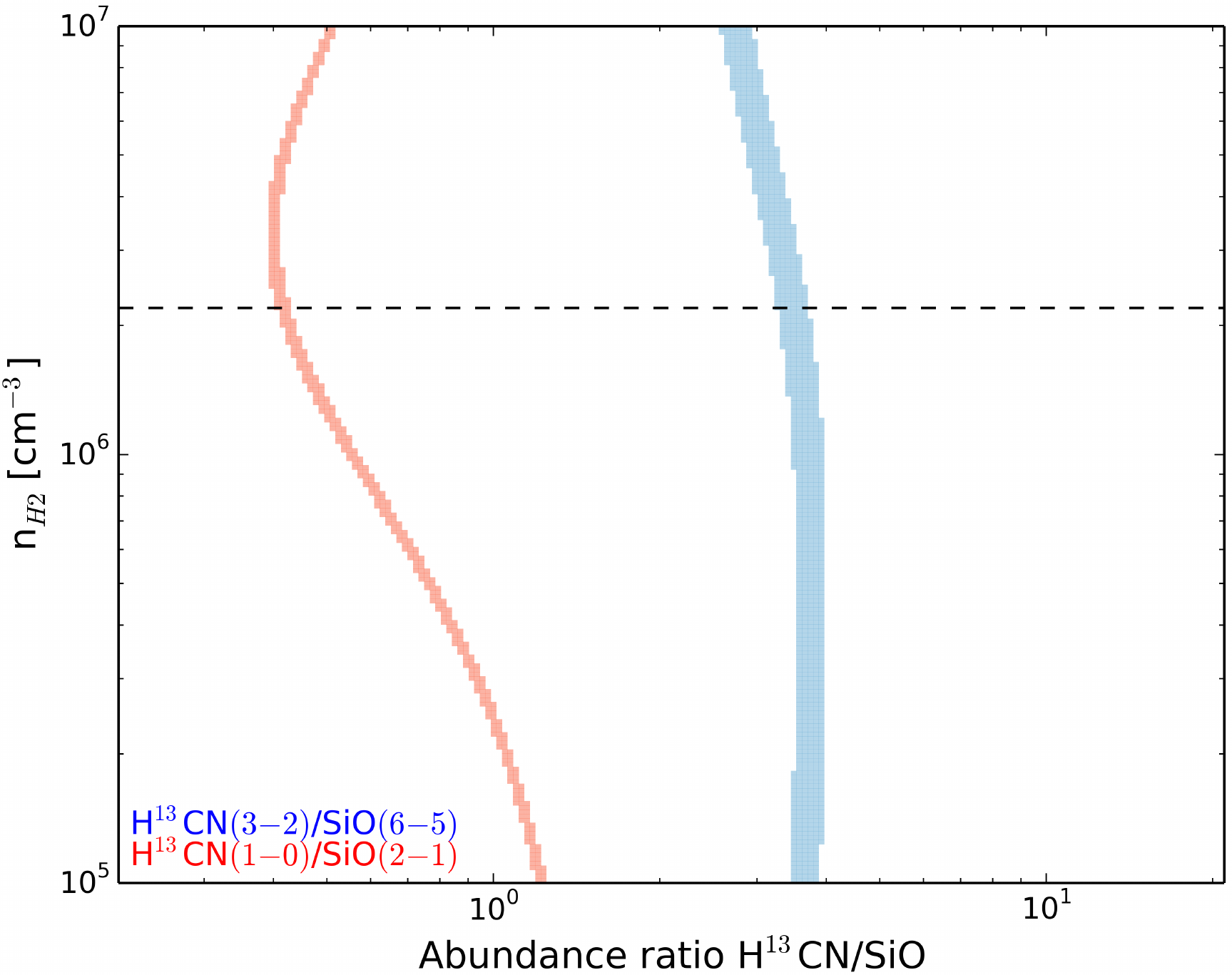}
	\hfill
	\includegraphics[width=0.49\textwidth]{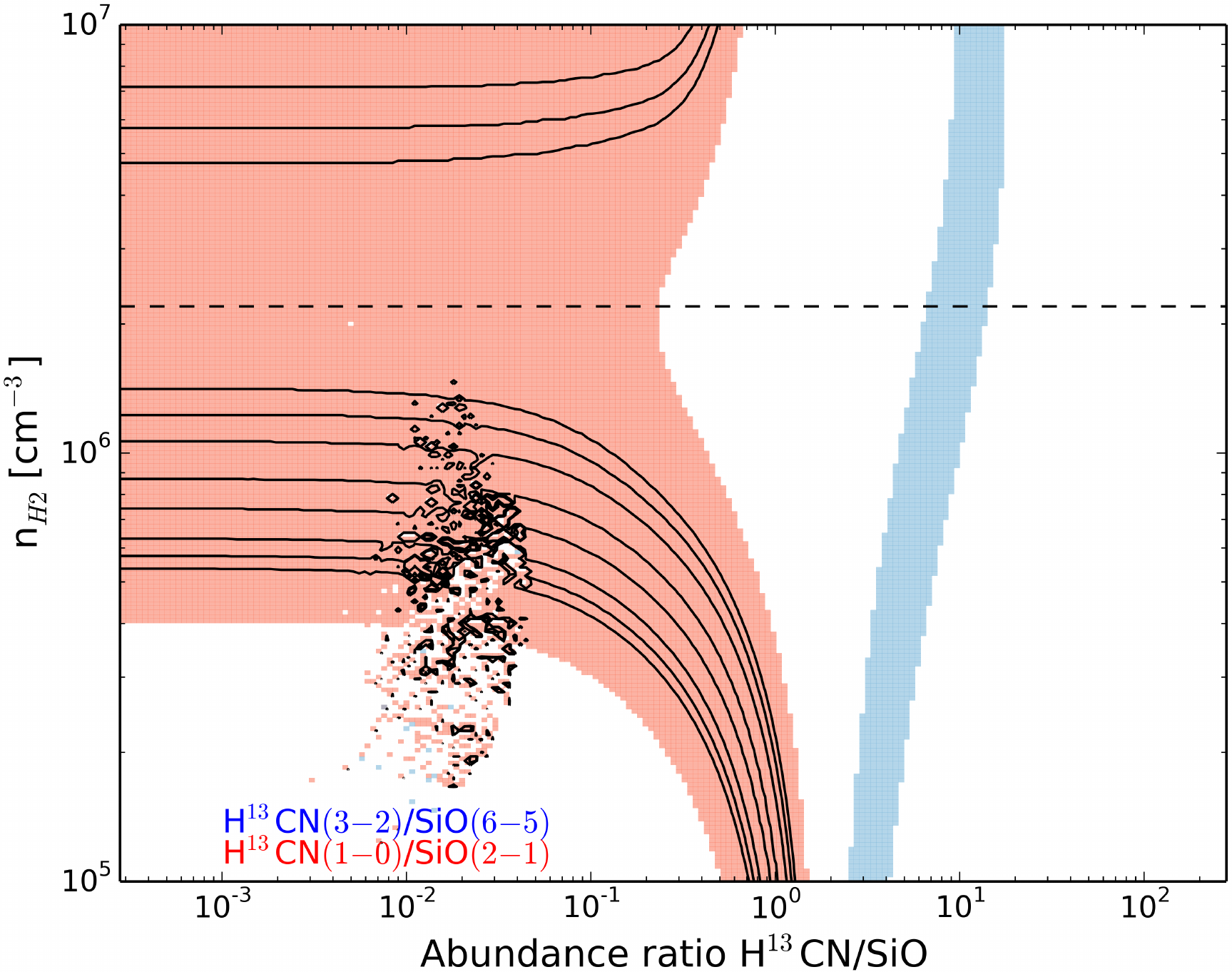}\\
	\includegraphics[width=0.49\textwidth]{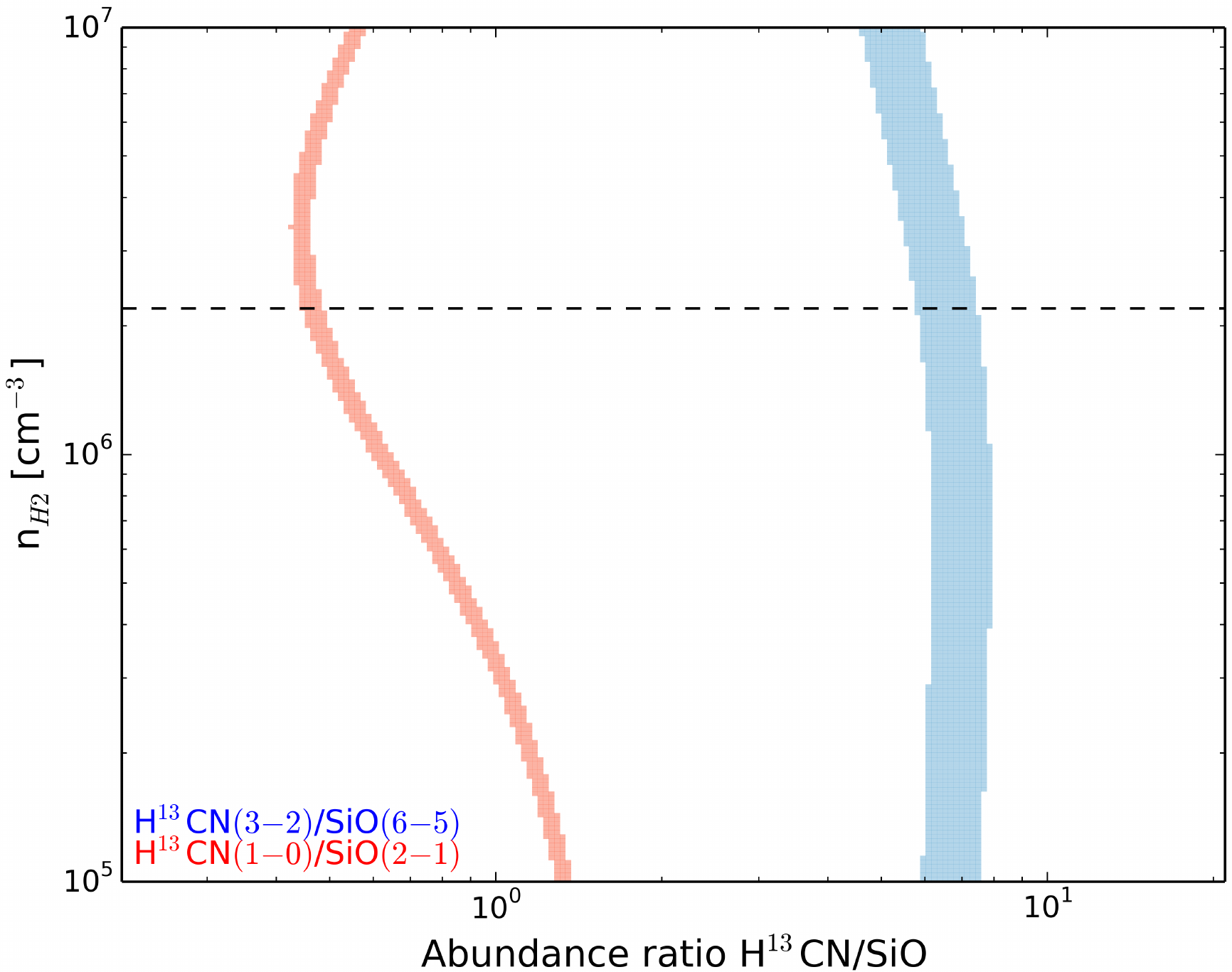}
	\hfill
	\includegraphics[width=0.49\textwidth]{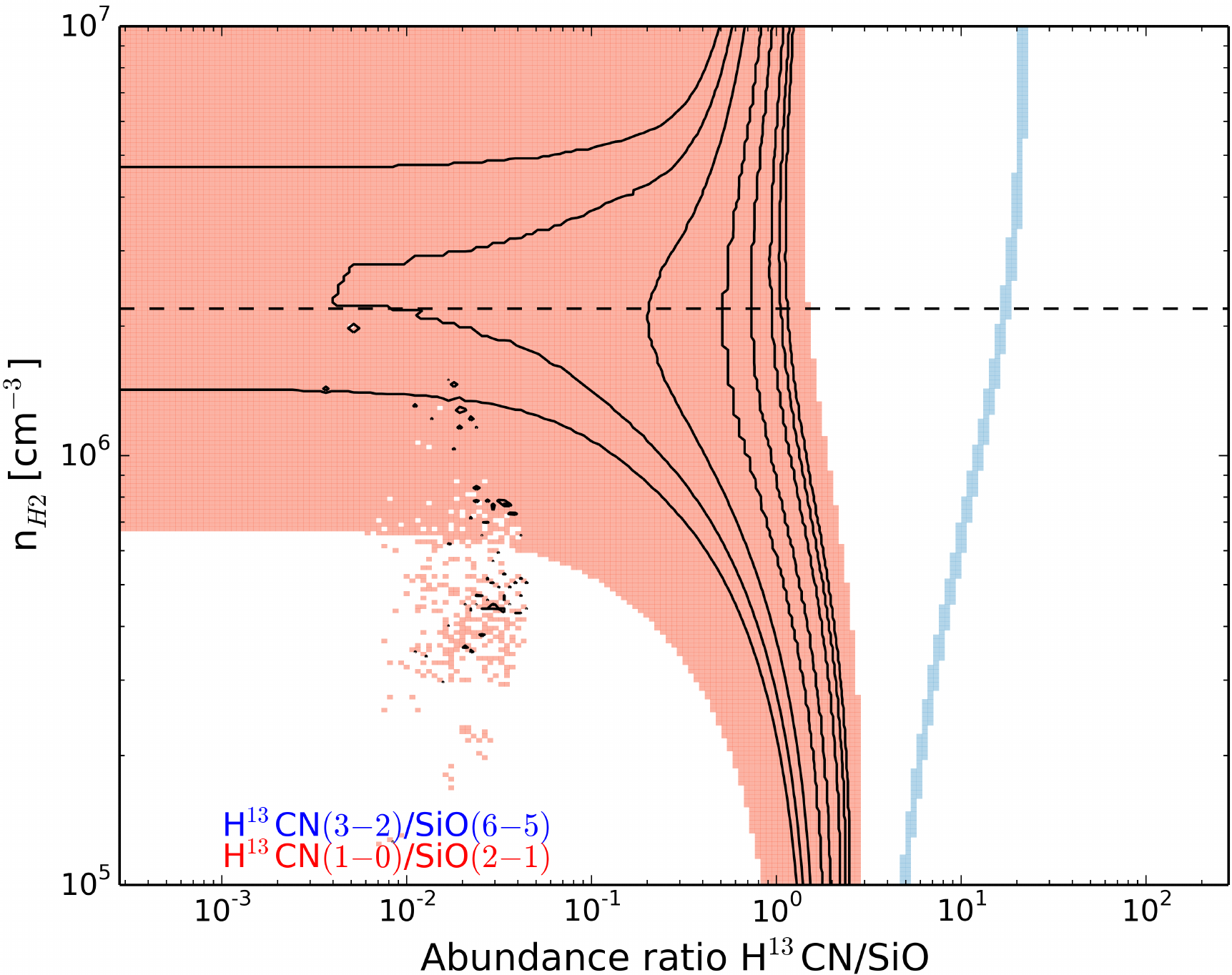}
	\caption{The $1\sigma$ regions for the modelled [H$^{13}$CN]/[SiO] abundance ratio in the WN (left) and EN (right) for both case 1 (top) and case 2 (bottom). The dashed horizontal line marks the best fitting $n_{\text{H}_2}$ from HCN modelling, and the EN plots show the $\chi^2$ contours for the 3.5\,mm component at 0.01, 0.1, 0.2 and 0.3. Numerical results of these models are shown in Table \protect\ref{tab:sio_abund}.}\label{fig:sio_abund}
	\end{center}
	\end{figure*}

We used the upper, middle and lower H$^{13}$CN column densities found for each nucleus, and restricted our parameters to $T_{\text{k}}=150$\,K, $T_{\text{bg}}=45$\,K and $\Delta v =100$\,km\,s$^{-1}$. The maximum and minimum values of the resultant SiO column density were used to estimate uncertainties. We ran grids for the parameter range $n_{\text{H}_2}$ for $10^5 - 10^7$\,cm$^{-3}$ in $\log_{10}$ steps of 0.01, while $N_{\text{SiO}}$ was varied to include the range permitted by the ratios. This was particularly significant for the {3.5\,mm ratio in the} EN, where the {ratio alone} cannot exclude extremely low [H$^{13}$CN]/[SiO] ratios. We used the $T_{\text{B}}{\rm d}v$ ratios, excluding the negative components of the SiO$(6 - 5)$ line. The results for both case 1 and 2 are shown in Fig.\ \ref{fig:sio_abund}. {Numerical results for the SiO fitting are collated in Table \ref{tab:sio_abund}.}

The results for case 1 and case 2 vary as expected, with higher [H$^{13}$CN]/[SiO] ratios for case 2, although the ratios are consistent within uncertainties. We find consistent [H$^{13}$CN]/[SiO] ratios for the two nuclei of $0.74^{+0.11}_{-0.08}$ and $0.28^{+0.50}_{-0.20}$ in the WN and EN respectively for case 1 and $0.87^{+0.11}_{-0.10}$ and $0.67^{+0.99}_{-0.14}$ in the WN and EN respectively for case 2. {This would suggest that the elevated [H$^{13}$CN]/[H$^{13}$CO$^+$] abundance ratio in the WN is due to either a process which is elevating the [H$^{13}$CN] and [SiO] abundances in tandem, or a process which is lowering the [H$^{13}$CO$^+$] abundance.}


	
	\begin{table*}
	\begin{center}
	\caption{sio abundance results}\label{tab:sio_abund}
	\begin{tabular}{l c c c c}\\ \hline\hline
	& \multicolumn{2}{c}{WN} & \multicolumn{2}{c}{EN}\\ 
	& [H$^{13}$CN]/[SiO] & $N_{\text{SiO}} /\Delta v  $ & [H$^{13}$CN]/[SiO] & $N_{\text{SiO}}/\Delta v  $\\
	& & $[10^{13}\text{cm}^{-2}\text{\,km}^{-1}\text{\,s}]$ & & $[10^{13}\text{cm}^{-2}\text{\,km}^{-1}\text{\,s}]$ \\ \hline\hline
	& \multicolumn{4}{c}{Case 1: no H$^{13}$CO$^+$}\\[0.1cm]
	1.2\,mm & $3.4^{+1.5}_{-0.8}$ & $5.6^{+1.7}_{-1.7}$ & $3.3^{+3.8}_{-1.2}$ & $5.8^{+3.3}_{-3.1}$\\[0.1cm]
	3.5\,mm & $0.42^{+0.04}_{-0.05}$ & $46^{+6}_{-4}$ & $0.15^{+0.35}_{-0.10}$ & $130^{+250}_{-90}$\\[0.1cm]
	Mean\footnote{{Since the varied parameter is the SiO column density, the mean here is the arithmetic mean for $N_\text{SiO}/\Delta v$, and the harmonic mean for the [H$^{13}$CN]/[SiO] abundance ratio.}} & $0.74^{+0.11}_{-0.08}$ & $25.5^{+3.8}_{-2.9}$ & $0.28^{+0.50}_{-0.20}$ & $70^{+130}_{-50}$ \\[0.1cm]
	& \multicolumn{4}{c}{Case 2: H$^{13}$CO$^+$}\\[0.1cm]
	1.2\,mm & $6.4^{+3.9}_{-2.1}$ & $3.0^{+1.4}_{-1.1}$ & $5.1^{+4.0}_{-1.6}$ & $3.7^{+1.7}_{-1.6}$\\[0.1cm]
	3.5\,mm & $0.47^{+0.05}_{-0.04}$ & $41^{+4}_{-4}$ & $0.36^{+0.08}_{-0.22}$ & $53^{+83}_{-10}$\\[0.1cm]
	Mean$^{\text{a}}$ & $0.87^{+0.11}_{-0.10}$ & $22.0^{+2.8}_{-2.4}$ & $0.67^{+0.99}_{-0.14}$ & $29^{+42}_{-6}$ \\ \hline
	\end{tabular}
	\end{center}
	\end{table*}
	
	\begin{table*}
	\begin{center}
	\caption{Abundance ratio comparison for $T_{\text{k}}=150$\,K, $T_\text{bg}=45$\,K and $T_{\text{k}}=450$\,K, $T_\text{bg}=110$\,K}\label{tab:T_comp}
	\begin{tabular}{l c c c c}\\ \hline\hline
	& \multicolumn{2}{c}{WN} & \multicolumn{2}{c}{EN}\\ 
	& \multicolumn{4}{c}{$T_{\text{k}}=150$\,K, $T_\text{bg}=45$\,K}\\[0.1cm]
	& [H$^{13}$CN]/[H$^{13}$CO$^+$] &  [H$^{13}$CN]/[SiO] & [H$^{13}$CN]/[H$^{13}$CO$^+$] &  [H$^{13}$CN]/[SiO]\\[0.1cm] \hline
	Case 1 & $>950$ & $0.74^{+0.11}_{-0.08}$ & $>185$ & $0.28^{+0.53}_{-0.20}$\\[0.1cm]
	Case 2 & $25^{+16}_{-4}$ & $0.87^{+0.11}_{-0.10}$ & $2.0^{+4.2}_{-1.5}$ & $0.67^{+0.99}_{-0.14}$\\[0.3cm]
	 & \multicolumn{4}{c}{$T_{\text{k}}=450$\,K, $T_\text{bg}=110$\,K}\\[0.1cm]
	& [H$^{13}$CN]/[H$^{13}$CO$^+$] &  [H$^{13}$CN]/[SiO] & [H$^{13}$CN]/[H$^{13}$CO$^+$] &  [H$^{13}$CN]/[SiO]\\[0.1cm] \hline
	Case 1 & $>640$ & $0.60^{+0.09}_{-0.07}$ & $>180$ & $0.171^{+0.003}_{-0.092}$\\[0.1cm]
	Case 2 & $23^{+17}_{-6}$ & $0.70^{+1.0}_{-0.09}$ &$2.0^{+3.3}_{-1.7}$ & $0.42^{+1.60}_{-0.17}$\\\hline
	\end{tabular}
	\end{center}
	\end{table*}
	
	\begin{table*}
	\begin{center}
	\caption{LVG model abundance results}\label{tab:lvg_results}
	\begin{tabular}{l c c c }\\ \hline\hline
	& Column densities & \multicolumn{2}{c}{Abundance ratios\footnote{Assuming $\Delta v=100$\,km\,s$^{-1}$.}}\\ 
	& $N_{X} / \Delta v  $ & [$X$]/[CO]\footnote{CO column densities are taken from the maximum likelihood estimates of \citet{Rangwala2011} as the sum of the cold and hot phase columns: $N_{\text{CO}} = 2.2\times10^{20}$\,cm$^{-2}$, i.e.\ [CO]/[H$_2$] = $2.2\times10^{-5}$.} & [$X$]/[H$_2$]\footnote{For $N_{\text{H}_2} = 1\times10^{25}$\,cm$^{-2}$.}  \\
	& $[10^{13}\text{cm}^{-2}\text{\,km}^{-1}\text{\,s}]$ &  &\\ \hline\hline
	HCN & $620^{+160}_{-160}$ & $2.8^{+0.7}_{-0.7}\times10^{-3}$ & $6.2^{+1.6}_{-1.6}\times10^{-8}$ \\[0.1cm]
	H$^{13}$CN$_{\text{WN}}$ & $19^{+22}_{-10}$ & $9^{+10}_{-5}\times10^{-5}$& $1.9^{+2.2}_{-1.0}\times10^{-9}$ \\[0.2cm]
	H$^{13}$CN$_{\text{EN}}$ & $37^{+57}_{-18}$ & $1.7^{+2.6}_{-0.8}\times10^{-4}$& $3.7^{+5.7}_{-1.8}\times10^{-9}$ \\[0.3cm]
	
	\multicolumn{4}{c}{Case 1: no H$^{13}$CO$^+$}\\\hline
	H$^{13}$CO${^+}_{\text{WN}}$ & $<0.02$ & $<9\times10^{-8}$  & $<2\times10^{-12}$\\[0.2cm]
	H$^{13}$CO${^+}_{\text{EN}}$ & $<0.2$ & $<9\times10^{-7}$ & $<2\times10^{-11}$ \\[0.2cm]
	SiO$_{\text{WN}}$ & $25.5^{+3.8}_{-2.9}$ &$1.16^{+0.17}_{-0.13}\times10^{-4}$ & $2.55^{+0.38}_{-0.29}\times10^{-9}$\\[0.2cm]
	SiO$_{\text{EN}}$ & $70^{+130}_{-50}$ & $3.2^{+5.9}_{-2.3}\times10^{-4}$ & $7^{+13}_{-5}\times10^{-9}$\\[0.3cm]
	
	\multicolumn{4}{c}{Case 2: H$^{13}$CO$^+$}\\\hline
	H$^{13}$CO${^+}_{\text{WN}}$ & $0.8^{+0.7}_{-0.4}$ & $3.6^{+3.2}_{-1.8}\times10^{-6}$  & $8^{+7}_{-4}\times10^{-11}$\\[0.2cm]
	H$^{13}$CO${^+}_{\text{EN}}$ & $19^{+66}_{-13}$ & $9^{+30}_{-6}\times10^{-5}$ & $1.9^{+6.6}_{-1.3}\times10^{-9}$ \\[0.2cm]
	SiO$_{\text{WN}}$ & $22.0^{+2.8}_{-2.4}$ &$1.00^{+0.13}_{-0.11}\times10^{-4}$ & $2.20^{+0.28}_{-0.24}\times10^{-9}$\\[0.2cm]
	SiO$_{\text{EN}}$ & $29^{+42}_{-6}$ & $1.32^{+1.91}_{-0.27}\times10^{-4}$ & $2.9^{+4.2}_{-0.6}\times10^{-9}$\\ \hline
	\end{tabular}
	\end{center}
	\end{table*}

\subsection{HC$^{15}$N$(3 - 2)$ or SO($6_6 - 5_5$)?}\label{subsec:hc15n}

We attempted to distinguish between HC$^{15}$N$(3 - 2)$ and SO($6_6 - 5_5$) by using our HCN parameter results as inputs to LVG models. Assuming [HCN]/[H$^{13}$CN] = 60 and [HCN]/[HC$^{15}$N] = 200, we predict an integrated brightness temperature ratio in the WN for HC$^{15}$N$(3 - 2)$/H$^{13}$CN$(3 - 2)$ of 0.4. Using the SO column density from \citet{Martin2011} of $\sim10\times10^{14}$\,cm$^{-2}$ the LVG model predicts an integrated ratio SO$(6_6 - 5_5)$/H$^{13}$CN$(3 - 2)$ = 0.006. Our observed value of $\sim2$, suggesting that the spectral feature may be dominated by HC$^{15}$N and not SO, or really is a blending of the two. However the $^{15}$N isotopic abundance is very uncertain, and could be $3\times$ lower. Furthermore, the beam size of \citet{Martin2011} is almost $300\times$ greater than ours: if their observed SO$(5_5 - 4_4)$ emission is as tightly centred about the nuclei as our SO$(6_6 - 5_5)$, then their LTE analysis could be heavily affected by beam dilution, underestimating the SO column density. Therefore we cannot exclude either possibility with these results, but if the spectral feature is dominated by HC$^{15}$N$(3 - 2)$, it is consistent with an abundance ratio [HCN]/[HC$^{15}$N] $\lesssim 200$. 

\bigskip

{The results of the above LVG analysis are collated in Table \ref{tab:T_comp}, where the results for $T_\text{k}=450$\,K and $T_\text{bg}=110$\,K are also shown. We found that while the higher $T_\text{k}$ and $T_\text{bg}$ lowered the absolute abundances by over a dex, the abundance ratios are practically unchanged and are consistent within uncertainties. Our LVG derived abundance ratios therefore appear to be quite robustly determined over a wide range in kinetic temperature. The column densities over $\Delta v$ and CO and H$_2$ abundance ratios for our $T_\text{k}=150$\,K and $T_\text{bg}=45$\,K results are given in Table \ref{tab:lvg_results}. The quoted uncertainties are propagated from the LVG modelling; the errors due to the uncertainty in $T_\text{k}$ are not included but, as noted previously, could be as large as a dex.}
	
\section{Discussion}\label{sec:discuss}

\subsection{Chemical Evidence for an AGN in the WN?}

An increasingly common strategy to identify the AGN contribution to the bolometric luminosities of ULIRGs has been to study their HCN/HCO$^+$ integrated intensity ratios \citep{Kohno2001,Imanishi2006,Papadopoulos2007,Krips2008,Imanishi2009}. The general trend, as described by \citet{Krips2008} is that of increasing ($J \rightarrow J-1)/(J-1 \rightarrow J-2)$ intensity ratios for HCN and HCO$^+$ transitions, as well as a decreasing HCN$(J \rightarrow J-1)$/HCO$^+(J \rightarrow J-1)$ ratio, for an increased starburst contribution. These ratios seem to follow a trend from PDR dominated emission in starbursts, to XDR dominated emission in the most AGN dominated galaxies. \citet{Krips2008} single dish analysis of Arp\,220 within this framework found it to not fit well with any of their other galaxies (starburst (SB) dominated, SB-AGN composite or AGN dominated), but that it was closest to NGC 6951, an SB dominated galaxy with a weak AGN contribution. This possibility of an AGN has been challenged however by \citet{Costagliola2011}, who found strong emission from HC$_{3}$N, a molecule thought to be readily destroyed in even hard UV environments, let alone XDRs, associated with the highest HCN/HCO$^+$ ratios. Instead, hot core chemistry models \citep{Bayet2006,Bayet2008} simultaneously explain both decreased HCO$^+$ and elevated HC$_{3}$N. Throughout all of this however, we must remember that we are averaging over the entire nuclear disk in our beam, covering dense clouds, diffuse gas, hot cores, PDRs and XDRs, all with potentially large cosmic ray contributions \citep[evident from excited OH$^+$ emission][]{Gonzalez-Alfonso2013}, due to the observed supernova rate of $\sim$4 a year, so the two phase LVG model we employ for H$^{13}$CN and H$^{13}$CO$^+$ will be an extreme simplification.

Most recently, \citet{Martin2014} have suggested that elevated [HCN]/[HCO$^+$] {abundance} ratios are due not to XDR or PDR chemistry, but rather extensive shock chemistry in the the turbulent regions of the CND. This would be consistent with the kinematics of the our 1.2\,mm H$^{13}$CN and SiO lines in the WN, which suggest an outflow originating close to the continuum centre, as well as the high densities and kinetic temperatures of the LVG fits. \citet{GarciaBurillo2014} observed an AGN driven molecular outflow in NGC 1068 with an elevated HCN$(4 - 3)$/HCO$^+(4 - 3)$ brightness temperature ratio $(\sim2.5)$ in the regions immediately surrounding the AGN (within 140\,pc), but significantly lower $(\sim1.3)$ directly towards the AGN. This is consistent with the elevated HCN/HCO$^+$ ratios being caused by AGN driven mechanically dominated regions (MDRs), and not XDRs. 

\citet{Meijerink2005,Meijerink2007,Meijerink2011} extensively modelled PDRs and XDRs over a range of densities and ionisation rates in an attempt to identify unique tracers of the two regions. For the high density environment of the HCN our [H$^{13}$CN]/[H$^{13}$CO$^+$] ratios are consistent with both PDR and XDR results. Additionally, \citet{Meijerink2011} studied the effects of cosmic rays and mechanical heating, and found evidence that strong mechanical heating can lead to very elevated HCN and HNC abundances ($10^{-4}$ and $10^{-6}$ respectively) and low [HNC]/[HCN] ratios ($1 - 0.01$). While our abundances are very uncertain, the HN$^{13}$C$(3 - 2)$/H$^{13}$CN$(3 - 2)$ line integrated ratios of $0.71\pm0.04$ and $0.66\pm0.04$ are consistent with shock-heating, but not with high density XDRs. 

\begin{figure}[!h]
\begin{center}
\includegraphics[width=0.49\textwidth]{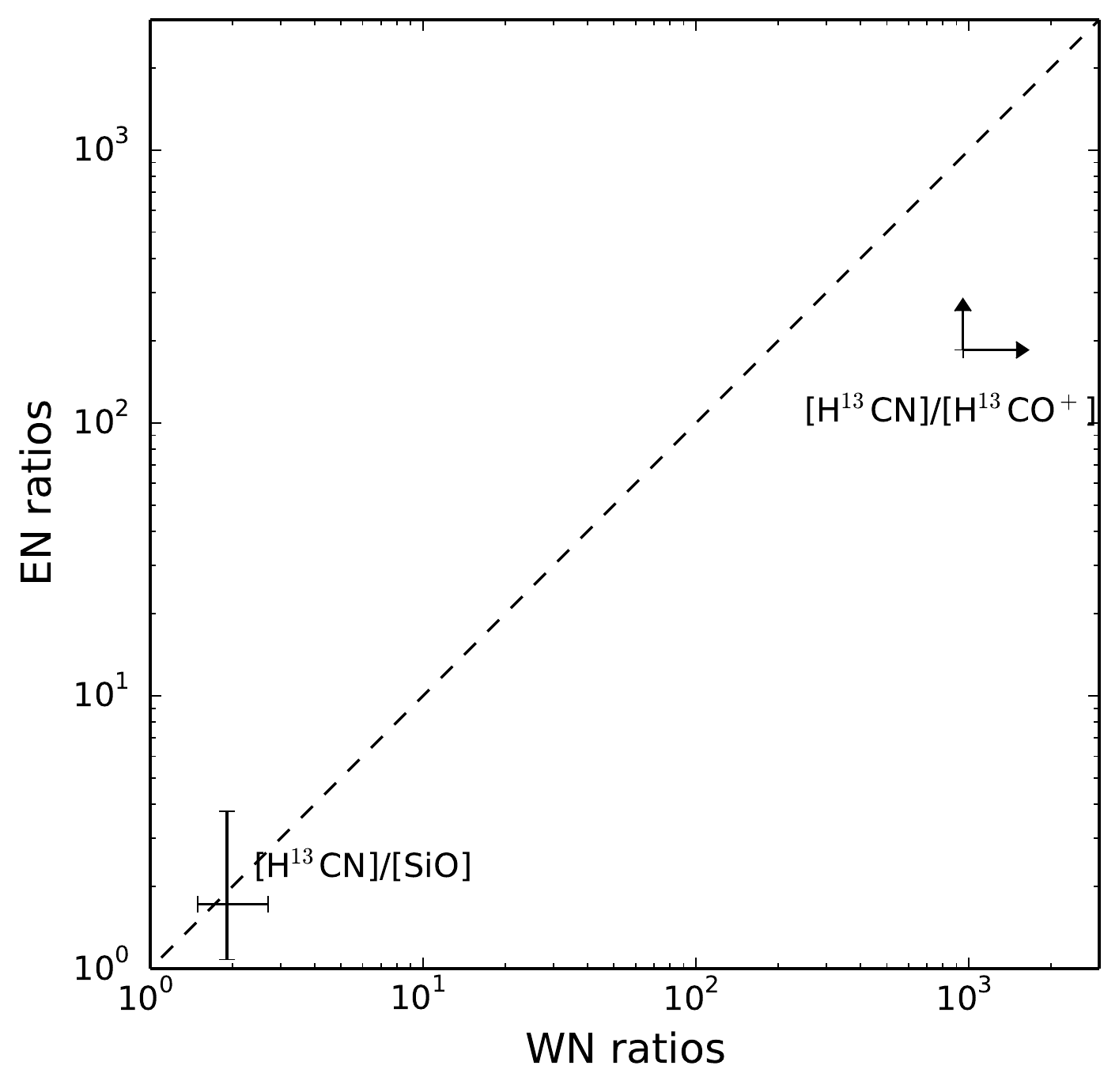}\\
\includegraphics[width=0.49\textwidth]{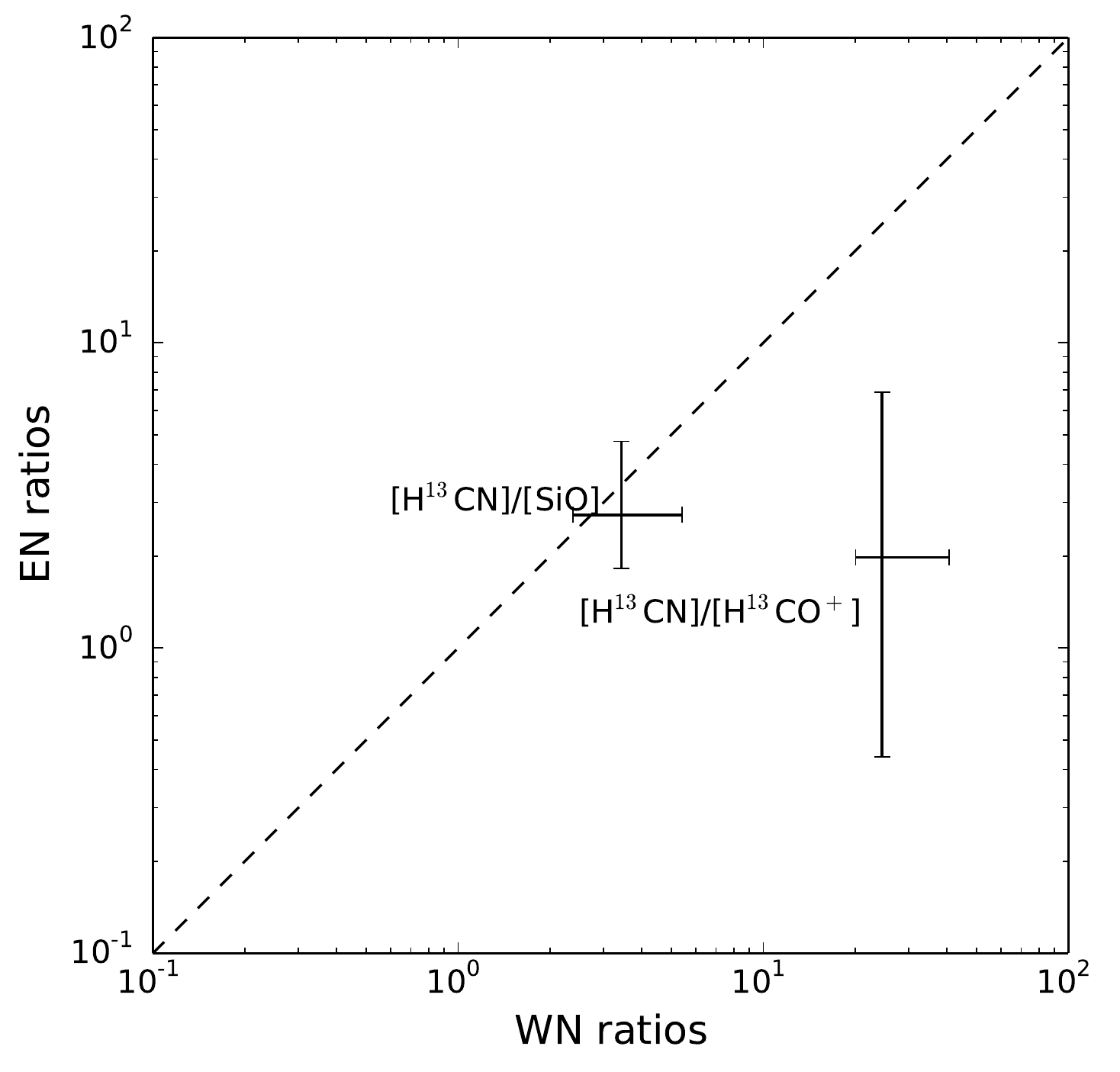}
\caption{The LVG derived abundance ratios [H$^{13}$CN]/[H$^{13}$CO$^+$] and [H$^{13}$CN]/[SiO] in the two nuclei, for case 1 (top) and case 2 (bottom). Both show a clear distinction between the two nuclei with [H$^{13}$CN]/[H$^{13}$CO$^+$] elevated in the WN. The dependent ratio [SiO]/[H$^{13}$CO$^+$] is similarly elevated in the WN.}\label{fig:abundances}
\end{center}
\end{figure}

Our LVG derived abundance ratios in case 1 and case 2 are shown in Fig.\ \ref{fig:abundances}. We omit the dependent ratio [SiO]/[H$^{13}$CO$^+$] from the plot as it can be calculated trivially from the other two, and lying so close to [H$^{13}$CN]/[H$^{13}$CO$^+$] including it would significantly obfuscate the figure.  We find an elevated [H$^{13}$CN]/[H$^{13}$CO$^+$] ratio in the WN in both case 1 and case 2. The case 1 ratios are unphysically high for the H$^{13}$CN derived $n_{\text{H}_2}$, but for $5\times10^4$\,cm$^{-3}$ the ratio of 44 in the WN is entirely physical. The case 2 ratios{, based on the upper limit of the H$^{13}$CO$^+$ contribution, yield lower limits on the [H$^{13}$CN]/[H$^{13}$CO$^+$] abundance ratios. Therefore the case 2 ratios} of $25^{+16}_{-4}$ and $2.0^{+4.2}_{-1.5}$ {in the WN and EN respectively,} confirm that the ratio is elevated in the WN {not only} with respect to the EN, {but also} with respect to the starburst dominated cases of \citet{Krips2008}. {We find a conservative $(3\sigma)$ lower limit on the ratios of 11 and 0.10 in the WN and EN respectively.} The WN [H$^{13}$CN]/[H$^{13}$CO$^+$] is consistent with [HCN]/[HCO$^+$] ratios found by \citet{Krips2008}, who found ratios between 0.01 and 50 (for starburst and AGN dominated galaxies respectively). Empirically therefore, our observations suggest that there is an energetically significant AGN in the WN of Arp\,220, while the EN is powered predominantly by a starburst. This is consistent with \citet{Engel2011} who found that the 10\,Myr old starburst is more prevalent in the EN. 

We note that it is entirely possible that we have case 1 in the WN and case 2 in the EN. The evidence for this is largely the kinematic similarity with HCO$^+(3 - 2)$ in the WN (which is far more convincing than in the EN) combined with the northwards offset of the case 2 H$^{13}$CO$^+(3 - 2)$ line in the EN, which is hard to explain in case 2 but in case 1 implies a symmetry in the SiO outflow with $\pm500$\,km\,s$^{-1}$ components. In the EN however the kinematic comparison to Sa09 makes it harder to exclude all H$^{13}$CO$^+(3 - 2)$ emission, favouring case 2.

The comparison with \citet{Krips2008} is entirely empirical, and cannot distinguish between an XDR and an MDR. Whether the trend of increasing [HCN]/[HCO$^+$] ratio with increasing AGN contribution is driven by an XDR or MDR does not affect the conclusion that an AGN is likely present in the WN. Nevertheless, the P Cygni profiles, high velocities, high gas kinetic temperatures and densities, as well as the low HN$^{13}$C$(3 - 2)$/H$^{13}$CN$(3 - 2)$ line integrated ratios, point towards extensive shock heating and shock chemistry in Arp\,220.

\subsection{Shocked Gas in a Molecular Outflow? - the origin of the species}\label{subsec:outflow}

Based on similar blueshifted absorption in \emph{both} nuclei despite their misaligned disks, Sa09 concluded that the P Cygni profiles were evidence of radial outflow in most directions, and not bound non-circular motion along the line of sight. Herschel observations have revealed P Cygni profiles in OH$^+$, H$_2$O and HF \citep{Rangwala2011}, as well as high velocity redwards components to the OH and H$_2$O lines (GA12). {The embedded P Cygni profile of H$^{13}$CN$(3 - 2)$ and the P Cygni profile of SiO$(6 - 5)$,} including the high velocity absorption (seen out to $-600$\,km\,s$^{-1}$ and troughing about $-500$\,km\,s$^{-1}$) add to the evidence for an outflow from the Arp\,220 WN. Furthermore, the high abundance of SiO and the high [H$^{13}$CN]/[H$^{13}$CO$^+$] ratio suggests that this outflow contains shocked gas. 

Notably, the high velocity blueshifted absorption seen in HCO$^+(3 - 2)$ (Sa09) and SiO$(6 - 5)$ has no analogue in blueshifted emission in any observed species, although Sa09 did see $\sim500$\,km\,s$^{-1}$ redshifted components in both HCO$^+(3 - 2)$ and CO$(3 - 2)$. As these are high high$-J$ transitions they require populated $J=2$ and $J=5$ levels respectively to absorb the continuum: future high resolution observations of SiO$(5 - 4)$ may reveal high velocity emission.

{As is discussed in detail in section \ref{sec:wn},} we suggest that H$^{13}$CN, with the small P Cygni component is prevalent throughout the CND, but also elevated in the outflow \citep[e.g.\ ][]{Martin2014}. SiO is almost exclusively present in the outflow, with a small CND component. H$^{13}$CO$^+$ would be present predominantly in the outflow (evidenced by the line profiles of Sa09), but its abundance is heavily suppressed.

\begin{figure}[!htb]
\begin{center}
\includegraphics[width=0.45\textwidth]{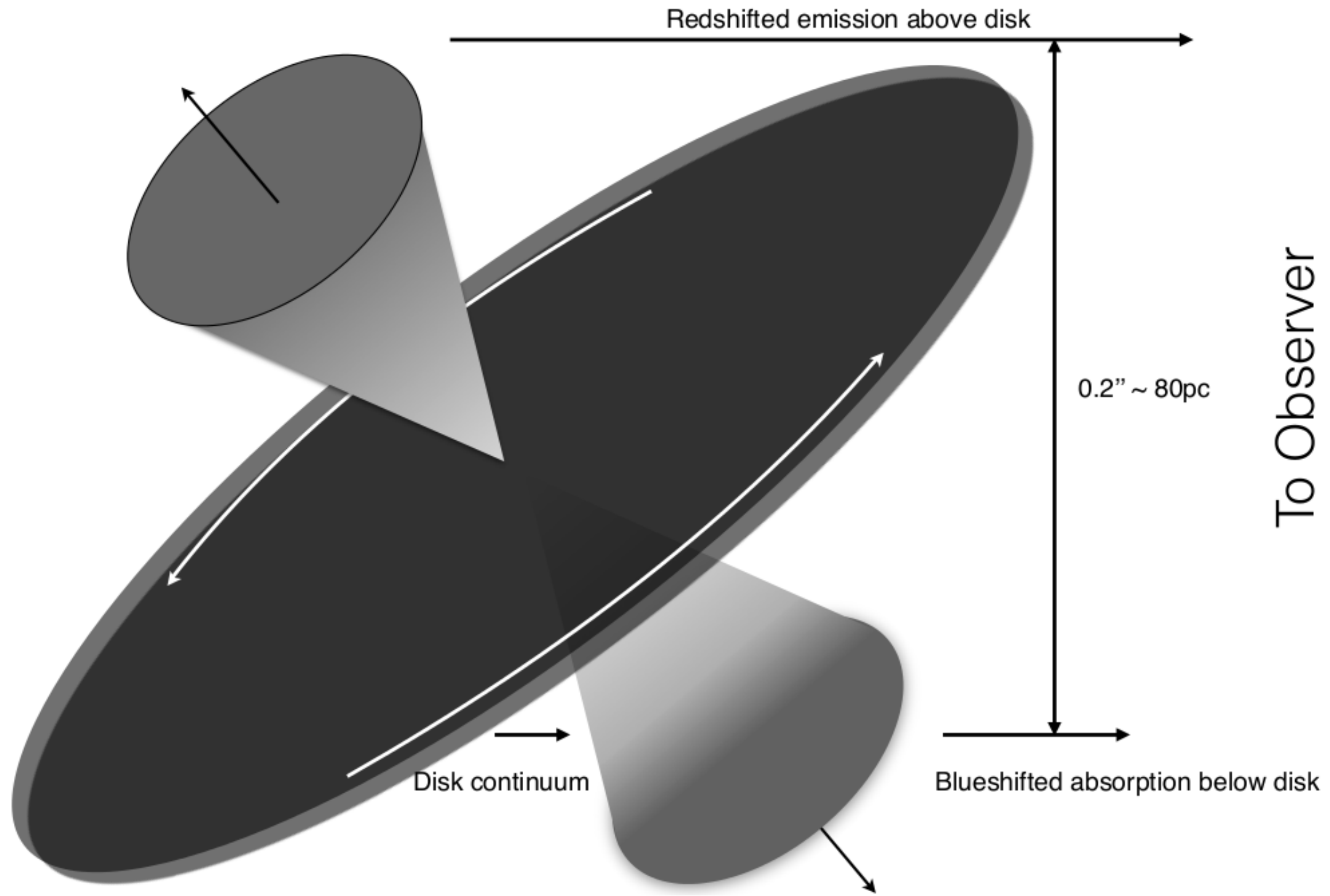}
\caption{Outline of a possible geometry for Arp\,220 WN. The highly schematic diagram shows the bipolar outflow scenario, where the north-south offsets of the emission and absorption peaks respectively is explained by a combination of dust opacity and the LoS column density being at a maximum away from the continuum centre. Redshifted emission, behind the disk, is significantly attenuated close to the continuum centre, but as it flows north and away from the disk is increasingly transmitted as thickness of intervening dust decreases. At $\sim40$\,pc north the light reaching us is maximal: further north the emitting column is decreasing or the excitation conditions are changing faster than transmitted fraction is increasing. South of the continuum centre, the absorption peaks at $\sim-40$\,pc because while the LoS column continues to increase, the continuum emission falls off rapidly further south.} \label{fig:geometry}
\end{center}
\end{figure}

\citet{Sakamoto1999} found, on the basis of 0.5$''$ CO$(2 - 1)$ observations, two counter rotating nuclear discs, each $\sim100$\,pc in diameter, embedded in a much larger rotating outer disk. They base the orientation of the WN disk (near-side southwards) on the sharp falloff of light in NICMOS images. More recent work \citep{Wilson2014} has constrained the continuum sizes to be $76\times<70$\,pc for the WN at $53^{\circ}$ inclination and $123\times79$\,pc for the EN. The bipolar outflow model presented below requires that the WN disk be oriented near-side northwards.

In stellar P Cygni profiles, the red emission and blue absorption {originate} along the same line of sight with gas leaving the star roughly isotropically. We see blue absorption across all of the southern half of the WN, i.e.\ south of the continuum peak, but it peaks $\sim40$\,pc south, while the emission peaks $\sim40$\,pc north. {In the $0$\,km\,s$^{-1}$ channel of Fig.\ \ref{fig:siomaps} the offset is even greater. This would suggest that while SiO is present mainly in the outflow, there is a disk component which is obscuring the outflow close to the continuum centre. While this could, hypothetically, be due to dust preferentially obscuring low velocity SiO, it is hard to envision the geometry of such a system that is consistent with the spatial profiles at higher velocities. }

{We propose} an outflow geometry similar to the one shown in Fig.\ \ref{fig:geometry}. In this toy model, an outflow leaves the nuclear disk roughly perpendicularly ($\sim37^{\circ}$ to the line of sight). As it advances it shocks the dusty medium, liberating SiO through grain sputtering. This leads to a maximum column depth being obtained offset from the disk centre. Since the SiO$(6 - 5)$ line is only just optically thick, the emission and absorption peaks coincide with the regions of maximum column depth, so we see emission and absorption peaks offset from one another and from the continuum. North of the disk, the redshifted emission behind the disk is attenuated by the optically thick dust between the outflow and the observer. As it flows north there is less intervening dust, increasing the transmitted fraction until there is a peak at $\sim40$\,pc north. Beyond this point, either due to column density or the gas properties the emission decreases rapidly. South of the disk, the fraction of the continuum absorbed by the outflow increases as the column density increases, until it reaches the edge of the disk and the continuum falls off rapidly, leaving nothing to be absorbed.

The requirement that the column density increases away from the continuum centre, which by number conservation requires SiO to be either swept up or sputtered from dust, would invalidate our abundance estimates of SiO as the SiO emission would be offset from the peak H$_2$ column density, implying a much greater SiO abundance than we found in section \ref{subsec:LVG}. GA12 found $3\times10^9$\,M$_\odot$ of dust in an optically thin, extended halo (650\,pc diameter) about both nuclei, confirming that there is dust available outside of the WN for this model. 

The lack of a P Cygni profile in the SiO$(2 - 1)$ line can be readily explained by the fainter continuum at 3.5\,mm as well as both the dust and the SiO line becoming optically thin at these longer wavelengths.

Sa09 noted that the low velocity of the main absorption HCO$^+(3 - 2)$ is evidence that any outflow is not driven by an AGN, but rather is either directly related to the nuclear starburst or to the merger super-wind. We see this absorption in SiO$(6 - 5)$ and H$^{13}$CN$(3 - 2)$, but given the high velocity SiO components in both HCO$^+$ and SiO and in light of the chemical evidence presented above, it could be that the outflow is only just beginning to be launched from the luminous core (be it starburst or AGN), and so has not {yet all} been accelerated to the velocities seen in more extreme outflows \citep[e.g.\ ][]{Cicone2014}. {For our proposed geometry projection effects are insignificant: correcting for the $\sim40^\circ$ angle of the outflow to the line of site only increases the velocities by a factor of 1.15.} We propose that the molecular outflow is driven by a nascent/heavily obscured AGN. The geometry of {the proposed }outflow prohibits an upper limit on the outflow age, it does provide a lower limit of $\sim6\times10^4$\,yr, assuming that the outflow was launched at $\sim500$\,km\,s$^{-1}$ and the 150\,km\,s$^{-1}$ absorption is from gas which has been slowed by shocks after escaping the WN. If, as we propose in the chemical discussion above, we have case 1 in the WN, then the SiO$(6 - 5)$ velocity slice is almost symmetric about $-25$\,km\,s$^{-1}$, with emission and absorption extending out to $+$ and $-550$\,km\,s$^{-1}$ respectively. 

The outflow structure will be investigated further in a future paper, where we will use H$_2$ kinematic models, and SiO shock chemistry models to better interpret these results. The profusion of supernova remnants and explosions (SNRs and SNe's respectively) observed in the radio \citet{Batejat2011} suggests that shocked gas and mechanically sputtered dust could be extremely prevalent in Arp\,220, and LVG codes, even with the addition of mid-IR pumping, may be insufficient to accurately model the abundances in shock-chemistry dominated regions.

\subsection{SiO Abundance}

Our intermediate SiO abundance is three orders of magnitude higher than would be expected for quiescent, Galactic, gas \citep{Martin-Pintado1992}, but one lower than seen in Galactic shocks. However, the extremely dusty Arp\,220 may have a higher quiescent gas phase silicate abundance than the Milky Way. \citet{Downes2007} found that if there is a buried AGN at the centre of the WN it must be surrounded by a very thick and dusty torus. 

Our abundances are consistent with other extragalactic studies of SiO$(2 - 1)$ and H$^{13}$CO$^+(1 - 0)$. \citet{GarciaBurillo2000,GarciaBurillo2001,GarciaBurillo2010} and \cite{Usero2006} found SiO abundances in the range $10^{-9} - 10^{-10}$ in the circumnuclear disks of NGC\,253, M\,82, NGC\,1068 and IC\,342 respectively. These SiO abundances were derived from the SiO/H$^{13}$CO$^+$ ratios and assumed H$^{13}$CO$^+$ abundances. For NGC\,1068 the abundance was also found from the SiO$(3 - 2)$/SiO$(2 - 1)$ ratio and was consistent with that derived from an assumed H$^{13}$CO$^+$ abundance. In NGC\,253 the abundance increases from $1\times10^{-10}$ in the central nuclear disk (CDN) to $5\times10^{-10}$ in the nuclear outflow. M\,82 has a particularly complex morphology with a low abundance in the galactic disk ($1\times10^{-11}$), increasing to $1\times10^{-10}$ in a supershell and to $1\times10^{-9}$ in a silicate chimney. NGC\,1068 has abundances in the range $1 - 5\times10^{-9}$ in the CND, while IC\,342 has a CND abundance of $2\times10^{-10}$ but this rises to $1\times10^{-9}$ in the spiral arms. Here, in case 2, we find abundances of $2.2^{+0.28}_{-0.24}\times10^{-9}$ and $2.9^{+4.2}_{-0.6}\times10^{-9}$ in the WN and EN CNDs respectively. These uncertainties are nominal, assuming the $T_\text{k}=150$\,K, $T_\text{bg}=45$\,K and $n_{\text{H}_2}=2.1\times10^6$\,cm$^{-3}$. For different conditions the abundances can vary by more than a dex.

Our SiO column density is $\sim2-7\times10^{16}$\,cm$^{-2}$, while \citet{Martin2011} found, based on LTE modelling of the SiO$(5 - 4)$ transition, $\sim1\times10^{14}$\,cm$^{-2}$, $100\times$ lower. This is particularly important since the absorption we see in the SiO$(6 - 5)$ line requires that the $J=5$ level is populated. This difference may be due to the large beam ($8.5''\times7.0''$, $300\times$ the area of our beam) of the SMA configuration used for the line survey of \citet{Martin2011}. If the SiO emission is more compact than other lines this would lead to significant dilution, and a reduced column estimate. Higher resolution observations of the SiO$(5 - 4)$ line combined with radiative excitation modelling are needed to obtain a better estimate of the column density.

\subsection{The Dense Gas Fraction}

\citet{Engel2011} found a dynamical mass within 100\,pc of the WN of $6.3\times10^9$\,$M_{\odot}$. Using the H$_2$ density found from our LVG models, including a factor of 1.3 for helium and assuming a single core and a WN disk $75\times70\times10$\,pc,  implies a gas mass $\gtrsim1.7\times10^{10}$\,$M_{\odot}$, about $3\times$ the dynamical mass, implying that the HCN and H$^{13}$CN disk emission is from multiple small, dense cores distributed across the nuclear disk. Repeating this exercise for the EN, dynamical mass $5.8\times10^9$\,$M_{\odot}$ leads to the same conclusion. \citet{Downes1998} suggested gas masses of $6\times10^8$\,$M_{\odot}$ and $1.1\times10^9$\,$M_{\odot}$ for the WN and EN respectively, which would suggest that the HCN dense cores cannot have a filling factor greater than 0.1 in the WN disk. 

A shell of gas about a single ``hot core'', powered by the 5\,pc radius 400\,K dust continuum found by GA12, is also possible on dynamical grounds, with a shell 0.1\,pc thick having a mass $\sim3\times10^6$\,M$_{\odot}$. 

\subsection{The HNC/HCN ratio}

\citet{Aalto2007} found the line integrated brightness temperature ratio HNC$(3 - 2)$/HCN$(3 - 2) = 1.9\pm0.3$. This was explained to be due to either mid-IR pumping, where HNC is preferentially pumped over HCN by two orders of magnitude, or XDR chemical effects. We found $T_\text{B}{\rm d}v$ ratios for HN$^{13}$C$(3 - 2)$/H$^{13}$CN$(3 - 2) = 0.71\pm0.04$ in the WN, and $0.66\pm0.04$ in the EN, and $T_{\text{B,peak}}$ ratios of $0.689\pm0.023$ and $0.453\pm0.010$ in the WN and EN respectively. Both measures present HN$^{13}$C as under-luminous {with respect to} H$^{13}$CN. 

The WN value is uncertain due to the missing blue component of the line, estimated generously to be $\lesssim25\%$ but not included in our results, but even allowing for a missing 40\% on the HN$^{13}$C$(3 - 2)$ line the ratio is still $\sim1$, as a very generous upper limit. The EN value on the other hand is much more reliable, due to the additional redshift of the EN shifting almost all of the line into our spectrum. This discrepancy between the optically thick HNC/HCN and their much less optically thick $^{13}$C isotopologues would suggest that the significantly different ratio is due to optical depth effects. Not only can we see deeper into the high density environments, but the molecules will be less self-shielded against mid-IR pumping. This would lead to us seeing a greater fraction of the HNC and HCN which is hidden in dense cores, and protected against mid-IR pumping. In the ``raisin roll'' model of \citet{Aalto2007}, the mid-IR field in Arp\,220 drives widespread HNC emission while the optical depth of the HCN emission, primarily from dense cores, reduces the observed HCN intensity below what would be expected from its true [HNC]/[HCN] abundance ratio and the HNC intensity, leading to the high observed HNC/HCN ratio. When we observe the isotopologues however, we see a truer representation of the [HNC]/[HCN] abundance ratio as we no longer have optically thick thermal emission from the dense cores. 

We note that there might be some concern with extending the logic applicable to the carbon-12 isotopologues to the carbon-13 isotopologues. Measurements of HN$^{13}$C transitions are few and far between, but data from \citet{Harris2008}, and the fact that the energy of the first excited bending mode (which is relevant for mid-IR pumping) has an energy proportional to $\rho^{-0.5}$ where $\rho$ is the reduced mass, suggest any differences should be on the order of 1\%. Any arguments based on the susceptibility to mid-IR pumping applicable to the HNC/HCN ratio should also apply to the HN$^{13}$C/H$^{13}$CN ratio.

\section{Conclusions}

We have presented spectral imaging of the prototypical ULIRG Arp\,220 at 3.5\,mm and 1.2\,mm from the PdBI. We observe complex molecular lines including H$^{13}$CN, H$^{13}$CO$^+$, HN$^{13}$C and SiO. These results are consistent with previous millimetre studies of Arp\,220, while also revealing shocked gas in the molecular outflow, building upon the molecular outflow observed by Sa09 in HCO$^+$. A chemical and kinematic analysis suggests that we have non-detections of H$^{13}$CO$^+$ in the WN, but detections in the EN. Further high resolution work is needed to study the kinematically complex gas in Arp\,220, in particular the SiO transition ladder at frequencies uncontaminated by H$^{13}$CO$^+$. Observations resolving the molecular emission within the disks are essential, and would allow us to identify whether the emission from different species is segregated to the extent we propose above. Chemical shock models, combined with kinematic and radiative transfer models may be able to shed more light on the environments traced by the complex line profiles we have observed.

\begin{itemize}
\item We have used HCN$(3 - 2)$ and HCN$(1 - 0)$ single dish data from the literature and our H$^{13}$CN$(3 - 2)$ and H$^{13}$CN$(1 - 0)$ interferometric data with LVG models to constrain the HCN environment in Arp\,220, finding dense ($n_{\text{H}_2}=2.1\pm0.6\times10^6$\,cm$^{-3}$), hot ($T_{\text{k}}=150 - 450$\,K) gas.
\item {Working in two limiting cases, where we attribute the reddest components of fitted line profiles to either SiO or H$^{13}$CO$^+$,} we find significant chemical differences between the two nuclei, {as well as an elevated [H$^{13}$CN]/[H$^{13}$CO$^+$] abundance ratio, in both cases}. {The lowest [H$^{13}$CN]/[H$^{13}$CO$^+$] abundance ratios correspond to the case where the line component is attributed to H$^{13}$CO$^+$, and we find } abundance ratios [H$^{13}$CN]/[SiO] = $0.87^{+0.11}_{-0.10}$ and $0.67^{+0.09}_{-0.14}$ and [H$^{13}$CN]/[H$^{13}$CO$^+$] = $25^{+16}_{-4}$ and $2.0^{+4.2}_{-1.5}$ in the WN and EN respectively. {Therefore the lower limit on the [H$^{13}$CN]/[H$^{13}$CO$^+$] abundance ratios, in the limit of a maximal H$^{13}$CO$^+$ contribution, are 11 and 0.10 in the WN and EN respectively.} If, as we believe is more likely, case 1 {(no H$^{13}$CO$^+$ detections)} is a more accurate description of the WN the [H$^{13}$CN]/[H$^{13}$CO$^+$] abundance ratio is significantly larger, most likely $>50$.
\item The elevated [H$^{13}$CN]/[H$^{13}$CO$^+$] ratio in the WN, compared to the EN and to SB galaxies, suggests (empirically) an energetically significant AGN contribution to the WN luminosity.
\item We find a ``normal'' $(0.1 - 1.0)$ brightness temperature ratio for HN$^{13}$C$(3 - 2)$/H$^{13}$CN$(3 - 2)$ in both nuclei, suggesting that the elevated HNC/HCN intensity ratio seen by \citet{Aalto2007} is due predominantly to mid-IR pumping, and not an XDR. Combined with the evidence for an AGN, this implies extensive mechanical heating and shock chemistry in Arp\,220.
\item We suggest that HCO$^+$, SiO$(6 - 5)$ and a fraction of H$^{13}$CN$(3 - 2)$ are tracing a dense molecular outflow with a large shocked component, and present a toy model geometry which can explain the north-south offset we observe in the SiO$(6 - 5)$ emission and absorption, which independently reproduces the size of the WN continuum source seen by \citet{Wilson2014}.
\end{itemize}


\acknowledgments{\footnotesize Acknowledgements: We thank the anonymous referee their extensive and patient comments, which greatly improved this paper. This research is supported by an STFC PhD studentship. AU and PP acknowledge support from Spanish grants AYA2012-32295 and FIS2012-32096. TRG acknowledges support from an STFC Advanced Fellowship. SGB acknowledges support from Spanish grants AYA2010-15169 and AYA2012-32295 and from the Junta de Andalucia through TIC-114 and the Excellence Project P08-TIC-03531. AF and SGB acknowledge support from Spanish MICIN program CONSOLIDER IMAGENIO 2010 under grant 'ASTROMOL' (ref.\ CSD2009-00038). EGA is a Research Associate at the Harvard-Smithsonian CfA, and thanks the Spanish Ministerio de Econom\'{\i}a-Competitividad for support under projects AYA2010-21697-C05-0 and FIS2012-39162-C06-01. Basic research in IR and millimetre wave astronomy at NRL is funded by the US ONR. We thank Kazushi Sakamoto for kindly providing the HCO$^+(3 - 2)$ spectrum used in Fig.\ \ref{fig:sakacomp}. We would also like to thank S.\ Viti, M.\ Matsura and Z.\ Zhang for informative and productive discussions. This work was based on observations carried out with the IRAM PdBI, supported by INSU/CNRS(France), MPG(Germany), and IGN(Spain), and includes observations made with the NASA/ESA Hubble Space Telescope, and obtained from the Hubble Legacy Archive, which is a collaboration between the Space Telescope Science Institute (STScI/NASA), the Space Telescope European Coordinating Facility (ST-ECF/ESA) and the Canadian Astronomy Data Centre (CADC/NRC/CSA). This publication also makes use of data products from the Wide-field Infrared Survey Explorer, which is a joint project of the University of California, Los Angeles, and the Jet Propulsion Laboratory/California Institute of Technology, funded by the National Aeronautics and Space Administration.}

\FloatBarrier


\bibliography{Arp220.bib}

\appendix

\section{A. Continuum Analysis}\label{sec:cont_details}
\FloatBarrier
{The GILDAS task UV\_FIT was used to fit two elliptical gaussians to the 1.2\,mm and 3.5\,mm continuum UV tables. Elliptical gaussians were chosen to facilitate meaningful comparisons with the literature. These produced good fits, with flux densities consistent with those obtained from the imaged data. Elliptical disks fit the data with the same rms.\ error, while point sources provide inferior fits. The residual UV table, after subtraction of the fitted components, was imaged and is consistent with noise. The peak intensity in the imaged residual data cube is 4\% of that of the imaged continuum, at the same level as the main beam side lobes, and is not associated with any structure in the map. The fitted parameters, with nominal errors, are presented in Table \ref{tab:continuum}, along with comparable continuum measurements from the literature.}

\subsection{Synchrotron subtraction}

\noindent\citet{Norris1988, Scoville1991} found radio synchrotron emission of 36\,mJy at 1.3\,cm with a spectral index of $-0.75$. We therefore predict a synchrotron flux density of $14\pm1$\,mJy for the two nuclei combined at 3.5\,mm and $6\pm1$\,mJy at 1.2\,mm, giving combined thermal flux densities of $14\pm2$\,mJy and $265\pm20$\,mJy at 3.5\,mm and 1.2\,mm respectively. Assuming the same non-thermal/thermal flux density ratio for both nuclei we obtain 8\,mJy and 6\,mJy for the WN and EN respectively at 3.5\,mm and 196\,mJy and 69\,mJy for the WN and EN respectively at 1.2\,mm (Table \ref{tab:continuum}). The 1.2\,mm non-thermal flux density is much less than the error in the absolute flux calibration.

\begin{table}
\begin{center}
\caption{Continuum data for Arp\,220 in the millimetre regime. Data from \citet{Martin2011} are not included as the resolution of the observations precludes individual nuclear size or position measurements.}\label{tab:continuum}
\begin{tabular}{l c c c c c} \hline \hline
Source&$\nu_0$&RA&Dec&Size\footnote{Numbers in parenthesis are the nominal uncertainties from UV fitting.}&$S_{\nu}$\footnote{Numbers in parenthesis are the estimated synchrotron subtracted flux densities. The errors are dominated by the 10\% estimated uncertainty in the absolute flux calibration.}\\ 
& [GHz] & & &[arcsec]&[mJy]\\\hline \hline
This work: WN&\multirow{2}{*}{86.6}&$15^{\text{h}}34^{\text{m}}57^{\text{s}}23$&$23^{\circ}30'11''50$&$0.53(0.06)\times0.27(0.05)$&$17(8)\pm2$\\[0.1cm]
This work: EN& &$15^{\text{h}}34^{\text{m}}57^{\text{s}}30$&$23^{\circ}30'11''32$&$0.43(0.09)\times0.32(0.07)$&$11(6)\pm1$\\[0.1cm]
\cite{Scoville1991}: both&110.3&$--$&$--$&$--$&$30\pm3$\\[0.1cm]
\citet{Downes2007}: WN&230&$15^{\text{h}}34^{\text{m}}57^{\text{s}}2226$&$23^{\circ}30'11''46$&$0.19\times0.13$&$106\pm2$\\[0.1cm]
\citet{Scoville1997}: both&229.4&$--$&$--$&$1.07\times0.63$&$192\pm20$\\[0.1cm]
This work: WN&\multirow{2}{*}{260}&$15^{\text{h}}34^{\text{m}}57^{\text{s}}22$&$23^{\circ}30'11''50$&$0.194(0.004)\times0.1546(0.0021)$&$200(196)\pm20$\\[0.1cm]
This work: EN& &$15^{\text{h}}34^{\text{m}}57^{\text{s}}29$&$23^{\circ}30'11''34$&$0.258(0.008)\times0.183(0.007)$&$71(69)\pm7$\\[0.1cm]
Sa09: WN&\multirow{2}{*}{273}&\multirow{2}{*}{$--$}&\multirow{2}{*}{$--$}&\multirow{2}{*}{$--$}&$200\pm20$\\[0.1cm]
Sa09: EN& & & & & $70\pm7$\\[0.1cm]
\citet{Aalto2009}:WN&\multirow{2}{*}{277}&$15^{\text{h}}34^{\text{m}}57^{\text{s}}230$&$23^{\circ}30'11''50$&$--$&$152\pm19$\\[0.1cm]
\citet{Aalto2009}:EN& &$15^{\text{h}}34^{\text{m}}57^{\text{s}}310$&$23^{\circ}30'11''40$&$--$&$57.6\pm16$\\[0.1cm]
\multirow{2}{*}{\citet{Sakamoto2008}: WN}&344.6&\multirow{2}{*}{$15^{\text{h}}34^{\text{m}}57^{\text{s}}215$}&\multirow{2}{*}{$23^{\circ}30'11''45$}&$0.28\times0.22$&$380\pm60$\\[0.1cm]
 &349.6& & &$0.16\times0.13$&$360\pm50$\\[0.1cm]
\multirow{2}{*}{\citet{Sakamoto2008}: EN}&344.6&\multirow{2}{*}{$15^{\text{h}}34^{\text{m}}57^{\text{s}}285$}&\multirow{2}{*}{$23^{\circ}30'11''27$}&$0.31\times0.25$&$200\pm30$\\[0.1cm]
 &349.6& & &$0.27\times0.14$&$190\pm30$\\[0.1cm]
\citet{Wilson2014}: WN&\multirow{2}{*}{691}&$15^{\text{h}}34^{\text{m}}57^{\text{s}}22$&$23^{\circ}30'11''5$&$0.21\times\leq0.19$&$1810\pm270$\\[0.1cm]
\citet{Wilson2014}: EN& &$15^{\text{h}}34^{\text{m}}57^{\text{s}}29$&$23^{\circ}30'11''3$&$0.34\times0.2$&$1510\pm230$\\ \hline
\end{tabular}
\end{center}
\end{table}

\subsection{SED fitting}

Using data from the literature including the fluxes in Table \ref{tab:continuum}, \citet{Eales1989} and \citet{Rangwala2011}, as well as Wide-field Infrared Survey Explorer (WISE) measurements of Arp\,220, we fitted two grey body models to the SED (Fig.\ \ref{fig:sakacomp}). The WISE flux densities were corrected assuming a 100\,K blackbody, as per \citet{Wright2010}. We fitted the function:

\begin{align}
S_{i,\nu} &= \left(1-e^{-\left(\frac{\nu}{\nu_{0,i}}\right)^{\beta_i}}\right)\frac{A_i\nu^3}{e^{\frac{h\nu}{kT_i}}-1},\\
S_{\nu} &= S_{1,\nu} +S_{2,\nu}, 
\end{align}

\noindent where $\nu_0$ is the turnover frequency where the dust becomes optically thin, $\beta$ is the dust emissivity spectral index, $T$ is the dust temperature and $A$ is an absolute normalisation. We performed a naturally weighted fit of these 8 free parameters to our SED, finding $\nu_{0,1} = 1.6\times10^3$\,GHz ($190\mu$m), $\beta_1=1.5$, $A_1=4.9\times10^{-32}$\,mJy\,Hz$^{-3}$ and $T_1=62$\,K for the main component, and $\nu_{0,2} = 1.0$\,GHz, $\beta_2=1.6$, $A_2=3.8\times10^{-37}$\,mJy\,Hz$^{-3}$ and $T_2=340$\,K. This fit has a reduced $\chi^2$ (45 degrees of freedom) of 6.8. We note that $\nu_{0,2}$ is unphysically low, suggesting a blackbody spectrum. This should not be interpreted literally, since we are combining four known dust components (GA12) into a simple 2 element model. As we are only interested in $\nu_{0,1}$ and  $\beta_1$ this is sufficient for our purposes. We find a {global} optical depth at 258\,GHz of 0.06. For comparison, GA12 developed a complex, 4 component model of the dust in Arp\,220 including a hot component, an extended component and models for the WN and EN. They found the greatest optical depth in the WN, with $\tau_{200\mu{\rm m}}=6-12$. Assuming ``worst case'' parameters: $\tau_{200\mu{\rm m}}=12$ and $\beta_{\text{WN}}=1 - 2$ gives $\tau_{258{\rm GHz}} = 2 - 0.4$ and $\tau_{86{\rm GHz}} = 0.7 - 0.04$. It seems reasonable then to assume that for our 3.5\,mm observations the dust is almost entirely optically thin, although it may still be optically thick at 1.2\,mm. Our optical depth is significantly lower ($7 - 33\times$ lower) than that obtained by GA12 for the WN. This is probably due to our much simpler fit and averaging across both nuclei. Our lower best fit temperature (62\,K cf.\ $90 - 130$\,K) suggests that the low optical depth, high luminosity extended dust component ($\tau_{200\mu{\rm m}}=0.17$, $T_{\text{d}}=90 - 40$\,K, diameter $\sim650$\,pc $\simeq1.7''$ (GA12)) could be having a significant effect upon our fit.

\begin{figure}
\begin{center}
\includegraphics[width=0.5\textwidth]{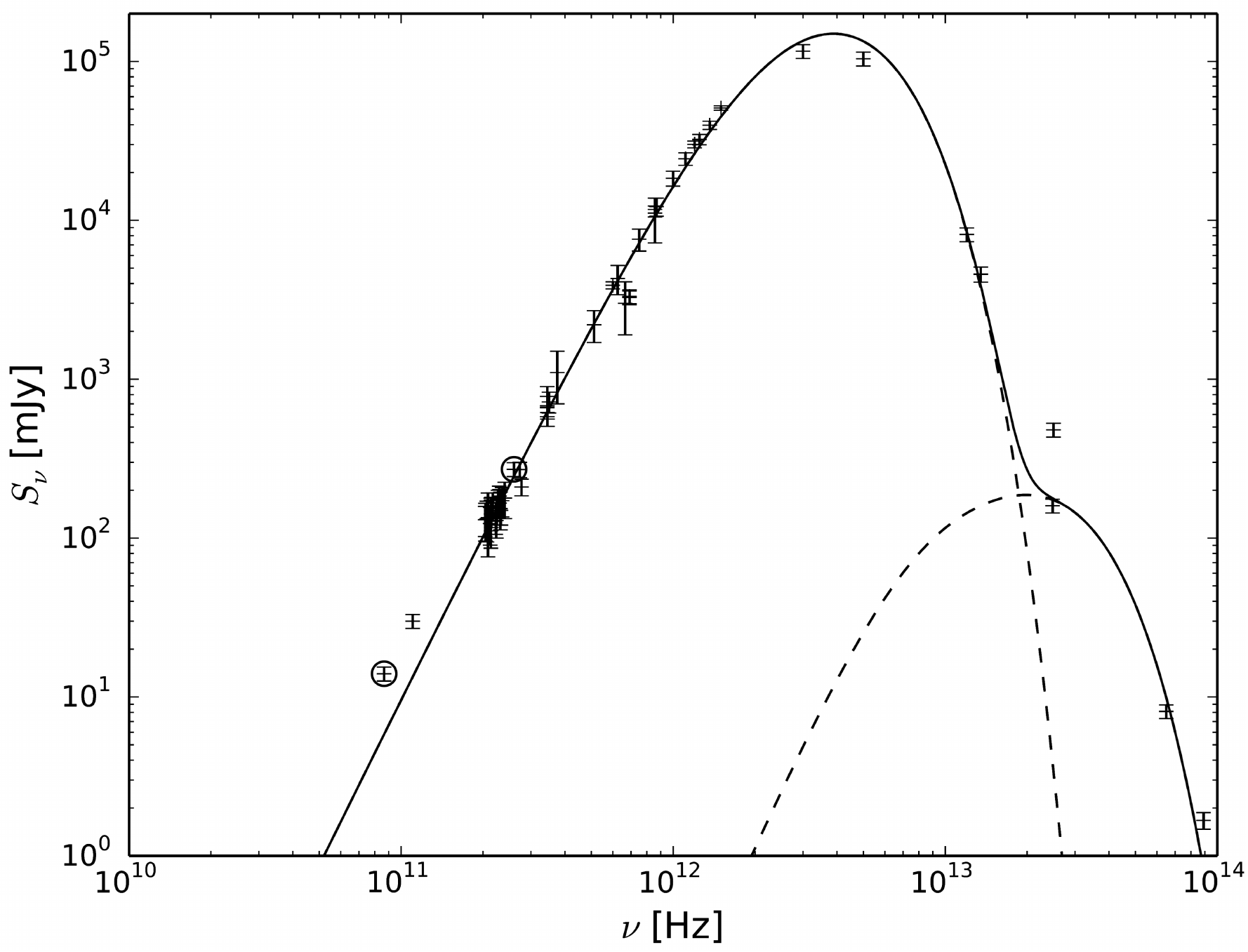}
\caption{Continuum fluxes for Arp\,220 from Table \protect\ref{tab:continuum}, \protect\citet{Martin2011}, \protect\citet{Rangwala2011} and WISE data. We fit the data with two grey bodies. Our continuum measurements are circled. We find best fit parameters $\nu_{0,1} = 1.6\times10^3$\,GHz ($190\mu$m), $\beta_1=1.5$, $A_1=4.9\times10^{-32}$\,mJy\,Hz$^{-3}$ and $T_1=62$\,K for the main component, and $\nu_{0,2} = 1.0$\,GHz, $\beta_2=1.6$, $A_2=3.8\times10^{-37}$\,mJy\,Hz$^{-3}$ and $T_2=340$\,K for the hotter component.}\label{fig:fluxcomp}
\end{center}
\end{figure}

\section{B. Line Fitting}\label{app:B}
{Since we have at least one pair of blended lines, and possibly two, we used multiple component Gaussian fits to parameterise the observed lines and to provide a handle for spectral deblending. Each of the boxes in the middle rows of Figs.\ \ref{fig:spec1}-\ref{fig:spec4} presents an individual fitting, which ranged from single to up to five components. We checked that randomised starting parameters ($\pm50$\%) converged to the same final fits, demonstrating that the fits are robust. The number of fitting components was chosen such that $\chi^2_\nu$ was closest to unity without having consecutive residuals above $\pm1\sigma$. The rms.\ uncertainty in each channel was included in the fitting.}

We include below in Table \ref{tab:line_fits} the details of the Gaussian fitting to the spectra. These Gaussians should not be interpreted literally as the line profiles, but rather as well defined parametric descriptions of lines' peaks and areas. Reduced $\chi^2$ is between 0.2 and 0.7 for all fits.

{Uncertainties in the fitting were derived from the covariance matrixes. For functions making use of more than one fitting parameter (such as the flux of a component or the total flux of a line summed over all components) the correlations in the covariance matrices were taken into account when calculating uncertainties.}

\begin{table}
\begin{center}
\caption{Details of Gaussian line fitting. Errors are determined from the fitting covariance matrices.}\label{tab:line_fits}
\begin{tabular}{l c c c c c} 
	\multicolumn{6}{c}{WN}\\ \hline \hline
Line & Component & $S_\nu$ & Centre\footnote{In the sky frame.} & FWHM & Area\footnote{Errors in parentheses include the 10\% calibration error. These errors also account for the correlations between the component peaks and widths.} \\
& & [mJy] & [GHz] & [km\,s$^{-1}$] & [Jy\,km\,s$^{-1}$] \\ \hline \hline
SO$(6_6 - 5_5)$ & 1 & $32\pm20$ & 253.69 & $350\pm300$ & $12\pm18(18)$\\
HC$^{15}$N$(3 - 2)$ & 1 & $30\pm60$ & 253.55 & $230\pm40$ & $6\pm15(15)$\\
SiO$(6 - 5)$ & 1 & $39.9\pm2.5$ & 255.589 & $223\pm23$ & $9,5\pm0.5(1.1)$\\
& 2 & $66.8\pm3.0$ & 255.800 & $153\pm11$ & $10.9\pm0.9(1.4)$\\
& 3 & $-20.8\pm3.8$ & 256.000 & $97\pm36$ & $-2.1\pm0.9(0.9)$\\
& 4 & $20.3\pm4$ & 256.099 & $90\pm40$ & $2.0\pm1.1(1.1)$\\
& 5 & $-15.4\pm2.7$ & 256.280 & $190\pm50$ & $-3.1\pm0.3(0.5)$\\
H$^{13}$CN$(3 - 2)$ & 1 & $173\pm19$ & 254.304 & $159\pm18$ & $22\pm6(6)$\\
& 2 & $69\pm4$ & 254.600 & $141\pm13$ & $10.3\pm1.5(1.8)$\\
& 3 & $110\pm15$ & 254.433 & $170\pm30$ & $20\pm7(7)$\\
HN$^{13}$C$(3 - 2)$ & 1 & $42\pm20$ & 256.789 & $330\pm300$ & $15\pm20(20)$\\
& 2 & $90\pm40$ & 256.617 & $240\pm40$ & $22\pm18(18)$\\ 
H$^{13}$CN$(1 - 0)$ & 1 & $12.1\pm1.0$ & 84.806 & $310\pm30$ & $4.0\pm0.07(0.4)$\\
SiO$(2 - 1)$ & 1 & $10.4\pm0.9$ & 85.288 & $410\pm40$ & $4.49\pm0.12(0.5)$\\
& 2 & $1.5\pm1.2$ & 85.137 & $220\pm220$ & $0.35\pm0.07(0.08)$\\ \hline

\multicolumn{6}{c}{EN}\\ \hline \hline
Line & Component & $S_\nu$ & Centre$^{\text{a}}$ & FWHM & Area$^{\text{b}}$ \\
& & [mJy] & [GHz] & [km\,s$^{-1}$] & [Jy\,km\,s$^{-1}$] \\ \hline \hline
SO$(6_6 - 5_5)$ & 1 & $15\pm4$ & 253.636 & $201\pm28$ & $3.2\pm1.2(1.2)$\\
HC$^{15}$N$(3 - 2)$ & 1 & $16.9\pm2.3$ & 253.460 & $170\pm60$ & $3.1\pm1.1(1.2)$\\
SiO$(6 - 5)$ & 1 & $12\pm3$ & 255.460 & $120\pm40$ & $1.63\pm0.19(0.25)$\\
& 2 & $20.9\pm2.5$ & 255.697 & $220\pm30$ & $4.87\pm0.19(0.5)$\\
H$^{13}$CN$(3 - 2)$ & 1 & $52.0\pm2.7$ & 254.256 & $182\pm12$ & $10.06\pm0.26(1.0)$\\
& 2 & $20\pm3$ & 254.463 & $125\pm26$ & $2.60\pm0.18(0.31)$\\
HN$^{13}$C$(3 - 2)$ & 1 & $23\pm4$ & 256.451 & $133\pm27$ & $3.2\pm1.1(1.2)$\\
& 2 & $18.2\pm2.3$ & 256.655 & $270\pm60$ & $5.2\pm0.6(0.8)$\\ 
H$^{13}$CN$(1 - 0)$ & 1 & $2.6\pm0.6$ & 84.751 & $180\pm50$ & $0.497\pm0.018(0.053)$\\
SiO$(2 - 1)$ & 1 & $0.9\pm0.5$ & 85.165 & $240\pm200$ & $0.22\pm0.06(0.06)$\\
& 2 & $3.2\pm0.6$ & 85.248 & $170\pm40$ & $0.56\pm0.16(0.17)$ \\
& 3 & $1.5\pm0.8$ & 85.339 & $90\pm60$ & $0.145\pm0.013(0.019)$\\ \hline
\end{tabular}
\end{center}
\end{table}

\section{C. H$^{13}$CO$^+(3 - 2)$ estimation}\label{app:h13co}

Quantitatively, we use the HCO$^+(3 - 2)$ data to estimate the predicted H$^{13}$CO$^+(3 - 2)$ contribution. From Sa09 (data kindly provided by Sakamoto) we have the peak brightness temperature, optical depth and FWHM of the best fit Gaussian to the emission region of the line profile. The optical depth is found for the absorbing column, where the observed fractional absorption,
\begin{equation}
f_a = \frac{I_{\text{out}}}{I_{\text{in}}} - 1,
\end{equation}
gives the optical depth of the absorption component:
\begin{equation}
\tau_a = -\log(1+f_a).
\end{equation}

For an integrated line emission intensity:

\begin{equation}
W = \int T_{\text{B}} \text{d}v \simeq 1.064T_{\text{B,peak}}v_{\text{FWHM}},
\end{equation}
we have an upper level column density:
\begin{equation}
N_u = \frac{8\pi k \nu^2 W}{hc^3A}\left(\frac{\Delta\Omega_a}{\Delta\Omega_s}\right)\left(\frac{\tau}{1-e^{-\tau}}\right),
\end{equation}
where $A$ is the Einstein coefficient and $\Delta\Omega_a$ and $\Delta\Omega_s$ are the beam and source sizes respectively. If we assume that [HCO$^+$]/[H$^{13}$CO$^+$]$\simeq 60$, that the two isotopologues trace similar regions and share similar excitation properties then if the source sizes are the same for both isotopologues or if they are beam filling sources we have:
\begin{equation}
W_{13} = \frac{1}{60}\frac{\nu_{12}^2}{\nu_{13}^2}\frac{\tau_{12}}{1-e^{-\tau_{12}}}\frac{1-e^{-\tau_{13}}}{\tau_{13}}W_{12}.
\end{equation}
Here the subscripts 12 and 13 refer to the H$^{12}$CO$^+$ and H$^{13}$CO$^+$ properties respectively, and $\nu$ is the frequency of the $(3 - 2)$ rotational transition. For the WN peak $T_{\text{B}}$ and FWHM we have $W_{12} \sim 2500$\,K\,km\,s$^{-1}$, so with $\tau_{12}=0.36$ and our fitted width of 180\,km\,$^{-1}$ to the emission region of the suspected H$^{13}$CO$^+(3 - 2)$ line we have a predicted peak $T_{\text{B}} = 0.27\text{\,K}\simeq2.9$\,mJy\,beam$^{-1}$. I.e.\ the peak contribution from H$^{13}$CO$^+(3 - 2)$ is expected to be $\simeq1\sigma$. It is worth noting that the HCO$^+(3 - 2)$ and SiO$(6 - 5)$ FWHMs are significantly larger (closer to 500\,km\,s$^{-1}$), suggesting a lower peak, but also that the optical depth has been derived from absorption, so the true optical depth is likely larger, raising the peak. The estimate is therefore quite uncertain, but demonstrates the possibility of H$^{13}$CO$^+(3 - 2)$  being seen in the WN. In the EN, Sa09 found a significantly higher HCO$^+(3 - 2)$ effective optical depth of 3.7. This yields an H$^{13}$CO$^+(3 - 2)$ peak intensity of 10.5\,mJy\,beam$^{-1}$ in the EN, which is readily detected, and is in fact very similar to the observed peak near the line centre.

\end{document}